\documentclass[10pt,oneside,a4paper]{amsart}
\usepackage{graphicx,epic}
\usepackage{amsfonts, amsmath, amssymb,color}

\newcommand{\1}{H^{1}\big(\R^3\big)}
\newcommand{\0}{{L^2(\Omega)}}

\newcommand{\B}{\mathbb{B}}

\newcommand{\s}{\int_{\mathbb{R}^3}}

\newcommand{\ys}{\left\langle}
\newcommand{\ym}{\right\rangle}

\newtheorem{thm}{Theorem}[section]
\newtheorem{cor}[thm]{Corollary}
\newtheorem{lem}[thm]{Lemma}
\newtheorem{prop}[thm]{Proposition}

\newtheorem{rem}[thm]{Remark}
\newcommand{\G}{\mathrm{I}\hspace{-2.0pt}\Gamma}% avant -2.6pt
\newcommand{\U}{\mathbb{U}}
\newcommand{\K}{\mathbb{K}}

\newcommand{\V}{\mathbb{V}}
\newcommand{\M}{\mathbb{M}}

\newcommand{\R}{{\mathbb{R}}}
\newcommand{\C}{\mathbb{C}}

\newcommand{\Span}{\mathrm{Span}}
\newcommand{\Ran}{\mathrm{Ran}}
\newcommand{\rank}{\mathrm{rank}}

\newcommand{\eps}{\epsilon}

\begin{document}
\numberwithin{equation}{section}
%%%%%%%%%%%%%%%%%%%%%%%%%%%%%%%
%
%\catchline{}{}{}{}{}
%
%%%%%%%%%%%%%%%%%%%%%%%%%%%%%%%%%%%%%%%%%%%%%%%%%%%%%%%%%%%%%%%%%%%%%%%%%%

\title[MCTDHF equations]{Setting and analysis of the multi-configuration time-dependent Hartree--Fock equations}

\author{Claude BARDOS}

\address{Laboratoire  Jacques-Louis Lions, Univ. Paris 7, 175 rue du Chevaleret \\
F-75013 Paris, 
France\\
W.P.I, c/o Fak. f. Mathematik, Univ. Wien - UZA 4, Nordbergstrasse 15 \\
Vienna, A-1090, Austria \\
claude.bardos@gmail.com
}
\author{Isabelle CATTO}
\address{CNRS, UMR7534, F-75016 Paris, France\\
\&  Univ. Paris-Dauphine, CEREMADE,  
Place du Mar\'echal de Lattre de Tassigny \\ 
F-75775 Paris Cedex 16, France\\
catto@ceremade.dauphine.fr}

\author{Norbert  J. MAUSER}

\address{Wolfgang Pauli Institute,  c/o Fak. f. 
Mathematik, Univ. Wien, Nordbergstrasse, 15\\ Vienna, A-1090, Austria \\
norbert.mauser@univie.ac.at}

\author{Saber TRABELSI}

\address{Fak. f. Mathematik, Univ. Wien, Nordbergstrasse, 15\\
Vienna, A-1090, Austria \\
Lab. Jacques-Louis Lions, Univ. Paris 7, 175 rue du Chevaleret \\
F-75013 Paris,
France\\
saber.trabelsi@ann.jussieu.fr }
\maketitle
\begin{abstract}
In this paper we   formulate and analyze  the   Multi-Confi\-gu\-ration  Time-Dependent  Hartree-Fock (MCTDHF) equations  for  molecular  systems with pairwise interaction. This is an approximation of  the $N$-particle time-dependent  Schr\"{o}dinger   equation which involves  (time-dependent) linear combination of (time-dependent) Slater determinants.   The mono-elec\-tronic  wave-functions satisfy nonlinear Schr\"{o}\-din\-ger-type  equations coupled to a  linear system of  ordinary differential  equations equations  for the expansion coefficients.  The invertibility of the one-body density matrix (full-rank hypothesis)   plays a crucial r\^ole in the analysis. Under the full-rank assumption   a fiber bundle structure shows up and produces unitary equivalence between different  useful representations of the approximation.  We   establish  existence and uniqueness of  maximal solutions to  the Cauchy problem    in the energy space as long as the density matrix is not singular for a large class of interactions  (including Coulomb potential). A sufficient condition  in terms  of  the energy of the initial data ensuring the global-in-time invertibility is provided (first result in this direction). Regularizing  the density matrix breaks down   energy conservation. However a global well-posedness for this system  in $L^2$ is  obtained  with Strichartz estimates. Eventually   solutions to this  regularized system are shown to converge to  the original one on the time interval when the density matrix is invertible. 
\end{abstract}


%\ccode{AMS Subject Classification: 22E46, 53C35, 57S20}
%%%%%%%%%%%%%%%%%%%%%%%%%%%%%%%%%
\section[Introduction]{Introduction}\label{sec:intro} 
The purpose of  the present paper is to lay out the mathematical analysis of  
the {\it multi-configuration time--dependent Hartree--Fock} (MCTDHF) approximation 
which is used  in quantum chemistry for the dynamics of few electron problems, 
or the interaction of an atom with a strong short laser-pulse \cite{Scrinzi, Zang1, Zang2} 
and \cite{KatoKono}. 
%This paper gives a comprehensive self-contained mathematical description of the  multi-configuration time-dependent Hartree-Fock (MCTDHF) equations. 
%Such models are used in many-body quantum physics and quantum chemistry to 
%approximate the solutions of the time-dependent $N$-particle linear Schr\"{o}dinger 
%equation with pairwise interaction. 
The MCTDHF models are natural generalizations 
of the  time-dependent Hartree-Fock (TDHF) approximation, yielding a hierarchy of models that, in principle, should converge to the exact model.

The physical motivation is  a  molecular quantum system  composed of a finite number $M$ of \textit{fixed} nuclei of masses $ m_1,\ldots, m_M>0 $
with charge $z_1,\ldots, z_M >0$ and a finite number $ N $ of electrons. Using atomic units, the $N$-body Hamiltonian of the 
electronic system submitted to the external potential due to  the nuclei is then the self-adjoint operator
\begin{equation}\label{Hamiltonien}
  \mathcal{H}_N = \sum_{1\leq i\leq N}\left(-\frac 12\Delta_{x_i} + U(x_i)\right)+
 V(x_1,\cdots,x_N)
\end{equation}
acting on the Hilbert space
$L^2(\Omega^{N};\mathbb{C})$ with pairwise interaction between the electrons
of the form 
\[
V(x_1,\cdots,x_N)=\sum_{1 \leq i<j\leq N}v(|x_i-x_j|), 
\] 
with $v$ real-valued and $v\geq 0$.  Here and below $\Omega$ is either the whole space $\R^3$ or a bounded domain in  $\R^3$  with  boundary conditions. 
The $N$ electrons
state is defined by a  \textit{wave-function} $\Psi=
\Psi(x_1,\ldots,x_N)$  in $L^2(\Omega^N)$  that is normalized by  $\|\Psi\|_{L^2(\Omega^{N})}=1 $.
%, for $|\Psi|^2$ is interpreted as the probability density of the $N$ electrons.  
To account for the Pauli exclusion principle which features
the fermionic nature of the electrons, the  antisymmetry condition  
% with respect to the electrons coordinates  
\begin{equation*}
  \Psi(x_1,\ldots,x_N)=\epsilon(\sigma)\Psi(x_{\sigma(1)},\ldots,x_{\sigma(N)}),
\end{equation*}
for every permutation $ \sigma $  of $\{1,\ldots, N\}$ is imposed to the wave-function $\Psi$. The space of
antisymmetric wave-functions will be denoted by $\bigwedge_{i=1}^N
L^2(\Omega)$.
In
(\ref{Hamiltonien}) and throughout the paper, the subscript $x_i$ of $-\Delta_{x_i} $ means
derivation with respect to the $i^{th}$ variable of the function
$\Psi$. Next,
\begin{equation*}
U(x):= -\sum_{m=1}^M \frac{z_m}{|x-R_m|}\end{equation*}
is the Coulomb potential created by $M$ nuclei of respective  charge $z_1,\cdots, z_M>0$  located at points 
$R_1, \cdots , R_M \in \mathbb{R}^3$ and $v(x)=\frac{1}{|x|}$ is the  Coulomb repulsive potential between the electrons.  Actually our whole analysis carries through to more general  hamiltonians  (possibly time-dependent) as explained in Section~\ref{sec:extension} below.

\vskip6pt
For nearly all applications, even with two interacting electrons the numerical treatment of the time-dependent Schr\"{o}dinger equation (TDSE)
\begin{equation}\label{exact}
  i\frac{\partial \Psi}{\partial t}= \mathcal{H}_N \,\Psi \,,\quad 
    \Psi(0) = \Psi^0,
    \end{equation} 
 is out of the reach of even the most powerful computers, and approximations are needed. 
% The time evolution of a quantum system starting from some initial data $\Psi^0 \in L^2(\Omega^N) $  
% is governed by By the Stone theorem which ensures the existence of an unitary group $
%\mathcal{U}(t)= \exp(-it\mathcal{H}_N) $ the Cauchy problem (\ref{exact}) is known to be well-posed and
% the unique global solution
%is given by $\Psi(t)=\mathcal{U}(t)\Psi^0 $ for all $t\in
%\mathbb{R}$.
\vskip6pt 
 Simplest elements of  $
 \bigwedge_{i=1}^N
L^2(\Omega) $  are  the
so-called \textit{Slater determinants}
\begin{equation}\label{HFnormal}
  \Psi(x_1,\ldots,x_N)=\frac{1}{\sqrt{N!}}\det\big(\phi_{i}(x_j)\big)_{1\leq i,j\leq N}
\end{equation}
constructed with any  orthonormal family  $\phi_i$ in   $\0\,.$ The factor
$ \frac{1}{\sqrt{N!}} $ ensures the normalization condition on the
wave-function. Such a Slater determinant will be denoted by
 $\phi_{1}\wedge \ldots \wedge \phi_{N}$. The  family of all Slater determinants   built from a complete  orthonormal set  of $\0$ is a complete  orthonormal set   of  $\bigwedge_{i=1}^N L^2(\Omega)$.  Algorithms based on the  restriction to a single  Slater determinant are called Hartree-Fock approximation (HF). On the other hand the basic idea of the multi-configuration methods is to use a finite linear combinations of such determinants constructed from a family of  $K(\geq N)$ orthonormal  mono-electronic wave-functions.

%For a mathematical theory of the use of the time-independent {\it multi-configuration Hartree--Fock} (MCHF) ansatz  in the computation of  so-called ground-- and bound  states  we  refer to \cite{Lebris, Fries, Lewin}. 
%Our goal in the present paper is to lay out the mathematical theory of the time--dependent problem, 
%the {\it multi-configuration time--dependent Hartree--Fock} (MCTDHF) 
%which is used for the dynamics of few electron problems, 
%\textit{e.g.} the formation of molecules in quantum chemistry, 
%or the interaction of an atom with a strong short laser-pulse \cite{Scrinzi, Zang1, Zang2} 
%and \cite{KatoKono}. 

One observes (this computation is done in Subsection~\ref{ssec:free}) that in the absence of pairwise interacting potentials any Slater determinant  constructed with orthonormal solutions $\phi_i(x,t)$ to the single-particle  time--dependent Schr\"{o}dinger equation gives an exact solution of the $N$-particle non interacting time--dependent Schr\"{o}dinger equation. Such $\phi_i(x,t)$ are called \textit{orbitals}  in the Chemistry literature. The same is true for any linear combination of Slater determinants with \textit{constant} coefficients. Of course, the situation  turns out to be completely different when pairwise interactions are added : a solution to TDSE starting with an initial data   
composed of one or a finite number of Slater determinants will not remain so for any time $t\not=0$. 
Such behavior (called ``explosion of rank") is part of the common belief, but is not shown rigorously as a property of the equations, to the best of our knowledge.
In the MCTDHF approach one introduces  time--dependent coefficients and time-dependent orbitals 
to take into account pairwise interactions  and to preserve the finite linear combination 
structure of Slater determinants in time. 
Using time-independent orbitals as it corresponds to a Galerkin-type approximation would
save the effort for the nonlinear equations, but requires a much larger number of relevant orbitals 
and hence the numerical cost is much higher. 
The motion of the electrons in the MCTDHF framework is then governed by a coupled system of 
$K$ nonlinear partial differential equations for the orbitals and 
$ {K\choose N}$ ordinary differential equations for the expansion coefficients (see for instance System~\eqref{working}).
\vskip6pt
Although MCTDHF is known for decades, the mathematical analysis has been tackled only recently. For a mathematical theory of the use of the time-independent {\it multi-configuration Hartree--Fock} (MCHF) ansatz  in the computation of  so-called ground-- and bound  states  we  refer to \cite{Lebris, Fries, Lewin}. A preliminary contribution was given by Lubich~\cite{Lubich2} and Koch and Lubich~\cite{Lubich1} for the time-dependent multi-configuration Hartree~(MCTDH) equations for bosons, for the simplified case of a regular and bounded interaction potential $v$ between the electrons and a  Hamiltonian without  exterior potential  $U$.  
The MCTDH equations are similar to MCTDHF  from the functional analysis point  of view, although more complicated from  the algebraic point of view, since more density-matrices have to be considered in the absence of \textit{a priori} antisymmetry requirements on the $N$-particle wave-function  (see also ~\cite{Koch2} for an extension to  MCTDHF equations).
 Using a  full-rank (\textit{i.e.} invertibility) assumption  on  the one-body density matrices, the authors proved  short-time existence and uniqueness of solutions in  the functions space  $H^2(\mathbb{R}^3)$ for the orbitals with the help of  Lie commutators techniques. 
Numerical algorithms are also proposed  and analyzed by the groups around Scrinzi (\textit{e.g.} ~
\cite{Zang1}) and Lubich, the proof of their  convergence generally requires the $H^2$-type  regularity assumptions (see \textit{e.g.} \cite{Lubich3}). 
\vskip6pt

We present  here well-posedness results for the MCTDHF Cauchy problem  in $H^1$, $H^2$ and $L^2$,
under  the full-rank assumption on  the  first-order density matrix and for the physically most relevant and mathematically most demanding case of  Coulomb interaction.   We also give sufficient conditions  for  global-in-time  full-rank in terms of  the energy of the initial data.  Eventually   solutions to a  perturbed  system with regularized density matrix are shown to converge to  the original one on the time interval when the density matrix is invertible.

\vskip6pt
This paper is organized as follows. In Section 2 we give a complete analysis of the {\it
ansatz} $\Psi$ associated to the multi-configuration Hartree-Fock
approximation. Essentially, this  ansatz  corresponds to a linear
combination of Slater determinants built from a vector of complex
coefficients $C$ and a set of orthonormal, square integrable functions
represented by a vector $\Phi=(\phi_1,\phi_2,\ldots,\phi_K)$, for $K\geq N$. The \textit{first-order density matrix}  is introduced and represented by a complex-valued
matrix $\G$ which corresponds to the representation  of the kernel of the first-order density matrix  in the orthonormal basis $\{\phi_1,\phi_2,\ldots,\phi_K\}$.
By abuse of language this   matrix $\G$ depending only on the expansion
coefficients $C$ is also called {\it density matrix}. Its  invertibility  is a crucial hypothesis which will be referred to as the {\it 
full-rank hypothesis}. Under this hypothesis, the corresponding set of  
pairs $(C,\Phi)$ is endowed with a structure of a {\it fiber bundle}. In
Section 3, two set of equivalent systems are presented. The first one,
$\mathcal{S}_0$, called \textit{variational system} is inspired by a variational
principle. The second one, $\mathcal{S}_\mathbf{H}$, will be referred to as
{\it working equations}.  In Section 4,  the system $\mathcal{S}_0$ is
used to prove the propagation
of the normalization constraints, the conservation of the total energy and
 an {\it a posteriori} error estimate for smooth solutions (if they exist). The system
$\mathcal{S}_\mathbf{H}$ is used to prove local existence, uniqueness and
stability with initial data in $H^m$ for $m\geq 1$. In particular the
space $H^1$ is used to balance the singularity of the potentials (of
Coulomb type) and we prove the local well-posedness using the Duhamel
formula. Next, the conservation of the total energy allows  to
extend  the local-in-time solution until  the associated density
matrix $\G$ becomes singular. Therefore, Section 5 is devoted to a
criteria  based on the conservation of the energy that guarantees the
global-in-time invertibility of the density matrix $\G$. To handle the
possible degeneracy of this matrix, a regularized problem is considered in
Section 6. For this  problem the conservation of the energy does not hold 
anymore. Hence, we propose an alternate proof, also valid for singular
potentials, but that is only based  on mass conservation. Such proof relies
on Strichartz estimates. Eventually, one expects that the solution of the
regularized problem converges towards the solution of the original one as
long as the unperturbed density matrix is invertible. The proof is a $H^1$ version of
the classical ``shadowing lemma" for ordinary differential equations. Finally, in  the last section we list  some extensions to time-dependent Hamiltonian including a laser field and/or a time-dependent external potential. The case of discrete systems is also discussed there. 
\vskip6pt
%The  time-dependent Hartree--Fock equations (TDHF)   for the motion of a  single determinant  turn out to be a special case of the MCTDHF equations that is therefore included in our analysis, although this case is  much simpler: since  the first-order density matrix is  the identity for all time,  the full-rank assumption is automatically satisfied. \vskip6pt
Some of the results presented here  have been announced in \cite{trabelsi1} and \cite{AML} and the details of the $L^2$ theory are worked out in \cite{Saber-L2}. 

\vskip10pt
 \noindent\textbf{Notation}.  $ \left\langle\cdot ,\cdot\right\rangle $ and $ \left\langle\cdot\vert\cdot\right\rangle$ respectively denote  the usual scalar products in $ \0 $ 
and  in $ L^2(\Omega^N)$, $\big(\cdot,\cdot)$ the scalar product in $\0^K$  and $a\cdot b$ the complex   scalar product of two complex vectors $a$  
and $b$. The bar denotes complex conjugation. We set $L_{\wedge}^2(\Omega^N):= \bigwedge_{k=1}^N\:\0$ where
the symbol $\wedge$ denotes the skew-symmetric tensorial product. Throughout the paper  bold face letters correspond to one-particle operators on $\0$, 
calligraphic  bold face letters to  operators on $L^2(\Omega^N)$, 
whereas ``black board"  bold face letters are reserved  to  matrices.  $\mathcal{L}(E;F)$ denotes the set of continuous linear applications from $E$ to $F$ (as usual $\mathcal{L}(E)=\mathcal{L}(E;E)$).

%
%%%%%%%%%%%%%%%%
\tableofcontents 
%%%%%%%%%%%%%%%%%%%%%%%%%%%%%%%%%%%%%%%%%%%%%%%%%%%%%%%%%%%%%%%%%%%%%%%%%%
% 
\section[Stationary Ansatz]{Fiber Bundle Structure of the  Multi-Configuration Hartree-Fock Ansatz}\label{sec:stationary}
%%%%%%%%%%%%%%%%%%%
\subsection{The MCHF {\it ansatz}.} For positive integers  $ N\leq K $, let $ \Sigma_{N,K} $ denote the set  of increasing mappings $ \sigma:\;\{1,\ldots,N\}\longrightarrow\{1,\ldots,K\} $  
\[
\Sigma_{N,K}=\Big\{\sigma=\{\sigma(1)<\ldots < \sigma(N)\}\subset\{1,\ldots,K\}\Big\},\qquad  \#\Sigma_{N,K}={K\choose N}:=r.
\]
For simplicity the same notation  is used for the mapping $ \sigma $ and its range $ \{\sigma(1)<\ldots < \sigma(N)\}$. Next we define 
\[
 \mathcal{F}_{N,K}:=S^{r-1}\times\mathcal{O}_{\0^K} 
 \]
with 
\begin{equation}\label{l2unitsphere}
\mathcal{O}_{\0^K}=\Bigl\{ \Phi=(\phi_1,\ldots,\phi_K )\in \0^K\,:\, \int_{\Omega}\phi_i\,
 \bar\phi_j\,dx=\delta_{i,j}\Bigr\},
\end{equation}
with $ \delta_{i,j}$ being the Kronecker delta and  with  $S^{r-1}$ being the  unit sphere in $\mathbb{C}^r$ endowed with the complex euclidean distance 
\begin{equation}\label{Crunitsphere}
S^{r-1}=\Bigl\{ C=(c_{\sigma})_{\sigma\in \Sigma_{N,K}} \in \mathbb{C}^{r}\,:\,\quad\Vert C\Vert^2=\sum_{ \sigma}|c_{\sigma}|^2=1\Bigr\}
\end{equation}
with  the shorthand  $ \sum_{ \sigma} $ for  $\sum_{ \sigma\in\Sigma_{N,K}} $. %The coordinates $c_\sigma$ are arranged in lexicographical order and 
{}To any $ \sigma \in \Sigma_{N,K} $ and $\Phi$ in $\mathcal{O}_{\0^K}$, we associate  the   Slater determinant
\[
\Phi_{\sigma}(x_1,\ldots,x_N)=\phi_{\sigma(1)}\wedge\ldots\wedge\phi_{\sigma(N)}=\frac{1}{\sqrt{N!}} \left|\begin{array}{ccc}
\phi_{\sigma(1)}(x_1)  &  \ldots &  \phi_{\sigma(1)}(x_N) \\
  \vdots &   &  \vdots \\
 \phi_{\sigma(N)}(x_1) &\ldots   &   \phi_{\sigma(N)}(x_N)
\end{array}\right|.
\]
%that is, the skew-symmetric function $\Phi_\sigma$ is the determinant built from the $ \phi_i$'s such that $ i\in \sigma$. 
%The vector $\Phi$ being in $\mathcal{O}_{\0^K} $, the factor $\frac{1}{\sqrt{N!}} $ ensures   the normalization $ \|\Phi_\sigma\|_{L^2(\Omega^N)}=1$. %\[
%\int_{\Omega^N}\:\Phi_\sigma(x_1,\ldots,x_N)\:\overline{\Phi}_\tau(x_1,\ldots,x_N) \:dx_1 \ldots dx_N=\delta_{\sigma,\tau}. 
%\]
The mapping
\begin{equation}\label{def-pi}
 (C,\Phi)  \longmapsto  \Psi=\pi_{N,K}(C,\Phi)= \sum_{\sigma} \:c_{\sigma}\:\Phi_\sigma.
 \end{equation}
is   multilinear, continuous and  even  infinitely differentiable  from $ \mathcal{F}_{N,K}$ equipped with the natural topology of $ \mathbb{C}^r \times \0^K$ into $ L_{\wedge}^2(\Omega^N)$. Its  range is denoted by  
\begin{equation*}
\mathcal{B}_{N,K}=\pi(\mathcal{F}_{N,K})=\Bigr\{ \Psi=\sum_{\sigma}c_{\sigma} \Phi_\sigma \,:\,\quad (C,\Phi) \in \mathcal{F}_{N,K}\Bigl\}.
\end{equation*}
When there is no ambiguity, we simply denote $\pi=\pi_{N,K}$.   The set  $\mathcal{B}_{N,N}$ is  the set of single determinants or \textit{Hartree--Fock states}. Of course $\mathcal{B}_{N,K}\subset \mathcal{B}_{N,K'}$ when $K'\geq K$ and actually 
\[
\lim_{K\to +\infty} \mathcal{B}_{N,K}=\Big\{\Psi\in L^2_\wedge\big(\Omega^N\big)\::\: \Vert\Psi\Vert=1\Big\}, 
\]
in the sense of an increasing sequence of sets, since   Slater determinants form an Hilbert basis of $L^2_\wedge\big(\Omega^N\big)$ (see ~\cite{Lowdin}). In particular, for $ \sigma,\tau \in \Sigma_{N,K} $, we have 
\begin{equation}\label{ps-sigma-tau}
\langle \Phi_\sigma\:\big\vert\:{\Phi}_\tau\rangle=\delta_{\sigma,\tau}. 
\end{equation}
Observe that without orthonormality condition the formula \eqref{ps-sigma-tau}  becomes  
\begin{equation}\label{ps-det}
\langle \phi_{1}\wedge\ldots\wedge\phi_{N}\:\big\vert\:\xi_{1}\wedge\ldots\wedge\xi_{N}\rangle=\det\left(\langle\phi_i;\xi_j\rangle\right)_{1\leq i,j\leq N}
\end{equation} 
for  $\Phi,\:\Xi\in \0^N$ which will be used below (see ~\cite{Lowdin}).
\vskip6pt
The set of multi-configuration ansatz  $\mathcal{B}_{N,K}$  is characterized   
in Proposition~\ref{car-ANK} in Subsection~\ref{densityop}   in terms of  the so-called first-order density matrix, and   its geometric structure is analyzed in Subsection~\ref{geodiff}.  
%%%%%%%%%%%%%%%%%%%%%
%%%%%%%%%%%%%%%%%%%%%
%%%%%%%%
\subsection{Density Operators}\label{densityop}
%\subsection{Algebraic structure}
For $ n=1,\ldots, N$ and for $\Psi\in L_\wedge^2(\Omega^N)$ with $\Vert \Psi\Vert=1$,  a trace-class self-adjoint operator  $\bigl[\Psi\otimes\Psi\bigr]_{:n}$, called   {\it $ n^{th} $ order  density operator}, is defined on $L^2_\wedge(\Omega^n)$ through its kernel $\bigl[\Psi\otimes\Psi\bigr]_{:n} $
\begin{equation}\label{kernal}
\bigl[\Psi\otimes\Psi\bigr]_{:n}(X_n,Y_n)
= { N \choose n } \int_{\Omega^{N-n}}\Psi(X_n,Z^N_n)\:\overline{\Psi}(Y_n,Z^N_n) \:dZ^N_n,
\end{equation}
for $1\leq n\leq N-1$ 
and 
\[
\bigl[\Psi\otimes\Psi\bigr]_{:N}(X_N,Y_N)= \Psi(X_N)\:\overline{\Psi}(Y_N)  ,\]
with  the notation
\[\begin{array}{ll}
X_n=(x_1,\ldots,x_n),&\quad Y_n=(y_1,\ldots,y_n),,\vspace{2mm}\\
Z_n^N=(z_{n+1},\ldots,z_N),&\quad dZ^N_n= dz_{n+1}\ldots dz_N,
\end{array}\]
and similarly for other capital letters. Our normalization follows  L\"owdin's   \cite{Lowdin}.  A simple calculation shows that, for $ 1\leq n \leq N-1 $,
\begin{equation}
\bigl[\Psi\otimes\Psi\bigr]_{:n}(X_n,Y_n)
=  \frac{n+1}{N-n}\:\int_{\Omega } \bigl[\Psi\otimes\Psi\bigr]_{:n+1}(X_n,z,Y_n,z) \:dz.\label{nn-1}
\end{equation}
In particular, given $1\leq n \leq p \leq N-1$, one can deduce the expression of $\bigl[\Psi\otimes\Psi\bigr]_{:n}$ from the one of $\bigl[\Psi\otimes\Psi\bigr]_{:p}$.  These operators satisfy:
%%%%%%%%%%%%%%%%%
\begin{prop}[\cite{Ando,Coleman1,Coleman2, Lowdin}]\label{prop-Dp} For every integer  $1\leq  n\leq  N$,  the $n$-th order density matrix is a trace-class self-adjoint operator on  $L_\wedge^2(\Omega^n)$ such that 
\begin{equation}\label{01}
0\leq \bigl[\Psi\otimes\Psi\bigr]_{:n}\leq 1,
\end{equation}
in the sense of operators, and 
\[
\mathrm{Tr}_{L^2(\Omega^n)} \,\bigl[\Psi\otimes\Psi\bigr]_{:n}= { N \choose n }
.\]
\end{prop}  
Actually, multi-configuration  ansatz correspond to  first-order density matrices with  \textit{finite rank}, and  we   have the following
%%%%%%%%%%%%
\begin{prop}\label{Lowdin}[L\"owdin's expansion theorem~\cite{Lowdin}; see also  \cite{Fries,Lewin}\/]\label{car-ANK} Let $K\geq N$, then \[
\mathcal{B}_{N,K}=\pi(\mathcal{F}_{N,K})=\big\{ \Psi\in L^2_\wedge(\Omega^N)\;:\; \Vert\Psi\Vert=1\quad \mathrm{and}\quad \mathrm{rank}\bigl[\Psi\otimes\Psi\bigr]_{:1}\leq K
\big\}. 
\]
\end{prop}
\vskip6pt\noindent 
More precisely, if $\Psi=\pi(C,\Phi)$ with $(C,\Phi)\in \mathcal{F}_{N,K}$,   then $\mathrm{rank}\bigl[\Psi\otimes\Psi\bigr]_{:1}\leq K$ and $\mathrm{Ran}\bigl[\Psi\otimes\Psi\bigr]_{:1}\subset \mathrm{Span}\{\phi_1,\cdots,\phi_K\}$.  If $\Psi\in \mathcal{B}_{N,K}$  and if $\mathrm{rank}\bigl[\Psi\otimes\Psi\bigr]_{:1}=K'$ with $N\leq K'\leq K$ and with $ \{\phi_1,\ldots,\phi_{K'}\}$ being an orthonormal basis of $\mathrm{Ran}\bigl[\Psi\otimes\Psi\bigr]_{:1}$, then $\Psi$ can be expanded as a linear combination of Slater determinants built from  $\{\phi_1,\cdots,\phi_{K'}\}$. The first-order (or one-particle) density matrix $\bigl[\Psi\otimes\Psi\bigr]_{:1}$ is often denoted by $\gamma_\Psi$ in the literature and  in the course of this paper.  According to   Proposition~\ref{prop-Dp} above  it  is a non-negative self-adjoint  trace-class operator on $\0$, with trace $N$ and with operator norm less or equal to $1$.  Therefore its sequence of  eigenvalues $\{\gamma_i\}_{i\geq 1}$  satisfies  $0\leq \gamma_i\leq 1$, for all $i\geq 1$, and $\sum_{i\geq 1} \gamma_i=N$. In particular,  at least $N$ of the  $\gamma_i$'s are not zero, and therefore $\mathrm{rank}\,\gamma_\Psi\geq N$, for any $\Psi\in L^2_\wedge(\Omega^N)$.  
\vskip6pt
Similarly, if $\Psi=\pi(C,\Phi)\in \mathcal{B}_{N,K}$,  the range of the operator $[\pi(C,\Phi)\otimes{\pi(C,\Phi)}]_{:n}$  is $\bigotimes_n\textrm{Span}\{\Phi\}$ and its kernel is   $\Big(\bigotimes_n\textrm{Span}\{\Phi\}\Big)^\bot$ with  $\mathrm{Span}\{\Phi\}:=\mathrm{Span}\{\phi_1,\ldots,\phi_K\}$.  Therefore the operator is represented by an Hermitian matrix in $\bigotimes_n\textrm{Span}\{\Phi\}$ whose entries turn out to depend only on the coefficients $C$ and the dependence is quadratic. For the first- and second- order density operators we have the explicit expressions
%%%%%%%%%%%%%%%%%%%%%%%%
\begin{prop}[\cite{Fries}, Appendix 1]\label{prop:coeff} Let $\Psi=\pi(C,\Phi)$ in $\mathcal{B}_{N,K}$, then the operator kernel of  the second-order density matrix kernel is given by  
\begin{equation}\label{densitymatrix2}
[\Psi\otimes\Psi]_{:2}(x,y,x',y')   = 
\sum_{i,j,k,l=1}^K \gamma_{ijkl}\,\phi_i(x) \:\phi_j(y)\:\overline{\phi}_k(x')\:\overline{\phi}_l(y')%\label{d2}
\end{equation}
with 
\begin{equation}\label{gammaijkl}
\gamma_{ijkl}=\frac 12\:(1-\delta_{i,j})(1-\delta_{k,l})\: \sum_{\substack 
{\sigma,\tau\:\vert \:i,j\in \sigma,\,k,l\in\tau
\\ 
\sigma\setminus\{i,j\}= \tau\setminus\{k,l\} 
}}
(-1)^{\sigma}_{i,j} (-1)^{\tau}_{k,l} \:{c}_{\sigma}\: \overline{c}_\tau,
\end{equation}
where for $i\neq j$, 
\[
 (-1)^{\sigma}_{i,j}=\frac{i-j}{|i-j|}\, (-1)^{\sigma^{-1}(i) +\sigma^{-1}(j)}. 
 \]
Similarly, the kernel of  the  first-order density matrix is  given by the formula
\[%\begin{equation}\label{densitymatrix1}
[\Psi\otimes\Psi]_{:1}(x,y)= \sum_{ i,j=1}^K \gamma_{ij} \: \phi_i (x)\, \overline{\phi}_j (y)
%\:=   \left[\G\:\Phi\right] \otimes \overline{\Phi }\label{partialtrace}\,.
\]%\end{equation}
with
\begin{equation}\label{gammaij}
\gamma_{ij}=\frac{2}{N-1}\sum_{k=1}^K\gamma_{ikjk}=
 \sum_{\substack{\sigma,\tau\:\vert\:i\in\sigma, \,j\in\tau\\ \sigma\setminus\{ i\}= \tau\setminus\{ j \}}} (-1)^{\sigma^{-1}(i)+\tau^{-1}(j)}\, c_\sigma\,\overline {c_\tau}
\end{equation}
and 
\begin{equation}\label{gammai}
\gamma_{ii}=
 \sum_{\sigma\:\vert\:i\in\sigma} \vert c_\sigma\vert^2.
\end{equation}
\end{prop}
%%%%%%%%%%%%%%%%%%%%%
The first-order density matrix allows to characterize the set $\mathcal{B}_{N,K}$ (see Proposition~\ref{car-ANK} above) whereas the second-order density matrix is needed to express expectation values of the energy Hamiltonian as soon as  two-body  interactions are involved.

Since the coefficients $\gamma_{ij}$ only depend on $C$, we denote by $\G(C)$  the $K\times K$  Hermitian matrix with entries  $\bar\gamma_{ij}$, $1\leq i,j\leq K$ (the adjoint of the matrix representation of the first-order density operator in $\mathrm{Span}\{\Phi\}$).  The matrix  $\G(C)$   is   positive, hermitian and  of trace $N$ with same eigenvalues as $\gamma_\Psi$ and same rank, and   there exists  a unitary $K\times K$ matrix $ U$  such that $ U\:\G(C)\:U^\star = \mathrm{diag}(\gamma_1,\ldots,\gamma_K)$ with $0\leq \gamma_k\leq 1$ and $\sum_{k=1}^K\gamma_k=N$. Hence, $\gamma_\Psi$ can also be  expanded as follows
\begin{equation}\label{naturaldensity}
\gamma_\Psi(x,y)= \sum_{i=1}^K\:\gamma_{i}\:\phi'_i(x)\:\overline{\phi'}_i(y),
\end{equation}
where $\Phi'=U\cdot \Phi$ with obvious notation and with $\{\phi'_1,\cdots,\phi'_{K}\}$ being an eigenbasis of $\gamma_\Psi$.    Note that  it is  easily recovered from \eqref{gammai}  that $0\leq \gamma_i\leq 1$ for $C\in S^{r-1}$. 

%
%Similarly $n$-th order eigenvalues and orbitals may be  defined for $n$-th order density matrices (see~\cite{Ando} for more details). Since we have restricted the analysis to the case of one- and two-body interactions, only the first- and second- order density matrices play a r\^ole here. 
%%%%%%%%%%%%%%%%%%%%%%
%%%%%%%%%%%%%%%%%%%%%
\begin{rem}\label{rem:HF} When $K=N$ (Hartree-Fock case), $\gamma_{\Psi}$ being of trace $N$ must be the projector on $\mathrm{Span}\{\Phi\}$; that is
\[
\gamma_\Psi(x,y)= \sum_{i=1}^N\phi_i(x)\:\overline{\phi}_i(y):=\mathbf{P}_\Phi(x,y), 
\]
with $\mathbf{P}_\Phi$ denoting the projector on $\Span\{\Phi\}$. In this case, \eqref{gammaij} and \eqref{gammaijkl} simply reduce to $\gamma_{ij}=\delta_{i,j}$ and $\gamma_{ijkl}=\frac 12\big(\delta_{i,k}\delta_{j,l}-\delta_{i,l}\delta_{j,k}\big)$; that is $\G(C)=\mathbb{I}_N$.
\end{rem}

%%%%%%%%%%%%%%%%%%%%
%\begin{rem}\label{Gamma-C} It is worth emphasizing the fact that the coefficients $\gamma_{ij}$ and $\gamma_{ijkl}$ only depend on the expansion coefficients $C$ and not on the orbitals, and that this dependency is quadratic.  This property actually holds true at any order $1\leq p\leq N-1$.  We shall rely on it in the proof of existence of solutions in Section~\ref{sec:analysis}.  \end{rem}
%%%%%%%%%%%%%%%%%%%%%%%

\vskip6pt
The representation of a wave-function $\Psi\in \mathcal{B}_{N,K}$ in terms of expansion coefficients $C$ and orbitals $\Phi$ is obviously not unique as it is already seen on the Hartree--Fock ansatz.   Indeed, if $\Psi^{HF}=\phi_1\wedge\cdots\wedge\phi_N= \psi_1\wedge\cdots\wedge\psi_N$, there exists a unique  $N\times N$ unitary transform $U$ such that $(\phi_1,\cdots,\phi_N)= (\psi_1,\cdots, \psi_N)\cdot U$.  The pre-image of $\Psi^{HF}$ by $\pi$  in    $\mathcal{F}_{N,N}$   is  the orbit of $(\phi_1,\cdots,\phi_N)$ under the action of   $ \mathcal{O}_N $, with $ \mathcal{O}_\ell $ being the set of $ \ell\times \ell $ unitary matrices. In the general case under the full-rank assumption the set $\mathcal{B}_{N,K}$ has a similar orbit-like structure as explained now. 

%%%%%%%%%%%%%%%%%%%%%%
%%%%%%%%%%%%%%
\subsection{Full-rank and fibration}\label{geodiff} 
%The first obvious fact  is
%%%%%%%%%%%%%%%%%%%%%%%%%%
%\begin{prop}\label{firstLemma}
%Let $ (C,\Phi) $ and $ (C',\Phi) $ in $\mathcal{F}_{N,K}$  such that $
%\sum_{\sigma}\: c_{\sigma}\:\Phi_\sigma=\sum_{\sigma}\: c'_{\sigma}\: \Phi_\sigma$. Then  $C=C'$.
%\end{prop}
%%%%%%%%%%%%%%%%%%%%%%%%%%
%\begin{proof}
%The family $ \{\phi_i\}_{1\leq i\leq K}$ being an orthonormal family in $ \0 $, the same holds  true for the family of determinants $\{\Phi_\sigma\}_{\sigma\in\Sigma_{N,K}}$ in $ \bigwedge_{k=1}^N\0 $. The claim follows obviously  for $c_\sigma=c'_\sigma=\langle\Psi\:|\:\Phi_\sigma\rangle$.
%\end{proof} 
We  introduce 
\[
\partial\mathcal{B}_{N,K}:=\big\{\Psi\in \mathcal{B}_{N,K}\,:\,\rank\,{\gamma_\Psi}=K\big\}
\]
and, by analogy,  
\[
\partial\mathcal{F}_{N,K}=\pi_{N,K}^{-1}(\partial\mathcal{B}_{N,K}):=\big\{(C,\Phi)\in \mathcal{F}_{N,K}\,:\,\rank\,\G(C)=K\big\}.
\]
$\partial\mathcal{F}_{N,K}$ is the  open subset of $\mathcal{F}_{N,K}$ corresponding to invertible  $\G(C)$'s (\textit{full-rank assumption}). \vskip6pt
Clearly $\partial\mathcal{B}_{N,N}=\mathcal{B}_{N,N}$ and $\partial\mathcal{F}_{N,N}=\mathcal{F}_{N,N}$; that is,  the full-rank assumption is automatically satisfied in the Hartree--Fock setting (see Remark~\ref{rem:HF}). 
\vskip6pt
On the opposite, it may happen that $\partial\mathcal{B}_{N,K}=\emptyset$ (in that case $\mathcal{B}_{N,K}= \mathcal{B}_{N,K-1}$). Indeed,  for  $K\geq N$ the   admissible ranks of  first-order density matrices must satisfy the relations \cite{Fries, Lewin}
\[
K\left\{
\begin{array}{lc}
=1&N=1\\
\geq 2,\;\mathrm{even}& N=2\\
\geq N,\neq N+1,&N\geq 3.
\end{array}
\right.
.
\]
%%%%%%%%%%%%%%%%%%%
{}From now on,  we only deal with pairs $(N,K)$ with $K$ admissible. We recall from \cite{Lowdin} the following
\vskip6pt
%%%%%%
 \begin{prop}\label{Unitarytransformation}
Let $ (C,\Phi) $ and $ (C',\Phi')$ in $\partial\mathcal{F}_{N,K}$ such that $\pi(C,\Phi)=\pi(C',\Phi') $. Then, there exists a unique unitary matrix $ U \in \mathcal{O}_K$ and a unique unitary matrix $d(U)=\overline{\mathbb{U}}\in \mathcal{O}_r$ such that 
\begin{equation*}
\Phi'=U\cdot\Phi, \quad C'=d(U)\cdot C
\end{equation*}
where, for every  $\sigma\in \Sigma_{N,K}$, 
\[
\Phi'_\sigma = \sum_{\tau} \mathbb{U}_{\sigma,\tau}\: \Phi_\tau .\]
Moreover, 
\begin{equation} \G(C')=U\:\G(C)\:U^\star.\label{densitytransform}
\end{equation}
\end{prop}
%%%%%%%
\begin{proof} Let $ (C,\Phi) $ and $ (C',\Phi') $ in $\partial\mathcal{F}_{N,K}$ such that $ \pi(C,\Phi)=\pi(C',\Phi')=\Psi\in  \partial\mathcal{B}_{N,K}$. From Proposition~\ref{car-ANK}, $\Span\{\Phi\}=\Span\{\Phi'\}=\Ran(\gamma_\Psi)$ with $\Phi$ and $\Phi'$ in $ \mathcal{O}_{\0^K}$. Therefore, 
there exists a unique unitary matrix $ U \in \mathcal{O}_K $ such that $ \Phi'=U\cdot\Phi$. Eqn. \eqref{densitytransform} follows by definition of $\G(C)$. Accordingly, there exists a unique unitary matrix  $ \mathbb{U}$ in $ \mathcal{O}_r$ that maps the family $ \{\Phi_\tau\}_{\tau \in \Sigma_{N,K}} $ to $ \{\Phi'_\sigma\}_{\sigma \in \Sigma_{N,K}} $.  More precisely, being  given $ \sigma \in \Sigma_{N,K} $, we have by a direct calculation 
(see also \cite{Lowdin})
\begin{equation}\label{def-U-Phi}
\Phi'_\sigma = \sum_{\tau} \mathbb{U}_{\sigma,\tau}\: \Phi_\tau
\end{equation}
where, for all $\sigma,\tau \in \Sigma_{N,K}$,  
\begin{align}\label{Udet}
\mathbb{U}_{\sigma,\tau } &= \left| \begin{array}{ccc}
 {U}_{\sigma(1),\tau(1)} &  \ldots &  {U}_{\sigma(N),\tau(1)} \\
  \vdots&  \vdots & \vdots  \\
  \vdots&  \vdots & \vdots  \\
  {U}_{\sigma(1),\tau(N)} &  \ldots &   {U}_{\sigma(N),\tau(N)}
\end{array}
\right|=\textrm{det}\big(U_{\sigma(j),\tau(i)}\big)_{1\leq i,j\leq N}\\
&= \textrm{det}\Big(\langle \phi'_{\sigma(j)};\phi_{\tau(i)}\rangle\Big)_{1\leq i,j\leq N}\nonumber.
\end{align}

By construction the $r\times r$ matrix $ \mathbb{U}$ with matrix elements $\mathbb{U}_{\sigma, \tau}$ is  unitary. By the orthonormality of the determinants, we have
\begin{equation}\label{def-U-c}
c'_{\sigma} =\Bigl\langle\pi(C,\Phi) \:\vert\:\Phi'_\sigma\Bigr\rangle\\
=\sum_{\tau} c_{\tau}\Bigl\langle\Phi_\tau \:\vert\:\Phi'_\sigma\Bigr\rangle \\
%&=& \sum_{\sigma,\kappa} \overline{\mathbb{U}}_{\tau,\kappa}\: c_{\sigma} \Bigl\langle\Phi_\sigma \:\vert\:\Phi_\kappa\Bigr\rangle \\
%&=& 
=\sum_{\tau}\overline{\mathbb{U}}_{\sigma,\tau} \:c_{\tau},
\end{equation}
whence the lemma with $d(U)=\overline{\mathbb{U}}$.
\end{proof}
Under the  full-rank assumption and given  $(N,K)$ admissible, the set $\partial \mathcal{B}_{N,K}$ is  a principal fiber bundle.   In differential geometry terminology, $\partial\mathcal{B}_{N,K}$ is called \textit{the base}, and, for any $\Psi\in \partial\mathcal{B}_{N,K}$,   the pre-image $\pi^{-1}(\Psi)$ is \textit{the fiber over $\Psi$}.  Proposition~\ref{Unitarytransformation} defines  a transitive  group action on $\partial\mathcal{F}_{N,K}$ according to
%\begin{equation}\label{Action}
%\begin{array}{l}
%(C',\Phi') = \mathcal{U}\cdot (C,\Phi) \quad\Longleftrightarrow\quad  C'=d(U)\cdot C \quad \text{and} \quad \Phi'=U\cdot \Phi, \vspace{2mm}
%\\
%\mathcal{U} := \big(d(U),U\big)  \in  \mathcal{O}_r\times \mathcal{O}_K. 
%\end{array}
%\end{equation}
\begin{equation}\label{Action}
\begin{split}
(C',\Phi') &= \mathcal{U}\cdot (C,\Phi) \quad\Longleftrightarrow\quad  C'=d(U)\cdot C \quad \text{and} \quad \Phi'=U\cdot \Phi,
\\
\mathcal{U} &:= \big(d(U),U\big)  \in  \mathcal{O}_r\times \mathcal{O}_K. 
\end{split}
\end{equation}
Indeed on the one hand,    it is clear  from  the expression for the matrix elements of $d(U)$ that $d(\mathbb{I}_K)=\mathbb{I}_r$. On the other hand from \eqref{def-U-Phi} and \eqref{def-U-c} it is easily checked that $d(UV)=d(U)\,d(V)$. Therefore couples of the form $\big (d(U),U\big)$ form a subgroup of $\mathcal{O}_r\times \mathcal{O}_K$ that  we denote by $ \mathcal{O}^{r}_K$. The action of $ \mathcal{O}^{r}_K$ is not free on $ \mathcal{F}_{N,K}$ itself --- this is illustrated in Remark~\ref{deg-HF} below  on the examples of Slater determinants in $ \mathcal{F}_{N,K}$ with $K>N$---,  but it is free on    $ \partial\mathcal{F}_{N,K}$ and transitive over  any fiber $\pi^{-1}(\Psi)$ for every $\Psi\in\partial\mathcal{B}_{N,K}$.   Therefore, the mapping $\pi$ defines a principal bundle with fiber given by the  group $\mathcal{O}^{r}_K$. We can define  local  (cross-)sections as  continuous maps $s: \Psi\mapsto (C,\Phi)$ from $ \partial\mathcal{B}_{N,K}$ to $ \partial\mathcal{F}_{N,K}$ such that $\pi\circ s$ is the identity. In particular, $\partial \mathcal{F}_{N,K}/\mathcal{O}^r_K$ is homeomorphic to $ \partial\mathcal{B}_{N,K}$. Since the map $\pi$ is  $C^\infty$,   one concludes from  the inverse mapping theorem that the above isomorphism is also topological.  In the Hartree--Fock case $K=N$ where the full-rank assumption is automatically fulfilled,   $ \pi_{N,N}^{-1}\big(\mathcal{B}_{N,N})$ is  a so-called Stiefel manifold.
%%%%%%%%%%%%%%%%
\begin{rem}\label{deg-HF}  The following example illustrates the necessity of the full-rank assumption. As
\[
K\leq K'\; \Longrightarrow  \; \mathcal{B}_{N,K}\subset \mathcal{B}_{N,K'},
\]
any Slater determinant $\Psi^{HF}=\phi_1\wedge\cdots\wedge\phi_N$ can also be seen as an element of  $\mathcal{B}_{N,K}$ for all $K\geq N$.   If $K>N$, the pre-image of $\Psi^{HF}$ by $\pi$ in $\mathcal{F}_{N,K}$ does not have a  similar orbit structure   as  shown on  the following example. Let    $C=(1,0,\ldots, 0)\in S^{r-1}$ where all coordinates but the first one are $0$ and let  $\Phi'=(\phi_1,\ldots, \phi_N, \phi_{N+1},\ldots,\phi_K)\in \mathcal{O}_{L^2(\Omega)^K}$ with    $\phi_i\in \Span\{\phi_1,\ldots,\phi_N\}^\bot$  for every $N+1\leq i\leq K$,  then $(C, \Phi')\in \mathcal{F}_{N,K}$ and $\Psi_{HF}=\pi (C,\Phi')$. There is no group-orbit structure  on $ \Big(\Span\{\phi_1,\ldots,\phi_N\}^\bot\Big)^{K-N}$.  
\end{rem}
%%%%%%%%%%%%
Having equipped $\partial\mathcal{B}_{N,K}$ with a  manifold structure we turn to the   study of its  the tangent space. 
\vskip6pt
Being  multi-linear with respect to  the variables $C$ and  $\Phi$, the application $\pi$ is clearly infinitely differentiable. Its  gradients 
\begin{equation*}
\begin{array}{cccc}
 \nabla \pi : & \mathbb{C}^r\times L^2(\Omega)^K  & \longrightarrow &  \mathcal{L}\left(\mathbb{C}^r  ; L^2(\Omega^N)\right)\times \mathcal{L}\left(L^2(\Omega)^K ; L^2(\Omega^N)\right)  \vspace{2mm} \\
     &\Psi=\pi(C,\Phi) &\longmapsto  & \nabla \:\Psi = (\nabla_C \:\Psi,\nabla_\Phi\:\Psi) 
 \end{array}
 \end{equation*}
are computed as follows for every $ (C,\Phi) \in \mathcal{F}_{N,K}$~:
\begin{itemize}
\item[i)] for any $\delta C$ in $\C^r$, 
\begin{equation}\label{deriv-C}
\nabla_C \Psi [\delta C]= \sum_{k=1}^{r} \delta c_k\, \frac{\partial  \Psi}{\partial c_{\sigma_k}}
= \sum_{k=1}^r \delta c_k\, \Phi_{\sigma_k},
\end{equation}
with $\Sigma_{N,K}=\{\sigma_1,\cdots,\sigma_r\}$, 
% and 
%\[\frac{\partial  \Psi}{\partial c_{\sigma_k}}\Big(\frac{\partial  \Psi}{\partial c_{\sigma_1} },\ldots,\frac{\partial  \Psi}{\partial c_{\sigma_r} } \Big)= ( \Phi_{\sigma_1},\ldots,\Phi_{\sigma_r}).\]
\item[ii)]  for any  $ \zeta=(\zeta_1,\ldots,\zeta_K) \in L^2(\Omega)^K$, 
\begin{equation}\label{deriv-Phi}
\nabla_\Phi \Psi \:[\zeta] =\sum_{k=1}^K \frac{\partial \Psi}{\partial \phi_k }[\zeta_k]=\sum_{\sigma\in \Sigma_{N,K}}c_\sigma\,\sum_{k=1}^K \frac{\partial \Phi_\sigma}{\partial \phi_k }[\zeta_k],
\end{equation}
with  
\begin{equation}\label{frechetderivatice}
\frac{\partial \Psi}{\partial \phi_k}[\zeta_j]= \sum_{i=1}^N \zeta_j(x_i)\:\int_{\Omega}\: \Psi(x_1,\ldots,x_N)\: \overline{\phi}_k(x_i) \:dx_i. 
\end{equation}
\end{itemize}
\begin{rem} For every $\sigma\in \Sigma_{N,K}$ and every   $1\leq k\leq K$,  we have 
\begin{equation}
\label{derivee-det}
\frac{\partial \Phi_\sigma}{\partial \phi_k}[\zeta]=\left\{
\begin{array}{cl}
\phi_{\sigma(1)}\wedge\cdots\wedge\phi_{\sigma(j-1)}\wedge \zeta\wedge \phi_{\sigma(j+1)}\wedge\cdots\wedge\phi_{\sigma(N)}& \mbox{ if } \sigma^{-1}(k)=j,\vspace{2mm}\\
0 & \mbox{  if }k\not\in \sigma.
\end{array}
\right.
\end{equation}
\end{rem}
\begin{rem} From \eqref{derivee-det} we recover the Euler Formula for homogeneous functions, that reads here 
\begin{equation}\label{compactpsi}
\Psi = \frac{1}{N}\sum_{k=1}^K \frac{\partial \Psi}{\partial \phi_k }[\phi_k] := \frac{1}{N}\:\nabla_\Phi\Psi\: [\Phi].
\end{equation}
\end{rem}
\vskip6pt
\noindent {F}rom the definition of the  adjoint $\nabla_\Phi \Psi^\star\in \mathcal{L}\big(L_\wedge^2(\Omega^N);\0^K\big)$   of the operator $\nabla_\Phi\Psi$ one has 
\begin{equation}\label{def-adjoint}
\forall \zeta\in \0^K,\, \forall \:\Xi \in L^2_\wedge(\Omega^N),\quad \langle\nabla_\Phi \Psi^\star[\Xi]\:;\:\zeta\rangle_{L^2(\Omega)^K}= \Bigl\langle \Xi\big\vert  \nabla_\Phi\Psi\:[\zeta]\Bigr\rangle_{L^2(\Omega^N)}, 
\end{equation}
with 
%  $\displaystyle\frac{\partial \Psi^\star}{\partial \phi_k}$ denotes the linear operator in $\mathcal{L}\big(L_\wedge^2(\Omega^N);\0\big)$ that is defined by
\[
\frac{\partial \Psi}{\partial \phi_k}^\star[\Xi](x)=N\int_\Omega\phi_k(y)\Big(
\int_{\Omega^{N-1}}\Xi(x,x_2,\ldots,x_N)\,\overline\Psi(y,x_2,\ldots,x_N)\,dx_2\cdots dx_N \Big)dy
\]
for all $1\leq k\leq K$, for any function $\Xi$ in $L_\wedge^2(\Omega^N)$. 
%%%%%%%
\vskip6pt
%%%%%%
It is also  worth emphasizing the fact  that changing $ (C,\Phi) $ to $ (C',\Phi') $ following the group action~\eqref{Action}, involves a straightforward change of ``variable" in the derivation of $ \Psi $; namely,  with a straightforward chain rule, 
\begin{equation}\label{transfo-grad}
\nabla_{C} \:\Psi=\mathbb{U}^\star\cdot \nabla_{C' } \:\Psi=\nabla_{C' } \:\Psi\cdot d(U),\quad \nabla_{\Phi} \:\Psi = \nabla_{\Phi' } \:\Psi\cdot U  
\end{equation}
and
\begin{equation}\label{chainrule}
[\nabla_{\Phi' } \:\Psi ]^\star = U \cdot [\nabla_{\Phi} \:\Psi ]^\star.
\end{equation}
%%%%%%%%%%%%%%%%%%%%%%%%%%%%%%%%%%%%%%%%%%%%%%%
The following further properties of  the functional derivatives of $ \Psi $ will help  to link  the full-rank assumption with the possibility   for $\pi$ to be  a local diffeomorphism in a  neighborhood of $\Psi_0=\pi(C_0,\Phi_0)\in \partial\mathcal{B}_{N,K}$. %%%%%%%
\begin{lem}\label{termsforVP}
Let   $ (C,\Phi) \in \mathcal{F}_{N,K}$ with $\Psi=\pi(C,\Phi)$. Then, for  all $\zeta \in \Span\{\Phi\}^\bot$, $\xi\in \0$  and  $\sigma,\tau \in \Sigma_{N,K}$,   we have
\begin{equation}\label{claim1}
\Bigl\langle\frac{\partial  \Phi_\tau}{\partial \phi_k}\:[\zeta]\:\Big\vert\: \Phi_\sigma \Bigr\rangle =0,
\end{equation}
and 
\begin{equation}\label{eq:termsforVP}\Bigl\langle \frac{\partial  \Psi}{\partial \phi_k} \:[\zeta]\:\Big\vert\: \frac{\partial  \Psi}{\partial \phi_l}  \:[\xi]\Bigr\rangle = \G_{lk} \, \left\langle \zeta,\xi\right\rangle, 
\end{equation}
for any $1\leq k,l\leq K$.
\end{lem}
%%%%%%%%%%%%%%%%%%%%%%%
\begin{proof}
The first  claim follows immediately in virtue  of \eqref{derivee-det} and \eqref{ps-det}. For the second claim we proceed as follows. Thanks to \eqref{derivee-det} again
\begin{align*}
\Bigl\langle \frac{\partial  \Psi}{\partial \phi_k} \:[\zeta]\:\Big\vert\: \frac{\partial  \Psi}{\partial \phi_l}  \:[\xi]\Bigr\rangle
 &= \sum_{\sigma,\tau\:\vert\:k\in\sigma,\,l\in\tau} c_\sigma\,\overline{c}_\tau\,\Bigl\langle \frac{\partial  \Phi_\sigma}{\partial \phi_k} \:[\zeta]\:\Big\vert\: \frac{\partial  \Phi_\tau}{\partial \phi_l}  \:[\xi]\Bigr\rangle  \\ 
 &= \sum_{\substack{\sigma,\tau\:\vert\:k\in\sigma,\,l\in\tau\\ \sigma\setminus\{k\}= \tau\setminus\{ l \}}} (-1)^{\sigma^{-1}(k)} (-1)^{\tau^{-1}(l)}\, c_\sigma\,\overline {c}_\tau\, \left\langle \zeta,\xi\right\rangle\\
&= \G_{lk}\:\Bigl\langle\zeta ,\xi\Bigr\rangle.
\end{align*}
We conclude with the help of \eqref{gammaij}.
\end{proof}
%%%%%%%%%%%%%%%%%%%%%%%%%%%%%%%%%%%%%%%%%%%%%%%

From  \eqref{compactpsi}, \eqref{deriv-C}, \eqref{deriv-Phi} and \eqref{frechetderivatice}, the tangent space of $\partial \mathcal{B}_{N,K}$ at $\Psi=\pi(C,\Phi)$ is given by 
\begin{equation}\label{plantgt}
\begin{split}
\mathbf{T}_\Psi \partial\mathcal{B}_{N,K}&=\bigg\{\delta\Psi = \sum_\sigma  \Phi_\sigma\, \delta c_\sigma + \frac1N\:\sum_\sigma\sum_{k=1}^K c_\sigma\,\frac{\partial\Phi_\sigma}{\partial \phi_k}[\delta\phi_k]\,\in L^2_\wedge(\Omega^N)\::
\\
\delta C=&\big(\delta c_{\sigma_1},\cdots,\delta c_{\sigma_r}\big)\in \C^r,\; \delta\phi_k\in \Span\{\Phi\}^\bot, \; \text{for every } 1\leq k\leq K \bigg\}. 
\end{split}
\end{equation}
%Note that in the above formula  the variations $ \delta\phi_i $ lie on different space variables. This implicit dependency is hidden in the expression of $\frac{\partial\Psi}{\partial \phi_i}$ according to \eqref{frechetderivatice}.
%%%%%%%%
Note that the tangent space \eqref{plantgt} only depends on the basis point $\Psi$ and not on the choice of coordinates $(C,\Phi)$  in  the corresponding fiber. In Physicists' terminology this is the space of allowed variations around $(C,\Phi)$ in $\mathcal{F}_{N,K}$ according to the constraints \eqref{l2unitsphere} and \eqref{Crunitsphere} on the expansion coefficients and the orbitals respectively. 
\vskip6pt
Let $\Psi_0=\pi(C_0,\Phi_0)$ be in $\mathcal{B}_{N,K}$ with  invertible $\G(C_0)$. Then the local mapping theorem at $(C_0,\Phi_0)$  allows to define a  so-called \textit{section} $\pi^{-1}: \Psi\mapsto (C,\Phi)$ as a $C^1$ diffeomorphism in the  neighborhood of $\Psi_0$. According  to   \eqref{plantgt}, we have to check that $(0,0)$ is the only solution in $\C^r\times   \Span\{\Phi_0\}^\bot$ to 
\begin{equation}\label{inverse}
 d\pi_{(C_0,\Phi_0)}(\delta C, \delta\Phi)= \sum_\sigma  \Phi_\sigma\, \delta c_\sigma + \frac 1 N\:\sum_{k=1}^K \frac{\partial\Psi}{\partial \phi_k}[\delta\phi_k]=0\,.
 \end{equation}
Indeed, on the one hand, if  we  scalar product the above equation with $\Phi_\tau$ for any  $\tau \in\Sigma_{N,K}$ we obtain $\delta C=0$ in virtue of the orthonormality of Slater determinants and \eqref{claim1}. On the other hand, for a given $1\leq l\leq K$ and  any $\xi\in \0$, the  scalar product of  \eqref{inverse}  with $\frac{\partial\Psi}{\partial \phi_l}[\xi]$ yields 
\[
\sum_{k=1}^K \Bigl\langle \frac{\partial  \Psi}{\partial \phi_k} \:[\delta\phi_k]\:\Big\vert\: \frac{\partial  \Psi}{\partial \phi_l}  \:[\xi]\Bigr\rangle =\sum_{k=1}^K  \G_{lk} \, \left\langle \delta\phi_k,\xi\right\rangle=\Big\langle \big(\G\delta\Phi\big)_l, \xi\Big\rangle=0
\]
thanks to \eqref{eq:termsforVP}.  Since $\xi$ is arbitrary in $L^2$  and since $\G$ is invertible this is equivalent to $\delta\Phi=0$, hence the result. The full-rank property is mandatory   for lifting  continuous paths $t\mapsto \Psi(t)$ on the basis $\partial\mathcal{B}_{N,K}$  to continuous paths $t\mapsto \big(C(t),\Phi(t)\big)$ on $\partial \mathcal{F}_{N,K}$. 
%%%%%%%%%%%%%%%%%
\subsection{Interpretation in terms of  quantum physics}

The wave-function $\Psi \in L^2(\Omega^N)$ with $\Vert\Psi\Vert=1$  is interpreted through the square 
of its modulus $|\Psi(X_N)|^2$ ($=\bigl[\Psi\otimes\Psi\bigr]_{:N}(X_N,X_N)$) 
that represents the density of probability of presence of  the $N$ electrons in $\Omega^N$.   
More generally, for all $1\leq n\leq N$,  the positive function $X_n\mapsto \bigl[\Psi\otimes\Psi\bigr]_{:n}(X_n,X_n)$ is in $L^1(\Omega^n)$ with $L^1$ norm equal 
to ${ N \choose n }$, and it is interpreted as ${ N \choose n }$ times the density of probability 
for finding $n$ electrons  located at $X_n\in \Omega^n$.  
Any set $\{\sigma(1),\ldots,\sigma(N)\}$ for $\sigma \in \Sigma_{N,K}$ is called  
a \textit{configuration} in quantum chemistry literature and this is where the 
terminology \textit{multi-configuration} comes from for wave-functions in $\mathcal{B}_{N,K}$.  
When $\{\phi_k\}_{1\leq k\leq K}$ is an orthonormal basis of $\mathrm{Ran}\bigl[\Psi\otimes\Psi\bigr]_{:1}$
each mono-electronic function $\phi_k$ is called an \textit{orbital} of $\Psi$. 
%For non-interacting electrons orbitals generate exact solutions of the $N$-body time-dependent Sch\"{o}dinger operator.  
When the orbitals are also  eigenfunctions  of $[\pi(C,\Phi)\otimes \pi(C,\Phi)]_{:1}$  
according to \eqref{naturaldensity} they are referred to  as \textit{natural orbitals }  
in the literature whereas the associated eigenvalues $ \{\gamma_i\}_{1\leq i \leq K   } $ are 
referred to as  \textit{occupation numbers}.  
Under the full-rank assumption, only occupied orbitals are taken into account. 
The  functions with $N-1$ variables  $\int_{\Omega}\: \Psi(x_1,\ldots,x_N)\: \overline{\phi}_k(x_i) \:dx_i$ that appear in formula \eqref{frechetderivatice}
 are known 
as  a {\it single-hole function} (see \textit{e.g.} \cite{Meyer,Scrinzi}).  
Finally,  the $K\times K$ matrix $\G(C)$  is called \textit{the charge- and bond matrix}  
(see L\"owdin~\cite{Lowdin}). 
\medskip

``Correlation" is a key concept for many-particle systems. 
Whereas the ``correlation energy" of a many particle wave-function associated to a many particle
Hamiltonian is a relatively well-defined concept, the intrinsic correlation of a many particle wave-function
as such is a rather vague concept, with several different definitions in the literature (see among others \cite{Grobe,Gottlieb2} and the references therein).
In \cite{Gottlieb2} Gottlieb and Mauser recently introduced a new measure for the correlation. 
This {\it non-freeness} is an entropy-type functional depending only on the density operator$ [\Psi\otimes \Psi]_{:1}$,
and defined as follows 
\[
\mathfrak{E}(\Psi) =-{\rm Tr}\biggl\lbrace[\Psi\otimes \Psi]_{:1}\log([\Psi\otimes \Psi]_{:1})\biggr\rbrace-{\rm Tr}\biggl\lbrace[(\mathbf{1}-[\Psi\otimes \Psi]_{:1})\log(\mathbf{1}-[\Psi\otimes \Psi]_{:1})\biggr\rbrace.
\]
Hence it depends on the eigenvalues of $  [\Psi\otimes \Psi]_{:1}$ in the following explicit way
\begin{equation*} 
\mathfrak{E}(\Psi) =- \sum_{i=1}^K \:\biggl (\gamma_i\log(\gamma_i) + (1-\gamma_i)\log(1-\gamma_i)\biggr)
\end{equation*}
It is a concave functional minimized for $ \gamma_i=0 $ or $1$. In the MCHF case this functional depends implicitly on $ K $ and $ N$ {\it via} the dependency on the $ \gamma_i'$s.  This definition of correlation has the  basic property that the correlation vanishes if and only if $ \Psi $ is a single Slater determinant. The simple proof is  based on the L\"owdin expansion theorem  (see Proposition~\ref{Lowdin} and Remark~\ref{rem:HF}).

The single Slater determinant case is usually taken as the definition of uncorrelated  wave-functions. The Hartree-Fock ansatz is not able to catch ``correlation effects". When there is no  binary interaction  the Schr\"odinger equation propagates Slater determinant (see Subsection~\ref{ssec:free}). However,   the interaction of the particles would immediately create ``correlations" in the time evolution even if  the initial data is a single Slater determinant, - 
however, the TDHF method forces the dynamics to stay on a manifold where correlation
is always zero.
\vskip6pt
Improving the approximation systematically by adding determinants  brings in 
correlation into the {\it  multi-configuration ansatz}. Now correlation effects of
the many particle wave-function can be included in the initial data and the effects
of dynamical ``correlation - decorrelation" can be caught in the time evolution. 
This is a very important conceptually advantage of MCTDHF for the modeling and
simulation of correlated few electron systems. Such systems, for example in ``photonics"
where an atom interacting with an intense laser is measured on the femto- or atto-second scale, 
are increasingly studied and have given a boost to MCTDHF (see \textit{e.g.}
\cite{Scrinzi},\cite{ScrinziNature}).

%%%%%%%%%%%%%%%%%%%%%%%%%%%%%%%%%%%%%%%%%
\section[Flow on the fiber bundle]{Flow on the Fiber Bundle}\label{sec:flow}
In this section, we consider a general  self-adjoint operator $\mathcal{H}$ in $L^2(\Omega^N)$. Most calculations here  stay at the  formal level with no consideration of functional analysis. Solutions are meant  in the classical sense and   in the domain of the operator $\mathcal{H}\,.$ In  Section~\ref{sec:analysis} below physical problems will be considered and details concerning proof of existence,  uniqueness of solutions  and blow-up alternatives in the appropriate functional spaces  will be given. 

From  this point onward, $T>0$ is fixed.  A key point of the time-dependent case is that the  set of ansatz $\mathcal{B}_{N,K}$ is not invariant by the Schr\"odinger dynamics. 
It is even  expected (but so far not proved to our knowledge) that the solution of the 
exact Schr\"odinger equation \eqref{exact} with initial data in $\mathcal{B}_{N,K}$ 
for some finite $K\geq N$ features an infinite rank at any positive time as long as 
many-body potentials are involved  (see \cite{Friesecke2} for related issues on the 
stationary solutions and Subsection~\ref{ssec:free} for the picture for non-interacting electrons).  We therefore have to rely on an approximation procedure that forces the  solutions to stay 
on the set of ansatz for all time. In Physics' literature, the MCTDHF equations are usually (formally) derived from the so-called \textit{Dirac--Frenkel variational principle} (see, among others, \cite{Dirac,Frenkel, Lubich1} and the references therein) that demands that for all $t\in [0,T]$, $\Psi=\Psi(t)\in \mathcal{B}_{N,K}$ and 
\begin{equation}\label{DF}
\Bigl\langle \: i\frac{\partial \Psi}{\partial t} - \mathcal{H} \Psi\:\Big\vert\:\delta \Psi\Bigr\rangle =0,\quad \text{for all}\quad \delta\Psi \in \mathbf{T}_\Psi \partial\mathcal{B}_{N,K}, 
\end{equation}
where  $\mathbf{T}_\Psi \mathcal{B}_{N,K}$ denotes the tangent space to the differentiable manifold $\partial\mathcal{B}_{N,K}$ at $ \Psi$. Equivalently, one solves  
 \begin{equation}\label{VP-argmin}
 \Psi(t)=\textrm{argmin}\Big\{ \Vert i\,\frac{\partial \Psi}{\partial t}-\mathcal{H}\,\Psi\Vert_{L^2(0,T;L^2(\Omega^N))} \::\:\Psi\in \mathcal{B}_{N,K}
\Big\}
\end{equation}
for every $T>0$ (see \cite{Lubich2}). A continuous flow  $t\mapsto \Psi(t)\in \partial \mathcal{B}_{N,K}$  on  $ [0,T]$ may be lifted by  infinitely many   continuous  flows $t\mapsto \big(C(t);\Phi(t)\big)$ foliating  the fibers  $\partial \mathcal{F}_{N,K}$ that  are related by the transitive action of a continuous family of unitary transforms. So called gauge transforms allow then to pass from one  flow $t\mapsto\big(C(t),\Phi(t)\big)$  to another (equivalent) flow $t\mapsto \big(C'(t),\Phi'(t)\big)$ such that $\Psi(t)=\pi\big(C(t),\Phi(t)\big)=\pi \big(C'(t),\Phi'(t)\big)$. 
This is illustrated   on Figure~\ref{fibration} below. 

One choice of gauge amounts to  imposing 
\begin{equation}\label{gauge0}
\big\langle \frac{\partial \phi_i}{\partial t},\phi_j\big \rangle=0\quad \textrm{ for all }1\leq i,j\leq K
\end{equation}
to the time-dependent orbitals. Formally the minimization problem \eqref{VP-argmin} under the constraints  $\Psi=\pi(C,\Phi)$, $(C,\Phi)\in \mathcal{F}_{N,K}$ along with \eqref{gauge0} leads to  the  following system of coupled differential equations
\[
\mathcal{S}_0:\quad \left\lbrace
\begin{array}{ll}
 &\displaystyle i\:\frac{d C}{dt}=\bigl\langle \mathcal{H}\:\Psi\:\vert\: \nabla_C\Psi\bigr\rangle,\\
 \vspace{2mm}
& \displaystyle i \:\G\big(C(t)\big)\:\frac{\partial \Phi}{\partial t} =  (\mathbf{I}- \mathbf{P}_\Phi) \:\nabla_\Phi \Psi^\star[\mathcal{H}\:\Psi],\\
\vspace{2mm}
&\big(C(0),\Phi(0)\big)=\big(C_0,\Phi_0\big),    
\end{array}
\right.
\]
for a given initial data $\big(C_0,\Phi_0\big)$ in $\mathcal{F}_{N,K}$. This system will be referred to as the \textit{variational system} in the following.

The operator $ \mathbf{P}_\Phi $ in $ \mathcal{S}_0$ denotes the projector onto the space spanned by the $ \phi_i'$s. More precisely
\begin{equation}\label{projector}
\mathbf{P}_\Phi  (\cdot) = \sum_{i=1}^K \bigl\langle\cdot\:,\:\phi_i\bigr\rangle \:\phi_i.
\end{equation}
Actually one checks that 
\[
\nabla_\Phi \Psi^\star[\mathcal{H}\:\Psi]=\nabla_{\bar{\Phi}}\Bigl\langle  \mathcal{H} \:\Psi\:\vert\:\Psi\Bigr\rangle. \]
Up to the   Lagrange multipliers associated to \eqref{gauge0} the right-hand side in the variational system corresponds to the Fr\'echet  derivatives of the energy expectation  $ 
\mathcal{E}(\Psi) = \Bigl \langle \mathcal{H}\:\Psi\:\vert\: \Psi\Bigr\rangle$ with respect to the conjugate (independent) variables $\bar C$ and $\bar \Phi$.

The variational system~$\mathcal{S}_0$  is   well-suited  for checking  energy  conservation and constraints propagation over the flow as shown in Subsection~\ref{ssec:conservation} below. However it  is  badly adapted for proving existence of solutions for the Cauchy problem or for designing numerical codes.   Equivalent representations of  the MCTDHF equations over different fibrations is made rigorous in Subsection~\ref{gauge-transforms}. In particular, we prove below that the variational system is unitarily (or gauge-) equivalent to System~\eqref{GT0} -- named \textit{working equations} -- whose mathematical analysis   in the physical case is the aim of Section~\ref{sec:analysis}.
%%%%%%%%%%%%%%%%%%%
\begin{rem} Since for every $\sigma\in \Sigma_{N,K}$,  $\displaystyle \frac{\partial \Psi}{\partial c_\sigma}=\Phi_\sigma$, the  system for the $c_\sigma$'s   can also be expressed as
\begin{equation}
i\:\frac{d c_\sigma}{dt}= \sum_{\tau} \:\langle \mathcal{H}\:\Phi_\tau \:\vert\: \Phi_\sigma\bigr\rangle \:c_\tau\label{gal}
.\end{equation}
This equation is then  obviously linear in the expansion coefficients. Furthermore,  when the $\phi_i $'s (or equivalently the $\Phi_\sigma$'s) are kept  constant  in time, (\ref{gal}) is nothing but  a Galerkin approximation to the exact Schr\"{o}dinger equation \eqref{exact}. The MCTDHF approximation then reveals as a generalization to a combination of time-dependent basis functions (with extra degree of freedom in the basis functions) of the Galerkin approximation. 
%However since the system $ \mathcal{S}_0$ is coupled, the Galerkin approximation cannot be seen as a special case of the MCTDHF unless $\nabla_\Phi\Psi^\star[\mathcal{H}\:\Psi]      
% \in \Span\{\Phi\}^\bot $ and $ \G(C)$ is of rank $K$ . \fbox{remarque cas discret}
\end{rem}
%%%%%%%%%%%%%%%%%%%%%%%%%%%%%%%%%%%%%%%%%%%%%%%%%%%%%%%%%%%%%%%%%%%%%%%%%%%%%%%%%%%%%%%%%%%%%%%%%%%%%%
\subsection{Conservation Laws}\label{ssec:conservation}
In this subsection, we assume  the full-rank assumption  on the time interval $[0,T)$; that is $\G\big(C(t)\big)$ is invertible for every $t\in [0,T]$. We check  here that  the expected  conservation laws (propagation of constraints, conservation of the energy) are granted by the variational  system.  Recall that to avoid technicalities all calculations in this section are formal but would be rigorous for regular  classical solutions.   We start with the following
%%%%%%%%
\begin{lem}[The dynamics preserves $\mathcal{F}_{N,K}$]\label{cons-contrainte}
Let  $ (C_0,\Phi_0) \in \mathcal{F}_{N,K} $ being the   initial data.  If there exists a solution to  the system $\mathcal{S}_0$ on $[0,T]$  such that $\mathrm{rank}\,\G\big(C(t)\big)=K$ for all  $t\in [0,T]$, then 
\[
\sum_{\sigma} |c_\sigma(t)|^2=1,\quad
\s\phi_i(t)\,\bar{\phi}_j(t)\,dx=\delta_{i,j},  \quad  \textrm{ for all  }t\in [0,T].
\]
 \end{lem}
 %%%%%%%%%%%%%%
\begin{proof} First we prove that $\sum_{\sigma}|c_\sigma(t)|^2=\sum_{\sigma}|c_\sigma(0)|^2$ for all $t\in [0,T]$. By  taking the scalar  product of the  differential equation  satisfied by $C$ in  $ \mathcal{S}_0 $ with  $ C$ itself, we get 
\[
\frac{d}{dt} |C(t)|^2 = 2\: \Re \big(\frac{d}{dt}C(t), C(t) \big)
=2 \:\Im\sum_{\sigma} \Bigl\langle\mathcal{H}\:\Psi\:\vert\:c_\sigma \:\Phi_\sigma\Bigr\rangle=2\: \Im \Bigl\langle\mathcal{H}\:\Psi\:\vert\:\Psi\Bigr\rangle= 0,
\]
thanks to the self-adjointness of  $\mathcal{H}$, where  $ \Re $ and $ \Im$ denote respectively real and  imaginary parts of a complex number.  From the other hand, the full-rank assumption allows to reformulate  the second equation in ($\mathcal{S}_0$)  as 
\begin{equation}\label{eq-Phi-FRA}
 i \:\frac{\partial \Phi }{\partial t} =  (\mathbf{I} -\mathbf{P}_\Phi ) \:\G(C)^{-1}\:\nabla_\Phi\: \Psi^\star[\mathcal{H}\:\Psi]      
. 
\end{equation} 
(Notice that $ \mathbf{P}_\Phi $ commutes with $ \G(C)^{-1}$.) By definition $\mathbf{I} -\mathbf{P}_\Phi  $ projects on the orthogonal  subspace of $\Span\{\Phi\}$, therefore  $\frac{\partial}{\partial t}\phi_i$   lives in $ \Span\{\Phi\}^\bot $ for all $t$.  Hence, 
\begin{equation}\label{orthvp}
\Bigl\langle\frac{\partial \phi_i(t)}{\partial t}  \:,\:{\phi}_j(t)\Bigr\rangle = 0.
\end{equation}
for all $1\leq i,j\leq K$ and  for all $t\in [0,T]$. This achieves the proof of the lemma.
\end{proof}
We now  check  that solutions to the variational system  indeed agree with  the Dirac-Frenkel variational principle.
%%%%%%%%%%%%%%%%%%
\begin{prop}[Link with the  Dirac-Frenkel variational principle]\label{VPformula}
 Let $ (C,\Phi)  \in \partial\mathcal{F}_{N,K}$ be a classical solution to  $ \mathcal{S}_0 $ on $[0,T]$. Then,  $ \Psi=\pi(C,\Phi)  $ satisfies 
 the Dirac--Frenkel variational principle \eqref{DF}.
\end{prop}
%%%%%%%%%%%%%%%%%%
\begin{proof}
We start with  the characterization \eqref{plantgt} of the elements in $\mathbf{T}_\Psi \partial\mathcal{B}_{N,K}$. Since the full-rank assumption is satisfied  on $[0,T]$, the orbitals satisfy \eqref{eq-Phi-FRA}, and therefore  $\frac{\partial \phi_k}{\partial t}\in \Span\{\Phi\}^\bot$   for all $t\in [0,T]$ and   $1\leq k\leq K$. Firstly, being given $ \sigma \in \Sigma_{N,K}$, we  have
\begin{align}
\Bigl\langle i\frac{\partial\Psi}{\partial t} - \mathcal{H} \Psi  \Big\vert\frac{\partial  \Psi}{\partial c_{\sigma}} \Bigr\rangle &=i\sum_\tau\frac{d c_\tau}{dt}\: \Bigl\langle   \Phi_\tau  \big\vert  \Phi_\sigma  \Bigr\rangle - \Bigl\langle \mathcal{H}\: \Psi  \big\vert  \Phi_\sigma \Bigr\rangle+ i\sum_\tau c_\tau\:\Bigl\langle  \frac{\partial \Phi_\tau}{\partial t}  \big\vert  \Phi_\sigma \Bigr\rangle \label{part1line2} \\ 
&=  i\:\frac{d c_\sigma}{dt} -\Bigl\langle \mathcal{H}\, \Psi  \big\vert \Phi_\sigma \Bigr\rangle=0, \nonumber
\end{align}
thanks to the equation satisfied by $c_\sigma$. Indeed, 
\[ \frac{\partial \Phi_\tau}{\partial t}=\sum_{k=1}^K \frac{\partial \Phi_\tau}{\partial \phi_k}\:\big[\frac{\partial \phi_k}{\partial t}\big]
\]
and therefore the sum in  \eqref{part1line2} vanishes thanks to Lemma~\ref{termsforVP}. Secondly, for every  $ 1\leq k \leq K$ and for any function $\zeta$ in $\Span\{\Phi\}^\bot$, we have
\begin{align}
\Bigl\langle i\frac{\partial \Psi}{\partial t}&- \mathcal{H}\Psi\big\vert \frac{\partial \Psi}{\partial \phi_k}[\zeta]\Bigr\rangle\nonumber\\
&=  i\:\sum_\sigma \frac{d c_\sigma}{dt}\Bigl\langle \Phi_\sigma\vert\frac{\partial \Psi}{\partial \phi_k} [\zeta] \Bigr\rangle  +i\:\sum_{j=1}^K\Bigl\langle \frac{\partial \Psi}{\partial \phi_j}\big[\frac{\partial \phi_j}{\partial t}\big]  \:\big\vert\: \frac{\partial \Psi}{\partial \phi_k}[\zeta] \Bigr\rangle  -\Bigl\langle \:\mathcal{H} \Psi\big\vert \frac{\partial \Psi}{\partial \phi_k}[\zeta]\Bigr\rangle \label{line1VPproof} \\ 
&= \Bigl\langle\:i\:\Big(\G\big(C(t)\big) \cdot \frac{\partial \Phi}{\partial t}\Big)_k -  \frac{\partial \Psi^\star}{\partial \phi_k}[\mathcal{H}\Psi ]\:,\:\zeta \: \Bigr\rangle \nonumber%\label{line2VPproof}
\\
&= -\Bigl\langle\mathbf{P}_\Phi\:\frac{\partial \Psi^\star }{\partial \phi_k}[\mathcal{H}\Psi]\:, \: \zeta \:\Bigr\rangle  =0.\label{line3VPproof}
\end{align}
Indeed, on the one hand, in virtue of  Lemma~ \ref{termsforVP}, the first term in the right-hand side of \eqref{line1VPproof} vanishes   whereas the second one identifies  with $ i \:\sum_{j=1}^K \G_{kj}(C)\,\big \langle \frac{\partial \phi_j}{\partial t}, \zeta\big\rangle=\Big\langle\Big(\G\big(C(t)\big) \cdot \frac{\partial \Phi}{\partial t}\Big)_k\:,\:\zeta\Big\rangle$  since $\frac{\partial \phi_j}{\partial t}$  and $\zeta$ both belong to $\Span\{\Phi\}^\bot$. On the other hand, the last line   \eqref{line3VPproof}  is obtained using the equation satisfied by $\Phi$  in  $ \mathcal{S}_0$ and by observing that   $ \mathbf{P}_\Phi \:\zeta=0$ since $\zeta\in \Span\{\Phi\}^\bot$. The proof is complete.
\end{proof}
Let us now recall the definition of the energy  
\begin{equation*}
\mathcal{E}(\Psi)=\mathcal{E}\big(\pi(C,\Phi)\big) = \Bigl \langle \mathcal{H}\:\Psi\:\vert\: \Psi\Bigr\rangle. 
\end{equation*}
It is clear that $\mathcal{E}(\Psi) $ depends on time {\it via} $ (C(t),\Phi(t))$. As a corollary  to Proposition~\ref{VPformula}  we have  the following
%%%%%%%%%%%%%%%
\begin{cor}[Energy is conserved  by the flow]\label{energyconservation}
Let    $ (C,\Phi)  \in \partial\mathcal{F}_{N,K}$ be a solution to  $ \mathcal{S}_0 $ on $[0,T]$  such  that  $\pi(C(t),\Phi(t))$ lies   in the  domain of $\mathcal{H}$ (or in the ``form domain" when $\mathcal H$ is semi-bounded) for all  $t$ in $[0,T]$. Then,   
\[
\mathcal{E}\big(\pi(C(t),\Phi(t))\big)=\mathcal{E}\big(\pi(C^0,\Phi^0)\big)\quad \textrm{ on } [0,T].\]
\end{cor}
%%%%%%%%%%%
\begin{proof}
Comparing with  \eqref{plantgt} we  observe that $ \frac{\partial \Psi}{\partial t}  \in\mathbf{T}_\Psi \partial\mathcal{B}_{N,K}$, for 
\[
 \frac{\partial \Psi}{\partial t} =\sum_{\sigma}\frac{dc_\sigma}{dt}\,\Phi_\sigma+\frac{1}{N}\,\sum_{\sigma}\sum_{k=1}^K c_\sigma\,\frac{\partial \Phi_\sigma}{\partial \phi_k}\Big[\frac{\partial\phi_k}{\partial  t}\Big],\]
with  $\frac{\partial\phi_k}{\partial  t}$  in $\Span\{\Phi\}^\bot$ whenever $\G(t)$ is invertible.  Then, applying Proposition~\ref{VPformula} to $\delta\Psi= \frac{\partial \Psi}{\partial t}$ one obtains
\begin{equation}
0= \Re \ys i\frac{\partial \Psi}{\partial t}-\mathcal{H}\:\Psi\:\big \vert\: \frac{\partial \Psi}{\partial t}\:\ym= -\:\Re  \Bigl\langle \mathcal{H}\Psi\:\vert\:\frac{\partial \Psi}{\partial t} \Bigr\rangle\\= -\frac{1}{2} \frac{d}{dt} \Bigl \langle \mathcal{H}\:\Psi\:\vert\: \Psi\Bigr\rangle.
\end{equation}
Hence the result.
\end{proof}
%%%%%%%%%%%%%%%%%%%%%%%%%%%%%%%%%%%%%%%%%%%%%%%%%%%%%%%%%%%%%%%%%%%%%%%%%%%%%%%%%%%%%%%%%%%%%%%%%%%%%%

%%%%%%%%%%%%%%%%%%%%%%%%%%%%
\subsection{An \textit{a posteriori} error estimate}
%%%%%%%%%%%%%%%%%%%%%%%%%%%%
We  establish an  error bound in $ L^2(\Omega)^N $  for the MCTDHF approximation compared with the exact solution to the linear TDSE~\eqref{exact}. Let us introduce the projection  $ \mathcal{P}_{\mathbf{T}_\Psi\partial\mathcal{B}_{N,K}} $ onto the tangent space $\mathbf{T}_\Psi\partial\mathcal{B}_{N,K} $  to $\partial \mathcal{B}_{N,K} $ at $ \Psi $. Then, we claim
%%%%%%%%%%%%%%
\begin{lem}\label{errorbound}
Given an initial data $ (C^0,\Phi^0) \in \partial\mathcal{F}_{N,K}$ and  an exact solution  $ \Psi_E $ to  the $ N$-particle Schr\"odinger equation~\eqref{exact}. Then, as long as $(C,\Phi)$ is a solution to $\mathcal{S}_0$ in $\partial\mathcal{F}_{N,K}$, we have for $\Psi(t) =\pi (C(t),\Phi(t))$ and $\Psi^0=\pi(C^0,\Phi^0)$ the estimate:
\[
\|\Psi_E-\Psi\|_{L^2(\Omega^N)} \leq \|\Psi_E(0)-\Psi^0\|_{L^2(\Omega^N)}  +\int_0^t \left\Vert(I- \mathcal{P}_{\mathbf{T}_\Psi\partial\mathcal{B}_{N,K}}) \:[\mathcal{H}\: \Psi(s)]\right\Vert_{ L^2(\Omega)^N}\: ds. 
\]
%Provided that the R.H.S is finite.
\end{lem}
%%%%%%%%%%%%%%
\begin{proof}
First,  Proposition~\ref{VPformula}  expresses the fact that $ \mathcal{P}_{\mathbf{T}_\Psi\partial\mathcal{B}_{N,K}}\big(i\frac{\partial \Psi}{\partial t} - \mathcal{H}\Psi\big) =0 $. Therefore the equation satisfied by the ansatz $\Psi$ is:
\begin{equation}
\label{equation appro}i\frac{\partial \Psi}{\partial t} - \mathcal{H}\Psi= (I-\mathcal{P}_{\mathbf{T}_\Psi\partial\mathcal{B}_{N,K}}) \:\Big[i\frac{\partial \Psi}{\partial t} -\mathcal{H}\:\Psi\Big]=  -(I-\mathcal{P}_{\mathbf{T}_\Psi\partial\mathcal{B}_{N,K}}) \:[\mathcal{H}\:\Psi],  
\end{equation}
since  $\frac{\partial \Psi}{\partial t} $ lives  in the tangent space $  \mathbf{T}_{\Psi}\partial\mathcal{B}_{N,K} $. Next, subtracting \eqref{equation appro} from \eqref{exact}, we get
\begin{equation}\label{error}
i\frac{\partial (\Psi_E-\Psi)}{\partial t} - \mathcal{H}(\Psi_E-\Psi)  = - (I-\mathcal{P}_{\mathbf{T}_\Psi\partial\mathcal{B}_{N,K}}) \:[\mathcal{H} \:\Psi]
\end{equation}
Then, we apply  the PDE above to  $\Psi_E-\Psi  $ and we integrate formally over $ \Omega^N $. The result follows by  taking  the imaginary of both sides  and by using  the self-adjointness of $ \mathcal{H}$.
\end{proof}
Roughly speaking, the above lemma tells that the closer is $ \mathcal{H}\:\Psi$ to the tangent space $T_\Psi\partial\mathcal{B}_{N,K} $, the better is the MCTDHF approximation. Intuitively, this is true for large values of  $K$. Let us mention  that this bound was already obtained in \cite{Lubich2} and it is probably far from being accurate. However if the MCTDHF algorithm is applied to a discrete model say of dimension $L$ then for $K$ large enough ($K\ge L$ ) this algorithm coincides with the original problem (see Subsection~\ref{sec:discrete}).

%\fbox{Ecrire l'\'equation satisfaite par $\Psi_{MC}$}
%%%%%%%%%%%%%%%%%%%%%%%%%%%%%%%%%%%%%%%%%%%%%%%%%%%%%%%%%%%%%%%%%%%%%%%%%%%%%%%%%%%%%%%%%%%%%%%%%%%%%%
\subsection{Unitary Group Action on the Flow}\label{gauge-transforms}
%%%%%%%%%%%%%%%
The variational system~$\mathcal{S}_0$  is  taylor-made  for checking  energy  conservation and constraints propagation over the flow. However it is  badly adapted for proving existence of solutions for the Cauchy problem or for designing numerical codes.   It is therefore convenient to have at our disposal several explicit  and equivalent representations of  the MCTDHF equations over different foliations  and to understand how they are  related. This is the purpose of this subsection. Proofs of technical lemma and theorems are postponed in the Appendix to facilitate straight reading.  \vskip10pt
%%%%%%%%%%%%%%%%%%%
We start with  the following (straightforward) lemma on regular flows of unitary  transforms :
%%%%%%%%%%%%%%%%%%%
\begin{lem}[Flow of unitary transforms]\label{utile} Let   $U_0\in \mathcal{O}_K$ and let  $t\mapsto U(t)$ be in $C^1\big([0,T); \mathcal{O}_K\big)$ with $U(0)=U_0$.  Then,  $t\mapsto M(t):=-i\,\frac{dU^*}{dt}\,U$ defines a continuous family of $K\times K$ hermitian matrices, and for all $t>0$, $U(t)$ is the unique    solution to the Cauchy problem
\begin{equation}
\left\{
\begin{aligned}
 i\:\frac{dU}{dt}&= U(t) M(t),\\
U(0)&=U_0.
\end{aligned}\right.
\label{polar}
\end{equation}
Conversely, if $t\mapsto M(t)$ is a continuous family of $K\times K$ Hermitian matrices  and if $U_0\in \mathcal{O}_K$ is  given, then (\ref{polar}) defines a unique $C^1$ family  of $K\times K$ unitary matrices.
\end{lem}
%%%%%%%%%%%%%%%%%%%%
The corresponding  flow for  unitary transforms on expansion  coefficients   is as follows:
%%%%%%%%%%%%%%%%%%%%
\begin{cor} \label{tutile} Let $(N,K)$ be an admissible pair,   let  $t\mapsto M(t)$ be  a continuous family of $K\times K$ Hermitian matrices  and let $U^0\in\mathcal{O}_K$. Then,  if  $t\mapsto U(t)$ denotes  the unique family of unitary $K\times K$ matrices that solves \eqref{polar}, the unitary $r\times r$ matrix  ${\mathbb U}$  given by \eqref{Udet} is the unique solution to the differential equation
\begin{equation}\label{basic1}
\left\{
\begin{aligned}
i\frac{d{\mathbb U}}{dt}&={\mathbb U}\, {\mathbb M },\\
 \mathbb{U}(0)&=d\big(U^0\big),
 \end{aligned}
 \right.
\end{equation} 
with
\begin{equation}
 {\mathbb M }_{\sigma, \tau}= \sum_{\substack{i\in\sigma, \,j\in\tau\\ \sigma\setminus\{ i\}= \tau\setminus\{ j \}}} (-1)^{ \sigma^{-1}(i) +\tau^{-1}(j)} M_{ij}.
 \label{truc2} 
 \end{equation}
 \end{cor}
 %%%%%%%%%%%%%%%%%%%%%%%
\vskip10pt
%%%%%%%%%%%%%%%%%%%%
The proof of this corollary is postponed to the Appendix. The   main result of this section is : 
%%%%%%%%%%%%%%%%%%%%%%%%%%
\begin{thm}[Flow of unitary equivalent foliations]\label{gauge} Let $U_0\in \mathcal{O}_K$  and  $(C_0,\Phi_0)\in \partial\mathcal{F}_{N,K}$. 
\vskip6pt
\noindent(i) Let $t\mapsto M(t)$ be  a continuous family of $K\times K$ Hermitian matrices on $[0,T]$  and let   $U(t)\in C^1\big([0,T); \mathcal{O}_K\big)$ be the corresponding solution to \eqref{polar}. Assume that there exists a solution $(C,\Phi)\in C^0\big(0,T;\partial\mathcal{F}_{N,K}\big) $ of $\mathcal{S}_0$ with initial data $(C_0,\Phi_0)$. Then, the couple $(C',\Phi')=\mathcal{U}(t)\cdot (C,\Phi)$ with $\mathcal{U}\in \mathcal{O}^r_K$ defined by \eqref{Action} and \eqref{Udet}  is solution to the system 
\begin{equation}\label{SM}
\left\lbrace
\begin{aligned}
  i\:\frac{d C'}{dt}&=\Bigl\langle  \mathcal{H} \:\Psi\:\vert\:\nabla_{C'}\Psi\Bigr\rangle\:- \mathbb M'\:C',    \\ 
 i \:\G(C')\:\frac{\partial \Phi' }{\partial t} &= (\mathbf{I}-\mathbf{P}_{\Phi'}) \:\nabla_{\Phi'}\Psi^\star[\mathcal{H}\:\Psi] + \G(C')\:M' \:\Phi'\\
 \big(C'(0),\Phi'(0)\big)&=\mathcal{U}_0\cdot (C_0,\Phi_0)
\end{aligned}
\right.
\end{equation}
with  $\Psi=\pi(C,\Phi)=\pi(C',\Phi')$, $\mathcal{U}_0=\big(U_0,d(U_0)\big)\in \mathcal{O}^r_K$ being defined by \eqref{Action} and with 
\[M'=UMU^\star, \quad \overline\M{}'=\U\M\U^\star
,\]
where $\mathbb{M}$ is  the $r\times r$ Hermitian matrix with entries given by \eqref{truc2}. 

\noindent(ii) Conversely, assume that there exists a solution $(C,\Phi)\in C^0\big(0,T;\partial\mathcal{F}_{N,K}\big) $ to  $\mathcal{S}_0$ with initial data $(C_0,\Phi_0)$ and let   $U(t)\in C^1\big([0,T); \mathcal{O}_K\big)$.  Then, the couple $(C',\Phi')=\mathcal{U}(t)\cdot (C,\Phi)$ with $\mathcal{U}\in \mathcal{O}^r_K$ defined by \eqref{Action} and \eqref{Udet}  is a solution to  System \eqref{SM} with $M(t)= -i\,\frac{dU^*}{dt}\,U$. 
\end{thm}
%%%%%%%%%%%%%%%%%%%%%%
\vskip10pt
\begin{rem}[Link with Lagrangian interpretation]  The  equations can  be  derived (at least formally) thanks to the Lagrangian formulation:  One writes  the stationarity condition for  the action 
\[
\mathcal{A}(\Psi)=\int_{0}^{T}\,\Big\langle i\frac{\partial  \Psi}{\partial t}-\mathcal{H} \Psi\big\vert \Psi\Big\rangle\,dt
\]
over   functions $\Psi=\Psi(t)$ that move on $\mathcal{F}_{N,K}$. The associated time-dependent Euler--Lagrange equations take the form \eqref{SM} with $\Psi=\pi(C,\Phi)$, $M$ an hermitian matrix and with $\mathbb{M}$ be the $r\times r$ hermitian matrix  linked to $M$ through Eqn. \eqref{truc2} above. As observed already by Canc\`es and Le Bris \cite{Cances},  even if they appear so, the Hermitian matrices $M$ and $\mathbb{M} $  should not be interpreted as time-dependent Lagrange multipliers associated
to the constraints $(C,\Phi)\in \mathcal{F}_{N,K}$ since the constraints on the coefficients and the orbitals are automatically propagated by the dynamics (see Lemma~\ref{cons-G}), but rather as degrees of freedom  within the fiber at $\Psi\,.$  In particular, this gauge invariance can be used to set  $M$   and $\mathbb{M}$ to zero
for all $ t$ as observed in Lemma~\ref{gauge}  and Eqn.~\eqref{usefulforTDHF} below, so that the above system can be transformed into the simpler system ($\mathcal{S}_0$) we started from. 
\end{rem}
%%%%%%%%%%%%%%%%%%%%%%%%%
%%%%%%%%%%%%%%%%%
As a first example of the change of gauge   one can  use the unitary transforms  to diagonalize the matrix  $\G$  for all time and therefore derive the evolution equations for natural orbitals following \cite{Meyer}
%%%%%%%%%%%%%%%%%%%%%%
\begin{lem}[Diagonal density matrix]\label{Mlemma}
Let $ (C,\Phi)$ satisfying $ \mathcal{S}_0$ with initial data $ (C_0,\Phi_0)$  and let $U_0\in \mathcal{O}_K$ that diagonalizes $\G(C_0)$. We assume  that  for all time the eigenvalues of $\G(C)$ are simple, that is  $\gamma_{i}\neq \gamma_{j}$ for  $ 1\leq i, j\leq K$  and $i\neq j$. Define a $K\times K$ Hermitian matrix by
\[
M_{ij} =\left\{
\begin{array}{ll}
\displaystyle\frac{1}{\gamma_{j}-\gamma_{i}} \left[\Bigl\langle\mathcal{H}\:\Psi\:\vert\:\frac{\partial \Psi}{\partial \phi_i}[\phi_j] \Bigr\rangle -  \Bigl\langle\:\frac{\partial \Psi}{\partial \phi_j}[\phi_i] \:\vert\:\mathcal{H}\:\Psi\Bigr\rangle\right] & \textrm{ if } i\neq j, \vspace{2mm} \\
0 &\textrm{ otherwise }, 
\end{array}
\right.
\]
and consider the family  $ t\mapsto U(t) \in \mathcal{O}_K $  that  satisfies \eqref{polar}  with $U(t=0)=U_0$. Then $ ({C}',{\Phi}') = \mathcal{U}(t)\cdot (C,\Phi) $ is solution to 
\[
\left\lbrace
\begin{aligned}
  i\:\frac{d C'}{dt}&=\Bigl\langle  \mathcal{H} \:\Psi\:\vert\:\nabla_{C'}\Psi\Bigr\rangle\:- \mathbb M'\:C',    \\ 
i \:\gamma_i(t)\:\frac{\partial \phi'_i }{\partial t} &= (\mathbf{I}-\mathbf{P}_{\Phi'}) \:\frac{\partial \Psi} {\partial \phi'_i}^\star[\mathcal{H}\:\Psi] + \gamma_i(t)\:M' \:\Phi'\\
 \big(C'(0),\Phi'(0)\big)&=\mathcal{U}_0\cdot (C_0,\Phi_0)
\end{aligned}
\right.
\]
with the notation of Theorem~\ref{gauge}. In particular,    $ {\G}(C')=\textrm{diag}\big({\gamma}_1(t),\ldots,{\gamma}_K(t)\big)$ for every $t$.
\end{lem}
%%%%%%%%%%%%%%%%%%%%%%
\begin{proof} Using the equation for the coefficients in  \eqref{SM} together with  \eqref{gammaij},  the evolution equation for the coefficients of the density matrix writes
\begin{align*}
i\:\frac{d \gamma_{ij}}{dt} &=\sum_{\substack{\sigma,\tau\::\:i\in\sigma,\:j\in\tau\\ \sigma\setminus\{i\} = \tau\setminus\{j\}} }(-1)^{\sigma^{-1}(i) + \tau^{-1}(j)} \bigl[\left\langle \mathcal{H}\:\Psi\:|\:c_\sigma\:\Phi_\tau\right\rangle - \left\langle c_\tau\:\Phi_\sigma \:|\:\mathcal{H}\:\Psi\right\rangle \bigr] \\
&+ \sum_{\substack{\kappa, \sigma,\tau\::\:i\in\sigma,\:j\in\tau\\\sigma\setminus\{i\} = \tau\setminus\{j\}}  }(-1)^{\sigma^{-1}(i) + \tau^{-1}(j)} \bigl[ \mathbb{M}_{\sigma,\kappa} \:c_\kappa \:\overline{c}_\tau - \mathbb{M}_{\kappa,\tau} \:\overline{c}_\kappa \:{c}_\sigma \bigr]\\
&= \Bigl\langle\mathcal{H}\:\Psi\:\vert\:\frac{\partial \Psi}{\partial \phi_i}\:[\phi_j] \Bigr\rangle -  \Bigl\langle\:\frac{\partial \Psi}{\partial \phi_j}\:[\phi_i] \:\vert\:\mathcal{H}\:\Psi\Bigr\rangle - \sum_{k=1}^K\Bigl\lbrace \:\G_{ik}\:M_{kj} - M_{ik}\:\G_{kj}\: \Bigr\rbrace
\end{align*}
Next, we require that 
\[ \gamma_{ij}(t)= \gamma_{i}\:\delta_{i,j},\quad \text{that is }\quad\frac{d \G_{ij}}{dt}=0\quad \forall\:  1\leq i\neq j \leq K.\]
Using the  above equation, a sufficient condition is  given by 
\begin{equation*}
M_{ij} = \frac{1}{\gamma_{i}-\gamma_{j}}\: \Bigl[\Bigl\langle\mathcal{H}\:\Psi\:\vert\:\frac{\partial \Psi}{\partial \phi_i}[\phi_j] \Bigr\rangle -  \Bigl\langle\:\frac{\partial \Psi}{\partial \phi_j}[\phi_i] \:\vert\:\mathcal{H}\:\Psi\Bigr\rangle \Bigr]
\end{equation*}
This achieves the proof.
\end{proof}
As a second application of Theorem \ref{gauge} we investigate  particular (stationary) solutions or standing waves. A standing wave for the exact Schr\"odinger equation is of the form $\Psi(t,x)=e^{-i\lambda\,t}\Psi(x)$ with $\lambda\in\R\,.$ In the same spirit we  look for solutions $(C',\Phi')$ of System \eqref{SM} with $(C',\Phi')=\mathcal{U}(t)\cdot (e^{-i\lambda\,t}\,C,\Phi)$, where $(C,\Phi)\in \partial\mathcal{F}_{N,K}$ is fixed, independent of time,  and $\mathcal{U}(t)\in\mathcal{O}^r_K$. Using the formulas \eqref{transfo-grad} and \eqref{chainrule} for the changes of variables,  we arrive at 
\[
\left\lbrace
\begin{aligned}
\Big( i\:\frac{d \overline{\mathbb{U}}(t)}{dt}+  \mathbb M\:\overline{\mathbb{U}}+\lambda\,\overline{\mathbb{U}}\Big)\, C&=\overline{\mathbb{U}}\:\Bigl\langle  \mathcal{H} \:\Psi\:\vert\:\nabla_{C}\Psi\Bigr\rangle,    \\ 
 \G(C)\:\Big(i \:U^\star\:\frac{d U}{d t} -U^\star\,M\,U\Big)\:\Phi&= (\mathbf{I}-\mathbf{P}_{\Phi}) \:\nabla_{\Phi}\Psi^\star\:[\mathcal{H}\:\Psi], \\
 U(0)&=\mathbb{I}_K.
\end{aligned}
\right.
\]
In the above system $\Psi=\pi(C,\Phi)$  and $\G(C)$ are independent of time and  $\G(C)$ is invertible.  We start with the equation satisfied by $\Phi$. Observing that the left-hand side lives in $\Span\{\Phi\}$ whereas  the right-hand side lives  in  $\Span\{\Phi\}^\bot$, we conclude that there are both equal to zero. Therefore, there exists a $K\times K$ matrix $\Lambda$ that is independent of $t$ and  such that 
\begin{equation}
\nabla_{\Phi}\Psi^\star[\mathcal{H}\:\Psi]=\nabla_{\overline{\Phi}}\Bigl\langle  \mathcal{H} \:\Psi\:\vert\:\Psi\Bigr\rangle=\Lambda\cdot\Phi.\label{sats} \end{equation}
Also since the left-hand side has to be independent of $t$ we get
\[
i \:\frac{d U}{d t} = M\,U.
\] 
Comparing now with the equation for the coefficients we  infer from Corollary~\ref{tutile} that 
\[
i\:\frac{d \overline{\mathbb{U}}(t)}{dt}=- \mathbb M\:\overline{\mathbb{U}},
\]
hence 
\begin{equation}\label{EL-C}
\nabla_{\overline{C}}\:\Bigl\langle  \mathcal{H} \:\Psi\:\vert\:\Psi\Bigr\rangle=\lambda\,C.\end{equation}
Equations \eqref{sats} and \eqref{EL-C}  are precisely the  MCHF equations  that are satisfied by  critical points of the energy. They were derived by Lewin~\cite{Lewin} in the Coulomb case. The real $\lambda$ is the Lagrange multiplier corresponding to the constraint $C\in S^{r-1}$ whereas the Hermitian matrix $\Lambda$ is the matrix of Lagrange multipliers corresponding to the orthonormality constraints on the orbitals. Existence of such solutions in physical case  is recalled in Section~\ref{sec:analysis}.
\vskip10pt
%%%%%%%%%%%%
The proof of Theorem~\ref{thm-gauge} is postponed in the Appendix and we rather state before some corollaries or remarks.    In Physics' literature  the MCTDHF equations are derived from the variational principle \eqref{DF} under the constraints $\Psi=\pi(C,\Phi)\in \mathcal{B}_{N,K}$  along with additional  constraints on  the time-dependent orbitals   
\begin{equation}\label{gaugeG}
\big\langle \frac{\partial \phi_i}{\partial t} ,\phi_j\big \rangle=\langle \mathbf{G} \phi_i,\phi_j \rangle\quad \textrm{ for all }1\leq i,j\leq K.
\end{equation}
In the above equation  $\mathbf{G}$ is an arbitrary self-adjoint operator on $\0$ possibly time-dependent named the \textit{gauge}.  In this spirit the variational system corresponds to $\mathbf{G}=0$.  Therefore  a gauge field  is chosen \textit{a priori}   and the corresponding equations are derived accordingly.   Both approaches are equivalent by observing that, to every Hermitian matrix $ M $, one can associate a self-adjoint operator $\mathbf{G} $ in $\0$ such that $ M_{ij} =\langle \mathbf{G} \phi_i,\phi_j \rangle$ by demanding that 
\begin{equation*} %\label{FindG}
\mathbf{G} \,\phi_i =  \sum_{j=1}^K M_{ij}\,\phi_j\textrm{ for all }\; 1\leq i\leq K. \end{equation*}
Conversely being given the family $t\mapsto M(t)$ in Theorem~\ref{thm-gauge}  it follows immediately from the system \eqref{SM} that  for all $1\leq i,j\leq K$, 
 \[
i\,\big\langle \frac{\partial\phi_i'}{\partial t},\phi_j'\big\rangle=M'_{ij}.
\]
provided $\G(C')=U\:\G(C)\:U^\star$ is invertible on $[0,T)$. We state below  Theorem~\ref{gauge} that is the  equivalent  formulation  of Theorem~\ref{thm-gauge} in terms of gauge.  It is based on above remarks together with the following :
%%%%%%%%%%%%%%%%%%%%
\begin{lem}\label{utilG} Let   $t\mapsto \mathbf{G}(t)$ be  a  family of self-adjoint operators on $\0$  and let  $\Phi= (\phi_1(t) ,\phi_2(t), \ldots, \phi_K(t) )\in \mathcal{O}_{\0^K}$ such  that such that  $t\mapsto \langle {\mathbf G}(t)\phi_i(t),\phi_j(t)\rangle $ is continuous on $[0,T]\,.$ Then the    matrix  $M$ with entries $M_{ij}(t)=\langle {\mathbf G}(t)\phi_i(t),\phi_j(t)\rangle $ is Hermitian and the Cauchy problem \eqref{polar} defines a globally  well-defined $C^1$ flow on the set of unitary $K\times K$ matrices. In that case, the unitary transforms $\mathbb{U}=d(U)$ solve the Cauchy problem \eqref{basic1} with   $\mathbb{M}$ in   \eqref{truc2} given by 
 \begin{equation}
 {\mathbb M }_{\sigma, \tau}= \sum_{i=1}^N \big\langle \mathbf{G}_{x_i}\,\Phi_\sigma\big|\Phi_\tau\big\rangle.
 \label{truc3} 
 \end{equation}
\end{lem}
%%%%%%%%%%%%%%%%%%%
\begin{rem} In Lemma~\ref{utilG} the functions $t\mapsto \phi_i(t)$ are continuous with values  in the domain  of $\mathbf{G}$. When $\mathbf{G}$ is bounded from below it is enough to assume continuity in the form-domain. 
When  ${\mathbf  G}$ is the Laplace operator or,   more generally a one-body time-independent Schr\"odinger operator, we simply assume that    $\phi_i\in H^1(\R^3)$ or   $\phi_i\in H^1_0(\Omega)$ when $\Omega$ is a bounded domain. (Other boundary conditions could of course be considered.) 
\end{rem}

%%%%%%%%%%%%%%%%%%%%%%%%%
\begin{thm}[Flow in different gauge]\label{thm-gauge}
Let $U_0\in \mathcal{O}_K$,   $(C_0,\Phi_0)\in \partial\mathcal{F}_{N,K}$ and let $t\mapsto\mathbf{G}(t)$ be a  family of self-adjoint  operators in $L^2(\Omega)$. Assume that there exists a solution $(C,\Phi)\in C^0\big(0,T;\partial\mathcal{F}_{N,K}\big) $ to  $\mathcal{S}_0$ with initial data $(C_0,\Phi_0)$ such that $t\mapsto \langle \mathbf{G}(t)\phi_i(t),\phi_j(t)\rangle$ is continuous on  $[0,T]$ for every $1\leq i,j\leq K$.   Define the  family of unitary transforms $U(t)\in C^1\big([0,T); \mathcal{O}_K\big)$ that satisfy \eqref{polar} with $M_{ij}=\langle \mathbf{G}\,\phi_i,\phi_j\rangle$ 
as in  Lemma~\ref{utilG}. Then  the couple $(C',\Phi')= \mathcal{U}(t)\cdot\big(C,\Phi\big)$ with $ \mathcal{U}(t)=\big(d(U(t));U(t)\big)$ defined by \eqref{Action} and \eqref{Udet}  is a solution to
\begin{equation*}
 (\mathcal{S}_{\mathbf{G}})
\left\{
\begin{aligned}
 i\:\frac{d C'}{dt}&= \Bigl\langle \mathcal{H}\:\Psi\:\vert\:\nabla_{C'}\Psi\Bigr\rangle-\Bigl\langle \sum_{i=1}^N\mathbf{G}_{x_i}\:\Psi\:\vert\:\nabla_{C'}\Psi\Bigr\rangle, \\
 i \:\G(C')\:\frac{\partial  \Phi'}{\partial t}&= \G(C')\:\mathbf{G} \:\Phi' + (\mathbf{I}-\mathbf{P}_{\Phi'}) \:\nabla_{\Phi'}\Psi^\star\bigl[\mathcal{H}\:\Psi- \sum_{i=1}^N\mathbf{G}_{x_i}\:\Psi\bigr],
 \\
  \big(C'(0),\Phi'(0)\big)&=\mathcal{U}_0\cdot (C_0,\Phi_0),\end{aligned}
\right.
\end{equation*} 
with $\Psi=\pi(C,\Phi)=\pi(C',\Phi')$, $\mathcal{U}_0=\big(U_0,d(U_0)\big)\in \mathcal{O}^r_K$ being defined by \eqref{Action} and with $\mathbb{M}$ being  the $r\times r$ Hermitian matrix given by \eqref{truc2}.
 \end{thm}
 %%%%%%%%%%%
 
%%%%%%%%%%%%
\begin{rem} Passing from $\mathcal{S}_0$ to $\mathcal{S}_\mathbf{G}$ amounts to change the operator $\mathcal{H}$ by  $\mathcal{H}- \sum_{i=1}^N\mathbf{G}_{x_i}$ in both equations  and by adding the  linear  term $\G(C')\:\mathbf{G} \:\Phi'$ in the equation satisfied by $\Phi'$. Note that whereas solutions to $\mathcal{S}_0$  in $\partial\mathcal{F}_{N,K}$  satisfy 
\[
i\:\big\langle \frac{\partial\phi_i}{\partial t},\phi_j\big\rangle=0,
\]
for all $1\leq i,j\leq K$, solutions to ($\mathcal{S}_{\mathbf{G}}$) satisfy
\begin{equation}\label{gauge-phi}
i\:\big\langle \frac{\partial\phi'_i}{\partial t},\phi'_j\big\rangle=\big\langle\mathbf{G}\,\phi'_i,\phi'_j\big\rangle.
\end{equation}
System ${\mathcal S}_{\mathbf G}$ corresponds to the choice of  gauge ${\mathbf G}\,.$
\end{rem}
%%%%%%%%%%%%%%%%%%%%%%%%%
This is  illustrated and and summarized on Figure~\ref{fibration} below. 

\vskip6pt
Theorem \ref{thm-gauge} and Lemma~\ref{utilG} provide with the  differential equation that  satisfies the  unitary matrix $ U(t) $ that   transforms $ \mathcal{S}_0 $ into $\mathcal{S}_{\mathbf{G}}$. A direct calculation shows that, given two self-adjoint one-particle operators $ \mathbf{G} $ and $ \mathbf{G}'$, the solution to
\begin{equation} \left\{
 \begin{aligned}
 i\:\frac{d U}{dt}  &= U\: M_{G\rightarrow G'} ,\\
  U(t=0)&=U^0
 \end{aligned} \right. \label{usefulforTDHF}
 \end{equation}
with  $\Big(M_{G\rightarrow G'}\Big)_{ij}=\bigl\langle (\mathbf{G}-\mathbf{G}')\:\phi_i,\phi_j\bigr\rangle$ maps a solution to $ \mathcal{S}_\mathbf{G} $ to a  solution to $\mathcal{S}_{\mathbf{G}'} $. In particular, if we prove existence of solutions for the system   $ \mathcal{S}_\mathbf{G} $ for some operator $\mathbf{G} $ then  we have existence of solutions for any system $ \mathcal{S}_{\mathbf{G}'}$. 
\begin{figure}[h]
\centering\includegraphics[width=9cm]{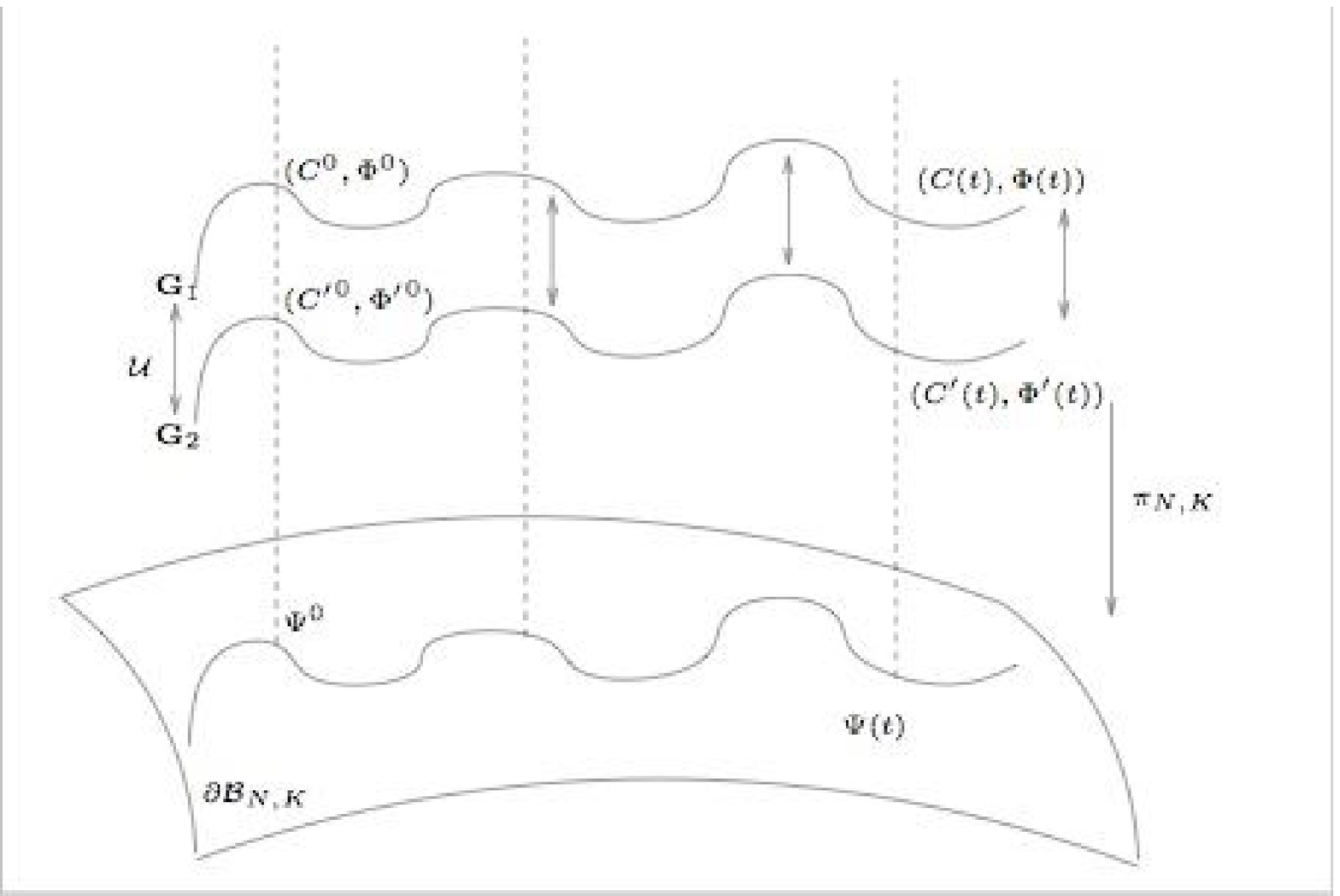}
\caption{Flow on the Fiber Bundle}
\label{fibration}
\end{figure}
Another  immediate  though crucial consequence of Theorem~\ref{thm-gauge} and Theorem~\ref{gauge} is given in Corollary~\ref{cons-G} below. It states that  for  any  choice of gauge  the constraints on the expansion coefficients and on  the orbitals are propagated by the flow and   the energy  is kept constant since it is the case for the system $\mathcal{S}_0$. Also   the rank of the first-order density matrices does not depend on the gauge. 
%%%%%%%%%
\begin{cor}[Gauge transforms and conservation properties] \label{cons-G} Let  $T>0$. Let $ \mathbf{G} $ be a self-adjoint (possibly time-dependent) operator  acting on $ \0$.  Assume that  there exists a solution to  the system $\mathcal{S}_G$ on $[0,T]$  such that $\mathrm{rank}\,\G\big(C(t)\big)=K$  and such that the matrix $t\mapsto \big\langle \mathbf{G}\phi_i,\phi_j\big\rangle _{1\leq i,j\leq K}$ is continuous. Then, for all  $t\in [0,T] $, 
\[
 (C(t),\Phi(t))\in  \partial\mathcal{F}_{N,K},\]
and the energy is conserved, that is 
\[
\mathcal{E}\big(\pi(C(t),\Phi(t))\big)= \mathcal{E}\big(\pi(C(0),\Phi(0))\big).
\]
In addition, $\Psi=\pi(C,\Phi)$ satisfies the Dirac-Frenkel variational principle \eqref{DF}.
\end{cor}
%%%%%%%%
\noindent\textit{Proof of Corollary~\ref{cons-G}.} {}By Theorem~\ref{thm-gauge} and its remark, if $ (C,\Phi) $ satisfies  $ \mathcal{S}_{\mathbf{G}}$ with initial data in $ \partial\mathcal{F}_{N,K}$,  there exists a  family of  unitary transforms $ U\in C^1\big(0,T;\mathcal{O}_K\big) $ such that $ (C,\Phi) = \mathcal{U}\cdot(C',\Phi') $ where $ (C',\Phi')$ satisfies $ \mathcal{S}_0$ with  same initial data. Since by Lemma~\ref{cons-contrainte}, $ \mathcal{S}_0$ preserves $ \mathcal{F}_{N,K}$,  so does $ \mathcal{S}_{\mathbf{G}} $  since $ U $ and $\mathbb{U}=d(U)$ are  unitary. Then, by  Lemma \ref{Unitarytransformation},   $ \pi(C,\Phi)=\pi(C',\Phi')=\Psi $, and  the energy is conserved by the flow since it only depends on $\Psi$. Eventually Eqn. \eqref{DF}  is satisfied  since $\mathbf{T}_\Psi \partial\mathcal{B}_{N,K}$ only depends on the point $\Psi$ on the basis $\mathcal{B}_{N,K}$ and not on the   pre-images in the fiber $\pi^{-1}(\Psi)$. \hfill$\Box$
\vskip6pt

So far  we have considered a generic Hamiltonian $ \mathcal{H}$ and we have written down an abstract  coupled system of  evolution equations for  this operator. In the  following subsection we turn  to the particular physical case of  $N$-body Schr\"odinger-type operators with pairwise interactions
%%%%%%%%%%%%%%%%%
\subsection{$N$-body Schr\"odinger type operators with pairwise interactions} 
 At this point, we consider  an Hamiltonian in $L^2(\Omega ^N)$ of the following form
\begin{equation}\label{def-HN}
 \mathcal{H}_N\:\Psi = \sum_{i=1}^N\mathbf{H}_{x_i}\:\Psi+ \sum_{1\leq i <j \leq N}\:v(|x_i-x_j |)\:\Psi.
 \end{equation}
In the above definition, $\mathbf{H}$ is a self-adjoint operator acting on $L^2(\Omega)$.  To fix ideas we take   $\mathbf{H}=-\frac 12\Delta+U$.  $ v$ is a real-valued  potential, and we  denote 
\[
V= \sum_{1\leq i <j \leq N}\:v(|x_i-x_j |)\,. 
\]
Expanding the expression of $ \mathcal{H}$ in the system $\mathcal{S}_0$ and arguing  as in the proof of Theorem~\ref{thm-gauge} we obtain
\begin{equation}\label{S0}
\mathcal{S}_0:\quad \left\lbrace
\begin{aligned}
 i\:\frac{d C}{dt}&= \Bigl\langle \sum_{i=1}^N\:\mathbf{H}_{x_i}\:\Psi\:\vert\:\nabla_C\Psi\Bigr\rangle + \Bigl\langle V\:\Psi\:\vert\:\nabla_C\Psi\Bigr\rangle \\
 i \:\G(C)\:\frac{\partial \Phi}{\partial t}&=  (\mathbf{I}-\mathbf{P}_{\Phi}) \:\nabla_\Phi  \Psi^\star\Big[V\:\Psi+ \sum_{i=1}^N\:\mathbf{H}_{x_i}\:\Psi\Big]\\
\big(C(0),\Phi(0)\big)&=\big(C_0,\Phi_0\big)\in \mathcal{F}_{N,K}.      
\end{aligned}
\right.
\end{equation}
%with $\mathbb{H}$ being the $ K\times K$ matrix such that  $\mathbb{H}_{ij} =\left\langle\mathbf{H}\:\phi_i\:,\:\phi_j\right\rangle$. 
Comparing with System $\mathcal{S}_{\mathbf{G}}$ in Theorem~\ref{thm-gauge}, one observes that the choice of gauge  $ \mathbf{G}=\mathbf{H}$ leads to the  equivalent system  
\begin{equation}\label{GT0}
\mathcal{S}_\mathbf{H}:\quad \left\lbrace
\begin{aligned}
 i\:\frac{d C}{dt}&= \biggl\langle \:V\:\Psi\:\:\vert\: \nabla_C\Psi\biggr\rangle, \\ 
 i \:\G(C)\:\frac{\partial \Phi}{\partial t} &= \:\G(C)\:\mathbf{H}\:\Phi + (\mathbf{I}-\mathbf{P}_\Phi) \:\nabla_\Phi \Psi^\star[V\:\Psi]\\
\big(C(0),\Phi(0)\big)&=\big(C_0,\Phi_0\big)\in \mathcal{F}_{N,K},      
\end{aligned}
\right.
\end{equation}
(provided $t\mapsto \langle\mathbf{H}\phi_i,\phi_j\rangle$ makes sense). From Corollary~\ref{cons-G} we know that  if the  initial data  in \eqref{GT0} lies  in $\mathcal{F}_{N,K}$ it persists for all time. This property   allows to recast System  \eqref{GT0} in a more tractable way where the equations satisfied by the orbitals  form  a coupled system of non-linear Schr\"odinger-type  equations. This new system that it is equivalent to System~\eqref{GT0} as long as the solution lies in $\mathcal{F}_{N,K}$ will be referred to as \textit{working equations} following  \cite{Scrinzi,Lubich1}. It  is  better adapted  for well-posedness analysis as will be seen in the forthcoming section. 
%%%%%%%%%%%%%
\begin{prop}[Working equations]\label{prop:working}  Let $(C,\Phi)$ be a solution to  \eqref{GT0} in $\mathcal{F}_{N,K}$, then it is a solution to
\begin{equation}\label{working}
\left\{
\begin{aligned}
i\frac{d C}{dt}&=  \K[\Phi]\:C,   \\ 
 i \:\G(C)\:\frac{\partial \Phi}{\partial t} &= \G(C)\:\mathbf{H} \:\Phi + (\mathbf{I}-\mathbf{P}_{\Phi})\:\mathbb{W}[C,\Phi]\:\Phi, \\
\big(C(0),\Phi(0)\big)&=\big(C^0,\Phi^0\big)\in \mathcal{F}_{N,K}\,,
% \\ &\\
% \Psi&=&\pi(C,\Phi).    
\end{aligned}
\right.
\end{equation}
where $\K[\Phi]$ (resp.  ${\mathbb{W}[C,\Phi]} $\/) is a $r\times r$  (resp. $ K\times K $) Hermitian matrix with entries
\begin{equation}\label{def-K}
\K[\Phi]_{\sigma,\tau}=\sum_{i,j\in\tau,\;k,l \in \sigma}\delta_{\tau\setminus\{i,j\},\sigma\setminus\{k,l\}}(-1)^{\tau}_{i,j}\;(-1)^{\sigma}_{k,l}\,D_v\big(\phi_{i}\:\bar\phi_{k}\:,\:\bar\phi_{j}\phi_{l}\big)
\end{equation}
and
\begin{equation}\label{def-W}
{\mathbb{W}[C,\Phi]} _{i j}(x) = 2\:\sum_{k,l=1}^K \gamma_{jkil} \,\big(\phi_k\,\bar\phi_l\star_\Omega v) %\int_\Omega \overline{\phi}_l(y) \:v(|x-y|)\: \phi_k(y)\,dy, 
\end{equation}
where here and below we denote 
\[
D_v(f,g) = \iint_{ \Omega\times \Omega} v(|x-y|)\:f(x)\:\overline{g}(y)\,dx dy,
\]
\[
f\star_\Omega v=\int_{\Omega}v(\cdot-y)\,f(y)\,dy
\]
and with the coefficients $\gamma_{ijkl}$ being defined by \eqref{gammaijkl} in Proposition~\ref{prop:coeff}. Conversely, any solution to \eqref{working} defines a flow on $\mathcal{F}_{N,K}$ as long as $\G(C)$ is invertible and is therefore a solution to \eqref{GT0}. 
\end{prop}
%%%%%%%%%%%%%%%%%%%%%%%%%%%
\begin{proof} We have to show that  for $\Psi=\pi(C,\Phi)$ in $\mathcal{B}_{N,K}$
\begin{equation}\label{pf-K}
\bigl\langle  V\:\Psi\:\:\vert\: \nabla_C\Psi\bigr\rangle= \nabla_{\overline{C}}\:\bigl\langle V\:\Psi\:\:\vert\:\Psi\bigr\rangle=\K[\Phi]\,C
\end{equation}
and 
\begin{equation}\label{pf-W}
\nabla_\Phi\Psi^\star[V\:\Psi] = \nabla_{\overline{\Phi}} \:\bigl\langle V\:\Psi\:\:\vert\:\Psi\bigr\rangle=  {\mathbb{W}[C,\Phi]} \: \Phi.
\end{equation}
%\begin{equation}\label{pf-W}
%(\mathbf{I}-\mathbf{P}_{\Phi}) \:\nabla_\Phi\Psi^\star[V\:\Psi]  =  (\mathbf{I}-\mathbf{P}_{\Phi}) \: {\mathbb{W}[C,\Phi]} \: \Phi.
%\end{equation}
We start from 
\[
\bigl\langle V\:\Psi\:\:\vert\:\Psi\bigr\rangle=\iint_{\R^3\times\R^3}[\Psi\otimes \Psi]_{:2}(x,y,x,y)\,v(|x-y|)\,dxdy 
\]
with 
\[
[\Psi\otimes\Psi]_{:2}(x,y,x,y)  = 
\sum_{i,j,k,l=1}^K \gamma_{ijkl}\,\phi_i(x) \:\phi_j(y)\:\overline{\phi}_k(x)\:\overline{\phi}_l(y)
\]
according to \eqref{densitymatrix2}. Since  only the coefficients $\gamma_{ijkl}$ depend on $C$ through Eqn. \eqref{gammaijkl} we first get 
\[
\nabla_{\overline{C}}\bigl\langle V\:\Psi\:\:\vert\:\Psi\bigr\rangle= \sum_{i,j,k,l=1}^K  \nabla_{\overline{C}}\big(\gamma_{ijkl}\big)\,D_v\big(\phi_{i}\:\bar\phi_{k}\:,\:\bar\phi_{j}\phi_{l}\big). 
\]
Hence \eqref{def-K} by using again Formula  \eqref{gammaijkl}.
\vskip6pt
We now turn to the proof of \eqref{pf-W} starting from 
\[
\bigl\langle V\:\Psi\:\:\vert\:\Psi\bigr\rangle=  \sum_{i,j,k,l=1}^K  \gamma_{ijkl}\, \iint_{\R^3\times \R^3} \phi_i(x) \:\phi_j(y)\:\overline{\phi}_k(x)\:\overline{\phi}_l(y)\,v(|x-y|)\,dxdy.
\]
Then, for every $1\leq p\leq K$ 
\begin{align*}
\frac{\partial}{\partial\overline{\phi}_p} \:\bigl\langle V\:\Psi\:\:\vert\:\Psi\bigr\rangle
&= \sum_{i,j,l=1}^K  \gamma_{ijpl}\,\big((\phi_j\overline{\phi}_l)\star v\big)\: \phi_i +  \sum_{i,j,k=1}^K  \gamma_{ijkp}\,\big((\phi_i\overline{\phi}_k)\star v\big)\: \phi_j \\
&=2\,  \sum_{i,j,l=1}^K  \gamma_{jipl}\,\big((\phi_i\overline{\phi}_l)\star v\big)\: \phi_j 
\end{align*}
by interchanging the r\^ole played by $i$ and $j$ in the first sum and by using  $\gamma_{ijkp}= \gamma_{jipk}$  and renaming $k$ as  $l$ in the second one.  Comparing with \eqref{def-W} we  find 
\[
\frac{\partial}{\partial\overline{\phi}_p} \:\bigl\langle V\:\Psi\:\:\vert\:\Psi\bigr\rangle=  2\,  \sum_{j=1}^K  \mathbb{W}[C,\Phi]_{pj}\,\phi_j. 
\]
To  achieve the proof of the proposition we now check that the system of equations in \eqref{working} preserves $\mathcal{F}_{N,K}$ as long as $\G(C)$ is invertible. The claim is obvious as regards the orthonormality of the  orbitals since $\mathbf{H}$ is self-adjoint and since $\mathbf{I}-\mathbf{P}_\Phi$ projects on $\textrm{Span}\{\Phi\}^\bot$. On the other hand, the equation on the coefficients leads to
\[
\frac{d}{dt} \Vert C(t)\Vert^2 = 2 \:\Im\sum_{\sigma,\tau}\mathbb{K}[\Phi]_{\sigma,\tau}c_\tau\,\overline{c}_\sigma= 0\]
since the matrix $\K[\Phi]$ is Hermitian. 
\end{proof}
\vskip10pt
We  treat apart in the last two  subsections the special cases of the linear  \textit{free  system} with no pairwise interaction   and of the \textit{time-dependent Hartree--Fock equations} for the evolution of a  single-determinant (TDHF in short) with pairwise interaction. 
%%%%%%%%%%%%%%%%%%%%%%%%%%%%%%%%%
%%%%%%%%%%%%%%%%%%%%%%%%%%%%%%%%%%%%%%%%%%%%%%%%%%%
\subsection{Interactionless Systems $v\equiv 0$}\label{ssec:free} 
In this section we consider systems for which  the binary interaction potential $ v$ is switched off. Then the system \eqref{GT0} becomes 
\begin{equation*}
\left\lbrace
\begin{aligned}
i\: \frac{d C}{dt}&= 0,\\
 i \:\G(C)\:\frac{\partial \Phi}{\partial t} &=\G(C)\:\mathbf{H}\:\Phi.
\end{aligned}
\right.
\end{equation*}
{}From the first equation the coefficients  $ c_\sigma$'s are constant during the evolution. In particular the  full-rank assumption is satisfied for all time whenever it is satisfied at start. In the latter case  the orbitals satisfy  $K$ independent linear Schr\"odinger equations 
\begin{equation}\label{free}
i \:\frac{\partial\Phi}{\partial t} =\mathbf{H}\:\Phi,
\end{equation}
and the $N$-particle wave-function $\Psi=\pi(C,\Phi)$ solves the exact Schr\"odinger equation
\begin{equation}\label{free-exact}
\left\{
\begin{aligned}
 i\frac{\partial\Psi}{\partial t}  &=\sum_{i=1}^N\mathbf{H}_{x_i}\:\Psi,  \\
\Psi(t=0)&=\pi(C_0,\Phi_0).       
\end{aligned}\right.
\end{equation}
Conversely, the unique solution to the  Cauchy problem \eqref{free-exact}  with $ (C_0,\Phi_0) \in \partial\mathcal{F}_{N,K}$  coincides with  $ \pi(C(t),\Phi(t)) \in \mathcal{F}_{N,K}$ where $\Phi(t) $ is the solution to  \eqref{free}. This is a direct consequence of the fact that the linear structure of \eqref{free-exact} propagates  the factorization of a Slater determinant.  In particular, this enlightens the fact that  the propagation of the full-rank assumption   is intricately related  to  the non-linearities created by the interaction potential  $ v$ between particles. 
%%%%%%%%%%%%%%%%%%%%%%%%%%%%%%%%%%%%%%%%%%%%%%%%%%%%%%%%%%%%%%%%%%%%%%%%%%%%%%%%%%%%%%%%%%%%%%%%
\subsection{MCTDHF ($K=N$) contains TDHF}
%The usual  time-dependent Hartree-Fock (TDHF) equations are obtained  in   the above  general setting for  $K=N$.  
%We consider the following single-determinant {\it ansatz} 
%\[ 
%\Psi^{HF}=\phi_{1}\wedge\ldots\wedge\phi_{N}
%\]
%for $\Phi=(\phi_1,\ldots,\phi_N)$  in $\mathcal{O}_{L^2(\Omega)^N}$. 
The TDHF equations write (up to a unitary transform)
\begin{equation}\label{TDHF}
i\:\frac{\partial \phi_i} {\partial t}=\mathbf{H}\:\phi_i + \mathcal{F}_\Phi\:\phi_i,
\end{equation}
for $1\leq i\leq N$, with $ \mathcal{F}_\Phi$ being  the self-adjoint operator on $\0$ that is defined by
\begin{equation*}
\mathcal{F}_{\Phi}\:w=\Big(\sum_{j=1}^N \int_\Omega v(|\cdot-y|)|\phi_j(y)|^2 dy\Big)\: w -\sum_{j=1}^N\Big( \int_\Omega v(|\cdot-y|)  \overline{\phi}_j(y) w(y) \:dy\Big) \:\phi_j.
\end{equation*}
The  global-in-time existence of solutions in the energy space $H^1(\Omega)^N$  goes back to Bove, Da Prato and  Fano \cite{Bove} for bounded interaction potentials and to  Chadam and Glassey \cite{Chadam} for the Coulomb potentials. They  also checked by integrating the equations    that  the TDHF equations propagate the orthonormality of the orbitals and 
that the Hartree--Fock energy is preserved by the flow. Derivation of the TDHF equations from the Dirac-Frenkel variational principle may be encountered in standard Physics textbooks (see {\it e.g.} \cite{McW}). Let us also mention the work \cite{Cances} by  Canc\`es and Le Bris  who have investigated  existence of solutions to  TDHF equations including time-dependent electric field and that are coupled with nuclear dynamics.
\vskip6pt 
By simply setting  $K=N$ in the MCTDHF formalism one gets
\begin{equation} \label{simplifications}
\# \Sigma_{N,K}= 1\:,\quad \G(t) =\mathbb{I}_N
\end{equation}
and 
\[
\quad \Psi(t):=C(t)\:\phi_{1}(t)\wedge\ldots\wedge\phi_{N}(t),\quad C(t)=e^{-i\theta_\Phi (t) }
\]
for some $\theta_\Phi\in \R$. In addition  according to Remark~\ref{rem:HF}, 
\begin{equation}\label{plusoumoins}
\gamma_{jkil}=\frac 12\:\big(\delta_{i,j}\:\delta_{k,l} - \delta_{i,k}\:\delta_{j,l}\big).
\end{equation}
Therefore with the definitions \eqref{def-K} and  \eqref{def-W}
 \begin{align*}
\K[\Phi]&= \big\langle\:V\:\phi_{1}\wedge\ldots\wedge\phi_{N}\:\vert\:\phi_{1}\wedge\ldots\wedge\phi_{N}\big\rangle\\
&=\sum_{i,j,k,l\::\:\{i,j\}=\{k,l\}}(-1)^{i+p_i(j)+k+p_k(l)}D_v(\phi_i\:\bar\phi_k; \bar\phi_j\:\phi_l)\\
&= \sum_{i=1}^N\big\langle \mathcal{F}_\Phi\phi_i,\phi_i\big\rangle
\end{align*}
and
\[
\sum_{j=1}^N{\mathbb{W}[C,\Phi}]_{ij}\:\phi_j=\mathcal{F}_\Phi \:\phi_i.\]
Eventually for $N=K$, according to \eqref{working},  the MCTDHF system in the working form turns out to be 
\begin{equation*}\mathcal{S}_{\mathbf{H}}(N=K)\quad \left\{\begin{aligned}
 \frac{d \theta_\Phi(t)}{dt}&= \sum_{i=1}^N\big\langle \mathcal{F}_\Phi\phi_i,\phi_i\big\rangle,\\ 
 i\:\frac{\partial \phi_i}{\partial t} &=\mathbf{H}\:\phi_i +(\mathbf{I}-\mathbf{P}_{\Phi})\:\mathcal{F}_\Phi \:\phi_i 
 \\ &=\mathbf{H}\:\phi_i + \mathcal{F}_\Phi \:\phi_i - \sum_{j=1}^N \langle \mathcal{F}_\Phi \:\phi_i,\phi_j \rangle\: \phi_j\end{aligned}\right.
\end{equation*}
with $
 \theta_\Phi(0)=0$ and  $\Phi(0)\in \mathcal{O}_{\0^N}\,.$
Comparing with \eqref{SM}, we introduce  the   $N\times N$ Hermitian matrix $M$ with entries $M_{ij}=- \langle \mathcal{F}_\Phi \:\phi_i,\phi_j \rangle$. According to Lemma~\ref{utile}  there exists a unique unitary matrix $ U(t)$ such that 
\begin{equation*}\ \left\{\begin{aligned}
i\:\frac{d U}{dt}  &= -\:U\: M,   
       \\
 U(t=0) &= \mathbb{I}_N.   
\end{aligned}\right.
\end{equation*}
In virtue of \eqref{Udet} the  unitary matrix $\mathbb{U}$ that transforms $\phi_1\wedge\ldots\wedge\phi_N$ into $(U\phi_1)\wedge\ldots\wedge (U\phi_N)$ is then simply a  complex number  of modulus $1$ ($\mathbb{U}=\mathrm{det}(U)$) that satisfies 
\begin{equation}\label{udoeinlem} \left\{\begin{aligned}
 i\:\frac{d \mathbb{U}}{dt}  &= -\mathrm{tr}(M)\:\mathbb{U},   \\
 \mathbb{U}(t=0) &= 1.   
\end{aligned}\right.
\end{equation}
Comparing \eqref{udoeinlem} with the equation satisfied by $\theta_\Phi(t)$ in $\mathcal{S}_{\mathbf{H}}(N=K)$ it turns out  that $\mathbb{U}=e^{i\theta_\Phi(t)}$. In that special case a change of gauge is simply a  multiplication by a  global phase factor. Applying Theorem~\ref{gauge},  the functions $\phi'_i$, $1\leq i\leq N$, defined by $\Phi'=U\Phi$ satisfy the standard Hartree--Fock equations \eqref{TDHF} and   $C'(t)=\overline{\mathbb{U}}\,C(t)=1$ for all time; that is $\Psi=\phi'_1\wedge\ldots \wedge\phi_N'$.   Being a special case of the MCTDHF setting we then recover ``for free" that the TDHF equations propagate the orthonormality of the initial data,  that they satisfy the Dirac-Frenkel variational principle and that the flow  keeps the energy constant.   
%%%%%%%%%%%%%%%%%%%%%%%%%%%%%%%%%%%%%%%%%%%%%%%%%%%%%%%%%%%%%%%%%%%%%%%%%%%%%%%%%%%%%%%%%%%
\section{Mathematical analysis of the MCDTHF Cauchy Problem}\label{sec:analysis}
%%%%%%%%%%%%%%%%%%%%%%%%%%%%%%%%%%%%%%%%%%%%%%%%%%%%%%%%%%%%%%
This section is devoted to the mathematical analysis of the  Cauchy problem for the $N$-body Schr\"odinger operator  with ``physical  interactions"
\begin{equation}\label{def-Uv}
U(x)= - \sum_{m=1}^M \frac{z_m}{|x-R_m|}\,,\quad \quad v(x)=\frac 1{|x|}\,,
\end{equation}
that is given by \eqref{GT0}:
\begin{equation*}
\mathcal{S}_\mathbf{H}:\quad \left\lbrace
\begin{aligned}
 i\:\frac{d C}{dt}&= \biggl\langle \:V\:\Psi\:\:\vert\: \nabla_C\Psi\biggr\rangle,   \\
 i \:\G(C)\:\frac{\partial \Phi}{\partial t} &=\:\G(C)\:\mathbf{H}\:\Phi + (\mathbf{I}-\mathbf{P}_\Phi) \:\nabla_\Phi \Psi^\star[V\:\Psi]\\
\big(C(0),\Phi(0)\big)&=\big(C_0,\Phi_0\big)\in \mathcal{F}_{N,K}.      
\end{aligned}
\right.
\end{equation*}
In this section $\Omega=\R^3$. According to  Proposition~\eqref{working}  solutions to $(\mathcal{S}_\mathbf{H})$   lie in $\mathcal{F}_{N,K}$ and they are therefore  solutions to 
\begin{equation}\label{workingcoulomb}
\left\{
\begin{aligned}
i\frac{d C}{dt}&=  \K[\Phi]\:C,   \\
 i \:\G(C)\:\frac{\partial \Phi}{\partial t} &= \G(C)\:\mathbf{H} \:\Phi + (\mathbf{I}-\mathbf{P}_{\Phi})\:\mathbb{W}[C,\Phi]\:\Phi, 
\\
\big(C(0),\Phi(0)\big)&=\big(C^0,\Phi^0\big)\in \mathcal{F}_{N,K}
% \\ &\\
% \Psi&=&\pi(C,\Phi).    
\end{aligned}
\right.
\end{equation}
with 
\begin{align*}
\K[\Phi]_{\sigma,\tau}&=\sum_{i,j\in\tau,\;k,l \in \sigma}\delta_{\tau\setminus\{i,j\},\sigma\setminus\{k,l\}}(-1)^{\tau}_{i,j}\;(-1)^{\sigma}_{k,l}\,D\big(\phi_{i}\:\bar\phi_{k}\:,\:\bar\phi_{j}\phi_{l}\big)\,,\\
{\mathbb{W}[C,\Phi]} _{i j}(x) &= 2\:\sum_{k,l=1}^K \gamma_{jkil} \,\big(\phi_k\,\bar\phi_l\star\frac1{|x|})\\
D(f,g) &= \iint_{ \R^3\times \R^3} \frac{1}{|x-y|}\, f(x)\:\overline{g}(y)\,dx dy\,.
\end{align*}
The above system is referred to  as the ``strong form" of the working equations. Let us emphasize again that it is equivalent to ($\mathcal{S}_\mathbf H$) provided $(C,\Phi)\in \mathcal{F}_{N,K}$.   The main sources of difficulties arise from the fact that the matrix $\G(C)$ may degenerate and from  the Coulomb singularities of the interaction potentials. Our strategy of proof works for more general  potentials $U$ and $v$. This is discussed in Section~\ref{sec:extension} below. 
\vskip6pt
The spaces $\C^r$ and $ W^{m,p}(\R^3)^K$  are equipped with  the Euclidian  norms for the vectors $C$ and $\Phi $ respectively 
\[\Vert C \Vert^2:= \sum_{\sigma\in\Sigma_{N,K}}|c_\sigma|^2 ,\quad \Vert \Phi \Vert_{W^{m,p}}^2 :=\sum_{i=1}^{ K} \Vert \phi_i \Vert^2_{W^{m,p}(\mathbb{R}^3)}\,. \]
Moreover, for a $ p\times p$ matrix $M$ we use the Frobenius norm
\[\Vert M \Vert = \sqrt{\sum_{ i,j=1}^{p}| M_{ij}|^2}.\]
We introduce the spaces   $X_m:=\C^r\times H^m(\R^3)^K$ for $m\in\mathbb{N}$   endowed with the   norms 
\begin{equation*}
  \Vert(C,\Phi)\Vert_{X_m}= \Vert C\Vert +
  \Vert\Phi\Vert_{H^m}\,.
\end{equation*}
The main result in this section is the following
%%%%%%%%%%%%%%%%%%%%%
\begin{thm}\label{H1}[The MCTDHF equations are well-posed] Let $m\geq 1$ and $ (C^0,\Phi^0)\in  \partial\mathcal{F}_{N,K} $ with  $\Phi^0$  
in $H^m(\R^3)^K $.  Then, there exists  a maximal existence time $ T^\star > 0 $ (possibly $+\infty$ but independent of $m$) such that:  

\vskip6pt\noindent (i)  The MCTDHF 
system \eqref{working} admits a unique
solution $(C,\Phi)$ with  
\[
C\in 
C^1\big([0,T^\star);\C^r\big), \qquad \Phi\in C^0\big([0,T^\star);H^{m}(\mathbb{R}^3)^K\big)\cap C^1\big([0,T^\star);H^{m-2}(\mathbb{R}^3)^K\big).
\]
This solution depends continuously on  the initial data $ (C^0,\Phi^0)$ in $X_m$. For every $0\leq t<T^\star$, 
\vskip6pt\noindent (ii)  $  
\big(C(t),\Phi(t)\big)\in  \partial\mathcal{F}_{N,K}$  \quad and \quad $
\G\big(C(t)\big)$ \textrm{ is invertible}.
\vskip6pt\noindent (iii)   The energy is conserved :
\[
\ys   \mathcal{H}_N\,\Psi(t)\big | \Psi(t)\ym
= \ys  \mathcal{H}_N\, \Psi^0\big | \Psi^0\ym\quad  \textrm{ with }\, \Psi=\pi(C,\Phi) \textrm { and } \Psi^0=\pi(C^0,\Phi^0).
\]
\vskip6pt\noindent (iv) The Dirac--Frenkel variational principle \eqref{DF} is satisfied.
\vskip6pt\noindent
\noindent (v)  When $T^\star<+\infty$ one has   
$$\limsup_{t\nearrow T^\star} \Vert \G(C(t))^{-1}\Vert=+\infty$$ and more precisely:
\[ 
 \int_0^{T^\star} \Vert \G\big(C(t)\big)^{-1}\Vert^{3/2}\,dt=+\infty\,.
 \]
% Even 
% \[
% (\sup_{t<T^*}|\!|\Phi|\!|_{H^1})\,( \int_0^{T^\star} \Vert \G\big(C(t)\big)^{-1}\Vert,dt)=+\infty\,.\]
% \fbox{Utile? Il manque la preuve}
\end{thm}
\vskip10pt
The global well-posedness in $H^1$ and $H^2$ of  the TDHF equations goes back to  Chadam and Glassey~\cite{Chadam}. Recently Koch and Lubich \cite{Lubich1} proved  local  well-posedness in  $H^2$  of  the MCTDH and MCTDHF equations     for regular  pairwise interaction potential $v$ and with $U\equiv 0$   by    using Lie commutators techniques. Our result extends both works.  The rest of the section is devoted to the proof of this theorem. The above system with the same notation is rewritten in the ``mild form" which makes sense as long as the matrix $\G(C(t))$ is not singular:
\begin{equation}\label{duhamel}
U(t)= e^{-it\mathcal A}U_0-i \int_0^t e^{-i(t-s)\mathcal A} \mathcal{L}\big(U(s)\big)\,ds
\end{equation}
with 
\begin{equation}\label{duhamel2}
U=\left(\begin{array}{c} C  \\ \Phi \end{array}\right), \;\mathcal A=\left(\begin{array}{c} 0  \\ {\mathbf H} \otimes \mathbb{I}_K\end{array}\right),\;  \mathcal{L}(U)=\left(\begin{array}{c} \K[\Phi]\,C
 \\ 
\G(C)^{-1}\,(\mathbf{I}-\mathbf{P}_{\Phi})\,\mathbb{W}[C,\Phi]\Phi \end{array}\right).
\end{equation}
%\begin{equation} \label{duhamel}
%\left\{
%\begin{array}{rcl}
%C(t) &=&\displaystyle C^0 -i\int_0^t \mathbb{K}[\Phi(s)]\,C(s)\,ds,
%\\
%& &
%\\
%\Phi(t) &=& \displaystyle  e^{it\mathbf{H}}\,\Phi^0-i\int_0^t e^{i(t-s)\mathbf{H}}\,\G(C(s))^{-1}\,(\mathbf{I}-\mathbf{P}_{\Phi})\:\mathbb{W}[C(s),\Phi(s)]\,\Phi(s)\,ds.
%\end{array}
%\right.
%\end{equation}
%with 
%System \eqref{duhamelc}--\eqref{duhamelphi} can also be written in synthetic  form:
The strategy of proof is  as follows. 

In Subsection~\ref{ssec:local} we show that the operator $\mathcal{L}$ is locally Lipschitz  continuous on $X_m$ for $m\geq 1$ in the neighborhood of any point $(C_0,\Phi_0)$ such that $\G(C_0)$ is invertible. Observe in particular that $\G(C)$  is a second-order homogeneous   function of the coefficients $C$ and therefore the invertibility of this matrix is a local property.   Standard theory of evolution equations  with locally Lipschitz non-linearities then guarantees local-in-time existence and uniqueness of a  mild solution in these spaces that is continuous with respect to the initial data  as long as the matrix $\G(C)$ remains invertible (see \textit{e.g} \cite{Segal, Pazy}). Next for initial data in $X_m$ with $m\geq 2$,  the corresponding mild  solution  in this space  is regular enough to be a strong solution to~\eqref{workingcoulomb} (see \cite{Pazy,Haraux}). As shown in the previous section (Proposition~\ref{prop:working}), the strong solution then remains on the constraints fiber bundle $\partial \mathcal F_{N,K}\, $  and it is therefore a solution to \eqref{GT0}. Furthermore using the gauge equivalence and Corollary~\ref{cons-G} one deduces that the energy of the solution  is conserved and that the Dirac-Frenkel variational principle is satisfied. Recall for further use  that the energy may be recasted in the following equivalent forms \cite{Fries, Lewin}
\begin{equation}
\begin{split}
\mathcal{E}(\Psi)&=\mathcal{E}\big(\pi(C,\Phi)\big)\\
&=\left(\Big( \mathbf{H}\:\G +
\frac{1}{2}\mathbb{W}[C,\Phi]\Big)\Phi,\Phi\right)_{L^2(\Omega)^K}\\
%&=&\int_{\R^3}\Big(\G(C)\,\Phi;\mathbf{H}\,\Phi\Big)\,dx+ \iint_{\R^3\times\R^3}\frac{[\Psi\otimes \Psi]_{:2}(x,y,x,y)}{|x-y|}\,dxdy\nonumber \\
&=\sum_{i,j=1}^K\gamma_{ij}\int_{\R^3}\left[\frac12\nabla \phi_i\cdot\nabla\bar\phi_j+U\, \phi_i\,\bar\phi_j\right]\,dx+\sum_{i,j,k,l=1}^K \gamma_{ijkl}\, D\big(\phi_l\,\bar\phi_i;\phi_k\,\bar\phi_j\big).
\end{split}
\label{CompactEnergy}\end{equation} 
In consequence for initial data in $X_m$, $m\geq 2$,   the norm of the vector $\Phi(t)$ remains locally bounded in $H^1$ (independently of the  $H^2$ norm). Therefore it is also a  strong solution in $H^1$    defined on the same  time interval  which depends only on the $H^1$norm and on $\G(C_0)$. Eventually using the density  of $ X_2\cap\partial \mathcal F_{N,K}$ in $ X_1\cap\partial \mathcal F_{N,K}$ and the continuous dependence with the initial data one obtains the local-in-time existence of a strong solution in $X_1\cap\partial \mathcal F_{N,K}$  with constant energy.  

In Subsection~\ref{ssec:maximal}, relying on  the conservation of the energy   we prove  the  existence of the  solution over a maximal time interval beyond which the density matrix degenerates. The equations themselves imply   the further regularity $C(t)\in  C^1\big([0,T^*), \C^r\big)$ and  $\Phi(t) \in  \times C^0([0,T^*),  H^{m}(\mathbb{R}^3)^K) \cap C^1([0,T^*),  H^{m-2}(\mathbb{R}^3)^K)\,.$ 
%%%%%%%%%%%%%%%%%%%%%%%%%%%%%%%%%
\subsection{Properties of the  one-parameter group and local Lipschitz properties of the  non-linearities}\label{ssec:local}
%%%%%%%%%%%%%%%%%%%%%%%%%%%%%%%%%
As in  Chadam and Glassey \cite{Chadam} for example, one checks that $ \big\{e^{it\mathcal A}\big\}_{t\in\R} $  is a one-parameter group of linear operators,    unitary in $X_0$ and uniformly bounded in time for $0\leq t\leq T$ in $X_1 $ and $X_2$.   
% In particular, for every $T>0$, there exists a positive constant $M_T$ (that only depends on $U$) such that 
%\begin{equation}\label{def:MT}
%\Vert  e^{i(t-s)H}\,\varphi\Vert_{H^1(\R^3)}\leq M_T\, \Vert \varphi\Vert_{H^1(\R^3)}, \quad \textrm{ for all } \varphi \in H^1(\R^3) \textrm { and } t,s \in [0,T].
%\end{equation} 
\vskip6pt
We now show that the   operator $\mathcal{L}$ in the right-hand side of \eqref{duhamel} is a locally bounded and locally Lipschitz continuous mapping in a small enough neighborhood of any $(C_0,\Phi_0)$ in $X_m$ such that $\G(C_0)$ is invertible for every $m\geq 1$.  The operator $\mathcal{L}$ reveals as  a composition of locally bounded and locally Lipschitz continuous mappings as now detailed. We first recall that invertible matrices form an open subset of $\mathcal{M}_{K\times K}(\C)$ and that the mapping $M\mapsto M^{-1}$ is   locally Lipschitz continuous  since 
 \begin{align*}
 \Vert M^{-1}-\tilde M^{-1}\Vert 
 &=  \Vert M^{-1}\,(\tilde M- M)\,{\tilde M}^{-1}\Vert\\
 &\leq\Vert M^{-1}\Vert\,\Vert{\tilde M}^{-1}\Vert\,\Vert M-\tilde M\Vert. 
 \end{align*}
In addition,  being quadratic,  the mapping $C\mapsto \G(C)$ is for any $m$ and independently of $m$ locally Lipschitz in $X_m$  in a small enough neighborhood of any $(C_0,\Phi_0)$ such that $\G(C_0)$ is invertible. The  same holds true for the mapping $C\mapsto \G(C)^{-1}$ by composition of locally bounded and locally Lipschitz functions. 
\vskip6pt
The operator $\mathbf{P}_{\Phi}$ is a sum of $K$ terms of the form $\langle \phi,\cdot\rangle_{L^2}\, \phi$ with $\phi$ in $H^m(\R^3)$.  Hence, for $m\geq 0$,  
\begin{equation}\label{borne-P}
\Vert \mathbf{P}_\Phi\Vert_{\mathcal{L}(H^m)} \lesssim  \Vert \Phi\Vert_{L^2}\:\Vert\Phi \Vert_{H^m}
\end{equation}
where  here and below $\lesssim$  is a shorthand for a bound with a universal  positive constant  that only depends on $K$ and $N$. Therefore  $\Phi\mapsto  \mathbf{P}_{\Phi}$ is locally Lipschitz from $X_m$ to  $\mathcal{L}(H^m)$ since it is quadratic with respect to $\Phi$. To deal with the other non-linearities  we start with  recalling a few  properties of the Coulomb potential taken  from~\cite[Lemma~2.3]{Chadam}. Their  proof is a straightforward application of Cauchy--Schwarz' and Hardy's inequalities and we skip it. Let $ \phi, \psi\in \1$, then  with $r=|x|$, $(\phi\psi) \star \frac{1}{r}\in W^{1,\infty}(\R^3)$, and we have
\begin{equation}\label{simple}
\|(\phi\psi) \star \frac{1}{r}\|_{L^\infty(\mathbb{R}^3)} \leq 2\,\|\nabla\phi\|_{\0}\|\psi\|_{\0}
\end{equation}
and 
\begin{equation*}
\Big\|\nabla\big((\phi\psi) \star \frac{1}{r}\big)\Big\|_{L^\infty(\mathbb{R}^3)} \leq 4\,\|\nabla\phi\|_{\0}\|\nabla\psi\|_{\0}.
\end{equation*} 
As a consequence of above inequalities and by an  induction argument that is detailed in \cite{Castella} for example, we have, for $\Phi\in H^m(\R^3)^K$ and for every $1\leq i,j,k\leq K$, 
\begin{equation}\label{Hm-Coulomb}
\big\|\big((\phi_i\,\phi_j) \star \frac{1}{r}\big)\,\phi_k\big\|_{H^m(\mathbb{R}^3)} \lesssim  \Vert \Phi\Vert_{H^{m'}}^2\,  \Vert \Phi\Vert_{H^m} \lesssim  \Vert \Phi\Vert_{H^m}^3
\end{equation}
with $m'=\max(m-1,1)$. First, recall from \eqref{def-W}, that $\mathbb{W}[C,\Phi]\,\Phi$ is a sum of terms of the form 
$\gamma_{jkil}\,\phi_j\,\big( \frac{1}{r}\star \phi_k\:\overline{\phi}_l\big)$ with the coefficients $\gamma_{jkil}$  depending  quadratically on $C$ according to \eqref{gammaijkl}. They are  therefore locally Lipschitz continuous with respect to $C$.  Gathering with \eqref{Hm-Coulomb} we have 
\begin{equation}
\left\|\mathbb{W}[C,\Phi]\:\Phi\right\|_{H^m}  \lesssim  \Vert C\Vert^2\,\|\Phi\|_{H^{m'}}^2\,\|\Phi\|_{H^{m}}\lesssim \Vert C\Vert^2\,\|\Phi\|_{H^{m}}^3\:.
\label{borneW}
\end{equation}
The mapping $(C,\Phi)\mapsto \mathbb{W}[C,\Phi]\:\Phi$ is then locally bounded in $X_m$ and being quadratic in $C$ and cubic in $\Phi$ it is locally Lipschitz continuous in $X_m$ by a standard polarization argument.  In particular, the first bounds reveals a linear dependence on the $H^m$ norm. Eventually, for every $1\leq i,j,k,l\leq K$, using \eqref{simple} and H\"older's inequality we obtain
\begin{equation*}
\left |D(\phi_j\:\overline{\phi}_i\:,\:\phi_k\overline{\phi}_l)\right|
\lesssim \Vert\Phi\Vert_{L^2}^3\,\Vert\Phi\Vert_{H^1}\lesssim \Vert\Phi\Vert_{H^m}^4, 
\end{equation*}
the last line being  a direct consequence of \eqref{Hm-Coulomb}. In particular this proves 
\begin{equation}\label{borne-K-L2}
\vert  \mathbb{K}[\Phi]\vert \lesssim  \Vert\Phi\Vert_{L^2}^3\,\Vert\Phi\Vert_{H^1},
\end{equation} 
\begin{equation}\label{borne-K}
\Vert  \mathbb{K}[\Phi]\,C\Vert \lesssim \Vert C\Vert\,\Vert  \Phi\Vert ^4_{H^m}
\end{equation}
and that $(C,\Phi)\mapsto \mathbb{K}[\Phi]\,C$ is locally Lipschitz continuous in $X_m$ since  according to \eqref{def-K}, $\mathbb{K}[\Phi]\,C$ is a finite sum of terms of this kind  up to some universal constant. 
\vskip6pt
For any $m\geq 1$  existence and uniqueness of a solution $(C(t),\Phi(t))$ to the integral equation \eqref{duhamel} in  a neighborhood of $(C^0,\Phi^0)$ in $\mathcal C^0(0,T;X_m)$ for $0< T$ small enough follows by Segal's Theorem~\cite{Segal}, which also ensures the continuity with respect to the initial data in $X_m$. 
\vskip10pt
We now turn to the existence of a maximal solution and to  the blow-up alternative  in $X_1$.
\subsection{ Existence  of the maximal solution and blow-up alternative}\label{ssec:maximal}
%%%%%%%%%%
To simplify notation, from now on we use the shorthand  $\G(t)$ for $\G\big(C(t)\big)$. Existence of a global-in-time  solution requires  to control uniformly  both the $H^1$ norm of $\Phi$ and  the norm of $\G^{-1}(t)$. With the conservation of the energy this turns to be equivalent to control only  the norm of $\G^{-1}(t)$. Let $T^*$ denotes the maximal existence time and  assume that $T^*<+\infty$.   We first show that 
\begin{equation}\label{explose-G}
\limsup_{t\uparrow T^*}\Vert \G(t){}^{-1}\Vert=+\infty.
\end{equation}
We argue by contradiction and assume that there exists a positive constant $M_0$  such that for all $t\in [0,T^*)$, $\;\Vert \G(t){}^{-1}\Vert\leq M_0$. 
%Thanks to \eqref{cons-normes}, we know  that $(C(t),  \Phi(t))$ stays  in $\mathcal{F}_{N,K}$ for all $t$ in  $[0,T^*)$. 
We now prove that there exists a positive constant $K_0$ such that
\begin{equation}\label{borneH1}
\forall t\in [0,T^*), \quad  \Vert \Phi(t)\Vert_{H^1}\leq K_0
.\end{equation}
Thanks to Lemma~\ref{energyconservation} and Corollary~\ref{cons-G},  the energy is preserved by the flow, and therefore using the expression \eqref{CompactEnergy}
\begin{align*} 
\Big(\mathbf{H}\:\G(t)\:\Phi(t),\Phi(t) \Big) &\leq \Big(
\mathbf{H}\:\G(t)\:\Phi(t),\Phi(t) \Big) + \frac{1}{2}\Big(\mathbb{W}[C,\Phi]\:\Phi(t) ,\Phi(t)\Big)\\
&=   \mathcal{E}\big(\pi(C,\Phi)\big)\:=\:\mathcal{E} \big(\pi(C^0,\Phi^0)\big)
\end{align*}
for all $0\leq t<T^*$ since, with $\Psi=\pi(C,\Phi)$, 
\begin{equation*}
\Big\langle V\,\Psi\big| \Psi\Big\rangle=\Big(\mathbb{W}[C,\Phi]\:\Phi(t) ,\Phi(t)\Big)\geq0  \end{equation*}
for $v\geq 0$. As in \cite{Lewin, Fries}, Kato's inequality then  yields  that 
\[
\Vert \sqrt{\G}\,\Phi\Vert_ {H^1}\leq M_1
\]
where $M_1$ is a positive constant   independent of $t\geq 0$. Now let $\mu(t)\in (0,1]$ be the smallest eigenvalue of the hermitian matrix $\G(t)$. Then according to the definition of the Frobenius norm 
\[
\Vert \G^{-1}(t)\Vert=\left(\sum_{k=1}^K\frac{1}{\mu_k(t)^2}\right)^{\frac 12}, 
\]
with $\{\mu_1(t),\cdots,\mu_K(t)\}$ being the eigenvalues of $\G(t)$, hence 
\[
\frac{1}{\mu(t)}\leq \Vert \G^{-1}\Vert\leq \frac{\sqrt{K}}{\mu(t)}\quad\textrm{and}\quad
\frac{1}{\sqrt{\mu(t)}}\leq \Vert \sqrt{\G}{}^{-1}\Vert\leq \frac{K^{1/4}}{\sqrt{\mu(t)}}
\]
for all $t\in [0,T^*)$. Therefore 
\begin{equation}\label{borne-mu}
\Vert \Phi\Vert_{H^1}  \leq \frac{K^{1/4}}{\sqrt{\mu(t)}}
\,\Vert\sqrt{\G} \Phi\Vert_{H^1} \leq K^{1/4}\,M_1\,\Vert \G{}^{-1}\Vert{}^{1/2}.
\end{equation}
In particular, this shows \eqref{borneH1} with $K_0=K^{1/4}\,M_1\, M_0^{1/2}$. Therefore, for any $t\in[0,T^*)$ arguing as  above, we may build a solution to the system on $[t,t+T_0]$ for $T_0>0$ that only depends on $M_0$ and $K_0$. Since $t$ is arbitrary close to $T^*$ we reach a contradiction with the definition of $T^*$. Hence \eqref{explose-G}. 
\vskip6pt
Now, taking the derivative with respect to $t$ of both sides of  $\G\,\G^{-1}= \mathbb{I}_K$, we get
\begin{equation}\label{edo-G}
\frac{d\G^{-1}}{dt}=-\G^{-1}\, \frac{d\G}{dt}\,\G^{-1},
\end{equation}
for all $t\in [0,T^*)$. From the expression of $\G$ in terms of $C$ and since $\Vert C\Vert=1$,   it holds
\begin{equation*}
\big\Vert \frac{d\G}{dt}\big\Vert \lesssim \big\Vert \frac{dC}{dt}\big\Vert
\lesssim  \big\Vert \Phi\big\Vert_{H^1}\lesssim\,\Vert \G{}^{-1}\Vert{}^{1/2}
\end{equation*}
in virtue of the bound \eqref{borne-K-L2} on $\mathbb{K}[\Phi]$ using the fact that $\Vert \Phi\Vert_{L^2}=K$. Inserting the last bound above in \eqref{edo-G} and integrating over $t$  yields 
\[
\Vert \G(t)^{-1}\Vert \leq \Vert \G(0)^{-1}\Vert + \mathrm{const.}\,\int_0^t\Vert\G(s)^{-1}\Vert^{3/2}\, ds,
\]
for all $t\in [0,T^*)$. Because of \eqref{explose-G}, this implies that $\int_0^{T^*}\Vert \G(s){}^{-1}\Vert{}^{3/2}\,ds=+\infty$. 
\vskip6pt
%%%%%%%%%%%
So far we have proved the local well-posedness of the MCTDHF equations in $X_m$ for every $m\geq 1$ and the existence of a maximal solution in $H^1$ until time $T^*$ when the density matrix becomes singular. We prove now that $T^*$ is the maximal time of existence regardless of the imposed regularity on the solution. Let $(C,\Phi)$  be a  solution in $X_2$, then it is in particular a maximal solution in $X_1$. We have to show that the $H^2$ norm of $\Phi$ cannot explode at  finite time $0<\tau<T^*$.  Indeed,  for any $\tau<T^*$, we have 
\begin{equation}\label{bd-local-G}
\max_{0\leq t\leq \tau}\Vert \G(t)^{-1}\Vert \lesssim 1\end{equation}
by definition of $T^*$, hence 
\begin{equation}\label{bd-local-H1}
\max_{0\leq t\leq \tau}\Vert \Phi(t)\Vert_{H^1} \lesssim 1.
\end{equation}
From  the Duhamel formula for the PDEs system \eqref{duhamel}--\eqref{duhamel2} and  using the bounds \eqref{borne-P} and  \eqref{Hm-Coulomb} together with $\Vert\Phi\Vert_{L^2}=1$ and $\Vert C\Vert=1$, we get for all $t\in [0,\tau] $
\begin{equation*}
\Vert  \Phi(t)\Vert _{H^2} \leq \Vert  \Phi^0\Vert _{H^2} + \textrm{C}\:\sup_{[0,\tau]}\int_0^t \Vert \Phi(s)\Vert _{H^2} \:ds
\end{equation*}
where $C$ is a positive constant that only depends on the local bounds \eqref{bd-local-G}  and  \eqref{bd-local-H1}. By  Gronwall's lemma we infer 
\[
\max_{0\leq t\leq \tau} \Vert  \Phi(t)\Vert _{H^2} \lesssim e^{C\tau}
\]
hence the conclusion. The proof for any $m\geq 2$ follows then by  a straightforward  induction argument using the corresponding bounds \eqref{borne-P} and  \eqref{Hm-Coulomb} by assuming that $\max_{0\leq t\leq \tau} \Vert  \Phi(t)\Vert_{H^{m-1}} \lesssim 1$.
\vskip10pt
The proof of  Theorem \ref{H1} is now complete.

%%%%%%%%%%

%%%%%%%%%%

%%% suite%%%%
\subsection{Existence of Standing wave solutions}
In the present case   the equations for the coefficients  write  \eqref{EL-C}  while Eqn.  \eqref{sats}  for the orbitals becomes:   
\begin{equation}\label{EL-Phi}
\G(C)\:\mathbf{H} \:\Phi + \mathbb{W}[C,\Phi]\:\Phi=\Lambda\cdot \Phi
\end{equation}
according to  Proposition~\ref{prop:working}. In \cite{Lebris} Le Bris has proved the existence of ground-states - that is,  minima of the energy over the set $\mathcal{F}_{N,K}$ -   for the physical Hamiltonian \eqref{Hamiltonien}, on the whole space $\R^3$, and  under the assumptions $K=N+2$ and $\sum_{m=1}^Mz_m> N-1$. Later on Friesecke extended this result to general admissible pairs $(N,K)$, under the same assumption on the nuclear charge. Finally Lewin proved the existence of infinitely many critical  points of the MCHF energy for any pairs $(N,K)$, hence the existence of infinitely many solutions to the coupled  system   \eqref{EL-Phi} -- \eqref{EL-C} that satisfy  the full-rank assumption. All these solutions then give rise to infinitely many standing waves of the MCTDHF system  and thereby to   particular global-in-time solutions. 
\vskip10pt

The conservation of the invertibility of the matrix $\G(t)$ being an essential issue in the MCTDHF setting  it is natural to give sufficient condition for such property.
%%%%%%%%%%%%%%%%%%%%%%
\section[Global-in-time-existence]{Sufficient condition for  global-in-time  existence}\label{sec:global}
%We denote  $ {\mathcal E} (\Psi) = \langle  \:\Psi\:\big|H_N\big| \: \Psi \rangle$ and , 
In this section we focus again on  the  $N$-body Schr\"odinger operator \eqref{Hamiltonien}  with  physical  interactions \eqref{def-Uv}. For any $K\geq N+1$ with fixed $N$,  we denote
\[
\mathcal{I}(K)=\inf\big\{{\mathcal E}(\pi(C,\Phi))\,:\, (C,\Phi) \in \mathcal{F}_{N, K}\big \}
\]
the ``$K$- ground state energy".  Obviously one has  
\begin{equation}\label{Energyorder}
\forall \, K'\le K \le  \infty, \quad \inf\sigma(\mathcal H_N)\le \mathcal{I}(K)\le \mathcal{I}(K'),
\end{equation}
with $\inf\sigma(\mathcal H_N)$ being the bottom of the spectrum of $\mathcal H_N$ on $L^2_\wedge(\Omega^N)$.  Recall that the maximal rank hypothesis corresponds to the following equivalent facts~:
\begin{itemize}
\item[(i)] The rank of the operator $[\pi(C,\Phi)\otimes{\pi(C,\Phi)}]_{:1}$ is equal to $K$;
\item[(ii)] The $K\times K$ matrix $\G(C)$ is invertible;
\item[(iii)] The smallest eigenvalue  of $\G(C)$ is strictly positive.
\end{itemize}
%Under the maximal rank hypothesis, any orthonormal basis $\{\phi_1' ,\ldots, \phi'_K\}$ which diagonalizes the operator $\gamma_{\pi(C,\Phi)}$ provides with a diagonal matrix $\G'$ and defines a unique unitary transform $(C,\Phi)\mapsto (C',\Phi')  \in \mathcal{F}_{N}^{K}$  such that one has $ \pi(C,\Phi)=\pi(C',\Phi') $. In this case the following relation holds 
%\begin{equation}\label{lienC}
%\gamma_i=\gamma_{ii}=\sum_{\sigma\:|\:i\in \sigma}|c'_\sigma|^2.
%\end{equation}
%More generally, 
%\begin{equation}\label{lienC-offdiag}
%\gamma_{ij}= \sum_{\substack{\sigma,\,\sigma'\:|\:i\in \sigma,\,j\in\sigma' \\ \sigma\setminus \{i\} = \sigma'\setminus \{j\}}}(-1)^{\sigma^{-1}(i)+\sigma'^{-1}(j)}\,\overline{c}_\sigma\,c_{\sigma'} \quad \text{for all} \quad 1\leq i,j\leq K.
%\end{equation}
%We also recall the formula for the coefficients of the two-body density matrix $[\pi(C,\Phi)\otimes \pi(C,\Phi)]_{:2}$
%\begin{equation}\label{formulA}
%\gamma_{ijkl}=\left\{
%\begin{array}{ll}
%0& \text{if } l=j \text{ or }{i=k},\\
%\displaystyle \sum_{\substack{\sigma,\,\sigma'\:|\:l,j\in\sigma,\,i,k\in\sigma'\\
%\sigma\setminus\{l,j\}=\sigma'\setminus\{i,k\}}}(-1)^\sigma_{l,j}\,(-1)^{\sigma'}_{i,k}\,{c_\sigma}\,\overline{c}_{\sigma'}&\text{otherwise }.
%\end{array}
%\right.
%\end{equation}
%%%%%%%%%%%%
Since this is satisfied for $K=N$ (Hartree--Fock case) and since $K$ must be admissible, we now assume that $K\geq N+2$. The main result of this section is the following:
\begin{thm}\label{mainth} Let   $(C^0, \Phi^0)\in \mathcal{F}_{N,K} $ be an  initial data in \eqref{working}  with $\G\big(C^0\big)$  invertible. Assume that $T^\star<+\infty$  then
\[
{\mathcal E} (\pi( C^0,\Phi^0)) \geq  \mathcal{I}(K-1).\]
\end{thm}
%%%%%%%
As an immediate  by-product we get a sufficient condition  ensuring the global-in-time invertibility of the matrix $\G\big(C(t)\big)$.
%, or equivalently, the non-existence of a finite time $T^*$ such that 
%\begin{equation}\lim_{t\rightarrow T*} \gamma_i(t)=0\, \quad \text{ for some }  \quad 1\leq i\leq K.\label{deg1}
%\end{equation}
%%%%%%%%%%
\begin{cor}  If $(C^0, \Phi^0)\in \partial\mathcal{F}_{N,K} $ satisfies 
\begin{equation} \label{CI}
\mathcal{I}(K)\leq  {\mathcal E} (\pi( C^0,\Phi^0)) < \mathcal{I}(K-1),
\end{equation} 
then $T^\star=+\infty$; that is, the maximal solution is global-in-time.
\end{cor}
\vskip6pt
%%%%%%%%%%%%%%%%%%
\begin{rem} The hypothesis   $\sum_{m=1}^M z_m\ge N$  in \eqref{def-Uv}  implies    the relation ${\mathcal I}(K)<{\mathcal I}(K-2)$ \cite{Lebris,Friesecke2}. Therefore  (\ref{CI}) can always  be satisfied by changing $K$ into $K-1$. 
%Hence, the global-in-time well-posedness of the associated Cauchy problems for such initial data.
\end{rem}
%%%%%%%%%%%%%%%%%%
\vskip6pt
\begin{rem} A key difficulty in the proof of above theorem is that  the energy functional $\Psi\mapsto \mathcal{E}(\Psi)$ is not weakly lower semi-continuous in $H^1(\R^{3N})$   while  it is  in $H^1(\Omega^{3N})$ for any bounded domain $\Omega$ as already observed by Friesecke~\cite{Fries}. When $\Omega$ is a bounded domain of $\R^3$ or when the potential $U$ is non-negative,  the proof of Theorem~\ref{mainth} is much easier thanks to the lower semi-continuity, and it is detailed in \cite{AML}.  In the general case the proof is  in the  very spirit of Lewin's one  for the convergence of critical points of the energy functional  \cite{Lewin}.
\end{rem}
\vskip6pt
\noindent\textbf{Proof of Theorem \ref{mainth}. } Let $(C,\Phi)$ be the maximal solution to   \eqref{working} on $[0,T^\star)$ with initial data $(C^0,\Phi^0)$  given by  Theorem~\ref{H1}. We  assume that $T^\star<+\infty$, then 
\[
\limsup_{t\nearrow T^\star}\Vert \G\big(C(t)\big)^{-1}\Vert=+\infty.
\]
Equivalently, with the eigenvalues of $\G(C)$ being arranged in decreasing order $0\le \gamma_K \le \gamma_{K-1}\le \ldots \le \gamma_1 \le 1$, this means 
\[
\liminf_{t\nearrow T^\star}\gamma_K(t)=0.
\]
Then there exists a sequence $ t_n$  converging to $T^\star$, a positive number $\beta$ and an integer $N+1\leq  m \leq K $ such that  
\begin{equation}\label{deg-1}
\lim_{n\rightarrow+ \infty} \gamma_m(t_n)=0\quad  \text{ and }\quad 0<\beta \le \gamma_{m-1} (t_n)\,.
\end{equation}
Indeed, since 
%\begin{equation}\label{trace-fix}
$\sum_{k=1}^K \gamma_k(t)=N, \quad\textrm{ for  all }t\in [0,T^\star)\,$,
%\end{equation}   
at least $N$ eigenvalues stay away from  zero when  $t$ goes to $T^\star$. We denote $C^n=C(t_n)$, $\Phi^n=\Phi(t_n)$, $\gamma_i^n=\gamma_i(t_n)$, $\G^n=\G\big(C(t^n)\big)$ and so on for  other involved quantities. 
\vskip6pt
For all $n\geq 1$, $(C^n, \Phi^n)\in \partial\mathcal{F}_{N,K}$. Thus   according to Proposition~\ref{Unitarytransformation}, there exists a unique  sequence of unitary transforms $\mathcal{U}^n\in \mathcal{O}^r_K$  that map $(C^n, \Phi^n)$ into $({C'}^n,{\Phi'}^n)$ with ${\Phi'}^n$ being an eigenbasis for the operator   $\gamma^n:=\gamma_{\Psi^n}$. In particular  the corresponding matrix $\G'^n:=\G(C'^n)$ is diagonal. 
In other words,  
\begin{align*}
\Psi^n:= \pi (C^n,\Phi^n)&\:=\:\sum_\sigma c_\sigma^n\,\Phi_\sigma^n\:=\: \sum_\sigma {c'_\sigma}^n\,{\Phi'_\sigma}^n\:=\:\pi({C'}^n,{\Phi'}^n),\\
 \gamma^n &= \sum_{i,j=1}^K \gamma_{ij}^n\,  \phi_i^n \otimes\,\overline{\phi_j}{}^n\:=\:\sum_{i=1}^K\gamma_{i}^n \,{\phi'_i}^n\otimes\,\overline{\phi'_i}{}^n .%\label{d1}
\end{align*}
Since the group of unitary transforms is compact, we may argue equivalently on the sequence $({C'}^n,{\Phi'}^n)$ that we keep denoting by $(C^n,\Phi^n)$ for simplicity. From \eqref{deg-1}
%Now, we claim the following 
%%%%%%%%%%%%%%%%%%%%%%%%%%%%%%%%%%%%%%%%%%%%%%
%\begin{lem} \label{nt} 
%%%%%%%%%
%Let  $ N+1\leq m\leq K$ and $ \beta >0$ and let $ (C^n,\Phi^n)  $ be a sequence in $\partial\mathcal{F}_{N,K}$ with 
%%that converges weakly towards $ (C^\star,\Phi^\star)$ in $\mathbb{C}^{r} \times L^2(\R^3)^K$. 
%$\G^n=\G(C^n)$  diagonal whose eigenvalues   satisfy 
\begin{equation}\label{deg}\left\lbrace
\begin{aligned}
%&0\leq  \gamma_K^n\le\ldots \le \gamma_{m}^n\:\le \beta\le \:\gamma_{m-1}^n \le \ldots \le\gamma_1^n \leq 1,\\
%&\\
\lim_{n\rightarrow +\infty } \gamma^n_i=0 &\text { for all}\quad \: m\leq i\leq K,\vspace{2mm}\\
\vspace{2mm}\liminf_{n\rightarrow +\infty } \gamma^n_i\geq \beta>0 & \text {  for all}\quad \: 1\leq i\leq m-1.    
\end{aligned}\right.
\end{equation}
Then, 
%\begin{equation}\label{limit-gammas}
%\text{for all } \:m\leq  i\leq K,\quad \displaystyle\lim_{n\to+\infty}\gamma_{i}^n=0.
%\end{equation}
%and
\begin{equation}\label{limitC}
\text{for all } \sigma\in\Sigma_N^K, \qquad \{m, \ldots, K\}\cap \sigma\neq \emptyset\quad \Longrightarrow \quad \lim_{n\to+\infty}c_\sigma^n=0,
\end{equation}
for  $\gamma_i^n=\sum_{i\in \sigma}  |c_\sigma^n|^2$ in virtue of \eqref{gammai}. In particular, the sequence $C^n\in S^{r-1}$ being   compact
\begin{equation}\label{compactC}
 \lim_{n\rightarrow +\infty }\sum_{\sigma \subset \{1,\ldots,m-1\}} |c_\sigma^n|^2=1.
\end{equation}
%%\begin{equation}\label{limitG}
%%\begin{array}{l} \G(C^n) \text{ converges to a  block-diagonal matrix   of rank } \leq m-1\text{  and of trace }N, \\
%%\text{ with eigenvalues }(\gamma_1^\star, \ldots,\gamma_{m-1}^\star,0,\ldots,0),
%%\end{array}
%%\end{equation} 
%\begin{equation}\label{limit-gamma}
%\text{for all } 1\leq i\leq m-1, \quad  \lim_{n\rightarrow +\infty }\sum_{\sigma\subset\{1,\ldots,m-1\}\;|\; i\in \sigma} |c_\sigma^n|^2=\gamma_i^*
%\end{equation}
%and 
%\begin{equation}\label{limitTrace}
%\sum_{i=1}^{m-1}\gamma_i^\star=N,
%\end{equation}
%from \eqref{trace-fix} 
 %\begin{equation}\label{limitPhi}
% \text{for all } m\leq i\leq K, \qquad \sqrt{\gamma_i^n}\,\phi_i^n\;\text{ converges strongly to }\;0 \text{ in } L^2(\R^3), 
%  \end{equation}
% \end{lem}
% %%%%%%%%%%%%%%%%%%%%%%
% \begin{proof}
% %%%%%%%%%%%
We  decompose 
 \[%begin{equation}
 \Psi^n=\pi(C^n,\Phi^n)= \Psi_n^++\Psi_n^-\]
with
\[
\Psi_n^-=\sum_{\sigma\cap \{m,\ldots, K\}\not=\emptyset}{c}^n_\sigma \:{\Phi}^n_\sigma, \quad 
\Psi_n^+=
 \sum_{\sigma\cap \{m,\ldots, K\}=\emptyset}{c^n_\sigma}\: {\Phi}_\sigma^n\,.
 \]
 %\label{decomposition} \end{equation}
Then
 \[
 \lim_{n\to +\infty}\big\Vert \Psi_n^-\big\Vert_{L^2(\R^{3N})}=0
 \]
as a consequence of \eqref{limitC} and since each determinant ${\Phi^n_\sigma}$ is normalized in $L^2(\R^{3N})$. Hence
\begin{equation}\label{limitPhi}
 \lim_{n\to +\infty}\big\Vert \Psi^n-\Psi_n^+\big\Vert_{L^2(\R^{3N})}=0.
 \end{equation} 
%\end{proof}
%%%%%%%%%%%%%%%%%%%%%%%%%%%%%%%%%%%%%%%%%%%%%
Since the MCTDHF flow keeps the energy constant, we have 
\[
\mathcal{E}\big(\pi(C^n,\Phi^n)\big)= \mathcal{E}\big(\pi(C^0,\Phi^0)\big), 
\]
for all $n\geq 1$. This property provides with additional  information on the sequence  $(C^n; \Phi^n)$.
%We first recall  that, for any $(C;\Phi)$ in $\mathcal{F}_{N,K}$ and for the particular Hamiltonian we are dealing with, the energy functional writes
%\begin{eqnarray*}
%\mathcal{E}\big(\pi(C,\Phi)\big)&=&\int_{\R^3}\Big(\G(C)\,\Phi;(-\frac12\,\Delta+U)\,\Phi\Big)\,dx+ \iint_{\R^3\times\R^3}\frac{[\Psi\otimes \Psi]_{:2}(x,y,x,y)}{|x-y|}\,dxdy\nonumber \\
%&=&\sum_{i,j=1}^K\gamma_{ij}\,\int_{\R^3}\left[\frac12\nabla \phi_i\,\nabla\bar\phi_j+U\, \phi_i\,\bar\phi_j\right]\,dx+\sum_{i,j,k,l=1}^K \gamma_{ijkl}\, D\big(\phi_l\,\bar\phi_i\,;\,\phi_k\,\bar\phi_j\big)
%\end{eqnarray*}
%with $(\cdot;\cdot)$ being the scalar product in $\mathbb{C}^K$. 
Using  the fact that the $\phi_i^n$'s  diagonalize $\gamma^n$, the energy \eqref{CompactEnergy} rewrites
\begin{align}
\mathcal{E}(\pi(C^n,\Phi^n))
&= \sum_{i=1}^K\gamma_i^n\,\int_{\R^3}\left[\frac12|\nabla \phi_i^n|^2+U\,|\phi_i^n|^2\right]\,dx+\sum_{i,j,k,l=1}^K \gamma_{ijkl}^n\, D\big(\phi_l^n\,\bar\phi_i^n\,;\,\phi_k^n\,\bar\phi_j^n\big)\nonumber\\
&\geq   \sum_{i=1}^K\gamma_i^n\,\int_{\R^3}\left[\frac12|\nabla \phi_i^n|^2+U\,|\phi_i^n|^2\right]\,dx\label{positif},
\end{align} 
where in  \eqref{positif} we used the positivity of the  two-body interaction potential $v$. By the Kato inequality, for any $0<\varepsilon<1$, there exists  $C_\varepsilon>0$ such that 
\[
|U|\leq -\varepsilon\,\Delta+C_\epsilon\]
in the sense of self-adjoint operators. Then
\[
\sum_{i=1}^K\gamma_i^n\,\int_{\R^3}U\,|\phi_i^n|^2\,dx\geq -\epsilon \left(\sum_{i=1}^K\gamma_i^n\,\int_{\R^3}|\nabla\phi_i^n|^2\,dx \right)-C_\epsilon\, N.\]
Therefore, inserting into \eqref{positif},
\[
\sum_{i=1}^K\gamma_i^n\,\int_{\R^3}|\nabla\phi_i^n|^2\,dx \leq cste.
\]
Thus, for all $1\leq i\leq K$, $\sqrt{\gamma_i^n}\,\phi_i^n$ is bounded in $H^1(\R^3)$. Then, from \eqref{deg} and  extracting subsequences if necessary,  we obtain the alternative
\begin{equation}\label{cvge0}
\text{for all } m\leq i\leq K, \quad \sqrt{\gamma_i^n}\,\phi_i^n \text{ converges  to } 0 \text{ weakly  in } H^1(\R^3) \text{ and strongly in } L^2(\R^3),
\end{equation}
and 
\begin{equation}\label{cvgephi}
\text{for all }1 \leq i\leq m-1, \,\phi_i^n \text{ is bounded in } H^1(\R^3). 
%\text{ and converges weakly to } \phi_i^\star \in  H^1(\R^3),
\end{equation}
%We denote by $\rho^n$ the electronic density and by $\gamma^n=\gamma_{\pi(C^n,%\Phi^n)}$. Then,
%\[\rho^n=\sum_{i=1}^K \gamma_i^n\,|\phi_i^n|^2,  \quad \int_{\R^3} \rho^n\,dx=N={\rm Tr}[\gamma^n].  \]
Since, under the hypotheses on $U$, the map  $\varphi\mapsto \int_{\R^3} U\,|\varphi|^2\,dx$ is weakly  lower semi-continuous on $H^1(\R^3)$, we deduce from \eqref{cvge0} that
\begin{equation}\label{liminf-kin}
\liminf_{n\to+\infty}  \sum_{i=1}^K\gamma_i^n\,\int_{\R^3}\left[\frac12|\nabla \phi_i^n|^2+U\,|\phi_i^n|^2\right]\,dx
\geq \liminf_{n\to+\infty}\sum_{i=1}^{m-1}\gamma_i^n\,\int_{\R^3} \Big[|\nabla \phi_i^n|^2+U\,|\phi_i^n|^2\Big]\,dx.
\end{equation}
We now check that
\begin{equation}\label{liminfA}
\liminf_{n\to+\infty}\sum_{i,j,k,l=1}^K \gamma_{ijkl}^n\, D\big(\phi_l^n\,\bar\phi_i^n\,;\,\phi_k^n\,\bar\phi_j^n)= \liminf_{n\to+\infty}\sum_{i,j,k,l=1}^{m-1} \gamma_{ijkl}^n\, D\big(\phi_i^n\,\bar\phi_l^n\,;\,\phi_k^n\,\bar\phi_j^n)
\end{equation}
by showing  that
\begin{equation}\label{limitD0}
\liminf_{n\to+\infty}\sum_{\stackrel{ i,j,k,l=1}{ \{i,j,k,l\}\cap \{m,\ldots,K\}\neq\emptyset}}^K \gamma_{ijkl}^n\, D\big(\phi_l^n\,\bar\phi_i^n\,;\,\phi_k^n\,\bar\phi_j^n)= 0.
\end{equation}
Let  $\{i,j,k,l\}\cap \{m,\ldots,K\}\neq\emptyset$. We assume  without loss of completeness that $i\geq m$. From  the expression \eqref{gammaijkl}  for  $\gamma_{iklj}^n$, we  observe that  
\begin{align}
\Big|\gamma_{ijkl}^n\Big|&\lesssim \min\big(\sqrt{\gamma_{i}^n}\,; \,\sqrt{\gamma_{j}^n}\big)\,\min\big(\sqrt{\gamma_{k}^n}\,; \,\sqrt{\gamma_{l}^n}\big)\nonumber\\
&\lesssim \min\big(\gamma_{i}^n\,; \,\gamma_{k}^n\,;\,\gamma_{j}^n\,; \,\gamma_{l}^n\big)^{1/2},
\label{borneA}
\end{align}
since $0\leq \gamma_{\cdot}^n\leq 1$. We thus  get 
 \begin{equation}\label{limitA}
 \text{if } \{i,j,k,l\}\cap \{m,\ldots, K\}\neq\emptyset, \quad \lim_{n\to +\infty} \gamma_{ikjl}^n=0, 
  \end{equation}
from \eqref{deg}. Then thanks to \eqref{simple} and  \eqref{borneA} 
 \begin{align*}
 |\gamma_{ijkl}^n\, D\big(\phi_l^n\,\bar\phi_i^n\,;\,\phi_k^n\,\bar\phi_j^n)|&\lesssim \sqrt{\gamma_i^n}\,\sqrt{\gamma_k^n}\,
 \|\nabla\phi_k^n\|_{L^2}\, \|\phi_i^n\|_{L^2}\,\|\phi_j^n\|_{L^2}\,\|\phi_l^n\|_{L^2}\\
 &\lesssim  \sqrt{\gamma_i^n}
 \end{align*}
since the $L^2$ norms of the orbitals equal $1$ and since in any case $\sqrt{\gamma_j^n}\,\nabla\phi_j^n$ is bounded in $L^2$ independently of $n$. Therefore each term which appears in the sum in  \eqref{limitD0}  converges to $0$ as $n$ goes  to infinity thanks to \eqref{cvge0}. Claim  \eqref{limitD0} then follows.  
\vskip6pt
Gathering together  \eqref{liminf-kin} and \eqref{liminfA} we have
\begin{eqnarray}
\lefteqn{\liminf_{n\to+\infty}\mathcal{E}\big(\pi(C^n;\Phi^n)\big)}\nonumber\\
&\geq&\! \!\!\!\!\liminf_{n\to +\infty}\Big[ \sum_{i=1}^{m-1}\gamma_i^n\!\!\int_{\R^3} \Big[|\nabla \phi_i^n|^2+U\,|\phi_i^n|^2\Big]\,dx+\!\!\!\!\sum_{i,j,k,l=1}^{m-1}\!\!\! \gamma_{ijkl}^n\, D\big(\phi_i^n\,\bar\phi_l^n\,;\,\phi_k^n\,\bar\phi_j^n)\Big].\label{liminf}
\end{eqnarray}
The point  now consists in   showing  that the right-hand side in \eqref{liminf} is bounded from below by $\displaystyle \liminf_{n\to+\infty}\mathcal{E}\big(\Psi_n^+\big)$. Indeed, let us set $\widetilde{\Psi}^n=\pi(\widetilde{C}^n,\widetilde{\Phi}^n)$ where    $\widetilde{C}^n=(c_\sigma^n)_{\sigma\subset \{1,\ldots, m-1\}}\in \mathbb{C}^{K\choose{m-1}}$  and $\widetilde{\Phi}^n=(\phi_1^n,\ldots, \phi_{m-1}^n)\in \mathcal{O}_{\0^{m-1}}$. There is a slight difficulty arising here   from the fact that  (with obvious notation) $\widetilde{\gamma}_{ij}^n$ is close but different from  $\gamma_{i}^n \,\delta_{ij}^n$ and similarly for $\widetilde{\gamma}_{ijkl}^n$ and $\gamma_{ijkl}^n$. (Also $\widetilde{C}^n$ is not normalized in  $\C^{m-1}$ (only asymptotically) but this will be dealt with afterwards.)
\vskip6pt
First we observe that because of  \eqref{gammaij}  for every $i,j\in \{1,\cdots, m-1\}$, 
\[
\gamma_{i}^n\,\delta_{ij}^n-\widetilde{\gamma}_{ij}^n
=\sum_{\substack{(\sigma \cup\tau)\, \cap \{m,\cdots, K\}\neq \emptyset \\i\in\sigma, \,j\in\tau\,, \sigma\setminus\{ i\}= \tau\setminus\{ j \}}} (-1)^{\sigma^{-1}(i)+\tau^{-1}(j)}\, c_\sigma^n\,\overline {c}_\tau^n
\]
goes to $0$ as $n$ goes to infinity thanks to \eqref{limitC}.  In addition, each term of the form $\int_{\R^3} \Big[\frac 12\nabla \phi_i^n\cdot\nabla \bar\phi_j^n +U\,\phi_i^n\cdot\bar\phi_j^n \Big]\,dx $   is bounded independently of $n$ for  $i,j\in \{1,\cdots, m-1\}$. Therefore
\begin{equation}\label{kin-ok}
\sum_{i=1}^{m-1}\gamma_i^n\,\int_{\R^3} \Big[\frac 12|\nabla \phi_i^n|^2+U\,|\phi_i^n|^2\Big]\,dx=\sum_{i,j=1}^{m-1}\widetilde{\gamma}_{ij}^n\,\int_{\R^3}\left[\frac12\nabla{\widetilde{\phi_i^n}}\,\nabla\bar{\widetilde{\phi_j^n}}+U\, \widetilde{\phi_i^n}\,\bar{\widetilde{\phi^n_j}}\right]\,dx+o(1).
\end{equation}
For the same reason, and with obvious notation,  for all $1\leq i,j,k,l\leq m-1$, 
\[
\lim_{n\to+\infty}\big|\gamma_{ijkl}^n-\widetilde{\gamma}_{ijkl}^n\big|=0\]
since according to \eqref{gammaijkl} the extra terms in these  differences only  involve coefficients $c_\sigma^n$ with $\sigma\cap \{m,\ldots,K\}\neq\emptyset$.  Again each term of the form  $D\big(\phi_i^n\,\bar\phi_l^n\,;\,\phi_k^n\,\bar\phi_j^n\big)$  is bounded independently of $n$ for $i,j,k,l\in\{1,\ldots,m-1\}$. Therefore 
\begin{equation}\label{pot-ok}
\sum_{i,j,k,l=1}^{m-1} \gamma_{ijkl}^n\, D\big(\phi_i^n\,\bar\phi_l^n\,;\,\phi_k^n\,\bar\phi_j^n)=\sum_{i,j,k,l=1}^{m-1} \widetilde{\gamma}_{ijkl}^n\, D\big(\phi_i^n\,\bar\phi_l^n\,;\,\phi_k^n\,\bar\phi_j^n)+o(1).
\end{equation}
Therefore, gathering together \eqref{liminf}, \eqref{kin-ok} and \eqref{pot-ok},
\begin{equation}\liminf_{n\to+\infty}\mathcal{E}\big(\pi(C^n;\Phi^n)\big)\geq \liminf_{n\to+\infty}\mathcal{E}\big(\pi(\widetilde{C}^n;\widetilde{\Phi}^n)\big)=\liminf_{n\to+\infty}\mathcal{E}\big(\widetilde{\Psi}^n).\label{liminf-tilde}
\end{equation}
Since $\widetilde{C}^n$ is not in $S^{{K \choose {m-1}}-1}$ (it is only the case  asymptotically), $(\widetilde{C}^n;\widetilde{\Phi}^n)$  is not in  $\mathcal{F}_{N,m-1}$, thus we cannot  bound  immediately $\mathcal{E}\big(\widetilde{\Psi}^n\big)$  from below by  $I(m-1)$. Anyway,  in virtue of  \eqref{compactC}, 
\begin{equation}\label{limit1}
\lim_{n\to+\infty}\|\widetilde{\Psi}^n\|^2=1.
\end{equation}
The energy being quadratic with respect to $\Psi$ 
\begin{equation}\label{quadratic}
\mathcal{E}\big(\widetilde{\Psi}^n\big)=\|\widetilde{\Psi}^n\|^2\, \mathcal{E}\Big(\frac{\widetilde{\Psi}^n}{\|\widetilde{\Psi}^n\|}\Big) \geq \|\widetilde{\Psi}^n\|^2\, I(m-1),
\end{equation}
for $\widetilde{\Psi}^n/ \|\widetilde{\Psi}^n\|\in \mathcal{F}_{N,m-1}$ for all $n\geq 1$.
Gathering together \eqref{liminf-tilde}, \eqref{limit1} and \eqref{quadratic} and taking the limit as $n$ goes to infinity we deduce 
\begin{equation}\label{liminfE}
\liminf_{n\to + \infty} \mathcal{E}\big(\pi(C^n,\Phi^n)\big)
%&\geq& \liminf_{n\to + \infty} \mathcal{E}\big(\pi(\widetilde{C}^n,\tilde{\Phi}^n)\big)\\
%&=&  \liminf_{n\to + \infty} \mathcal{E}\big(\pi(\widetilde{C}^\star,\tilde{\Phi}^n)\big)\\
\geq  I(m-1).
\end{equation} Hence the theorem.
\hfill $\Box$
%%%%%%%
\begin{rem} When $\Omega$ is a bounded domain of $\R^3$,  any sequence in $\mathcal{F}_{N,K}$ is relatively compact in $\C^r\times L^2(\Omega)^K$ thanks to the Rellich theorem. On the other hand, the energy functional $\Psi\mapsto \mathcal{E}(\Psi)$ is weakly lower semi-continuous in $H^1(\Omega^{3N})$. Therefore it is easily checked in that case that
\[
\liminf_{n\to +\infty}\mathcal{E}\big(\pi(C^n;\Phi^n)\big)\geq \liminf_{n\to +\infty}\mathcal{E}\big(\pi(C^\star;\Phi^\star)\big)\geq I(m-1)
\]
with $(C^\star;\Phi^\star)\in \mathcal{F}_{N,m-1}$ being the weak limit of the sequence $(\widetilde{C}^n;\widetilde{\Phi}^n)$ introduced in the above proof.
\end{rem}
%%%%%%%
\vskip10pt
\begin{rem}[Stability, Consistency and Invertibility of the density matrix $\G$] The main factor in the instability of the working  equations or any gauge-equivalent system, is the inverse of the density matrix. In the present   section,  criteria for the global invertibility of $\G(C)$ have been  given. These criteria do not provide with an uniform estimate for $\Vert \G^{-1}\Vert$, and furthermore increasing the consistency of the MCTDHF approximation leads to an increase of the number $K$ of orbitals. As usual consistency and stability are both necessary and antinomic. Indeed, the most obvious observation is that one always has 
\[
\Vert \G^{-1}\Vert\geq \frac K N,
\]
for $\G$  has at most $K$ positive eigenvalues whose sum equals $N$. Therefore the smallest can  be at most  $N/K$. These considerations lead either to a limitation on $K$ or to a regularization or a ``cut-off'' of $\G^{-1}$. In fact the ``consistency" in the sense of numerical analysis is obtained  with fixed $N$ by letting $K$ go to infinity. This is basically different from the idea (in spirit  of statistical mechanics) of letting $N$ go to infinity \cite{Bardos2}.
\end{rem}
 
%%%%%%%%%%
\section[$L^2$ analysis]{Stabilization of  $\G$  and existence of $L^2$ solutions}\label{sec:L2}
%%%%%%%%%%
In the above analysis, both for existence of maximal  solutions and for global invertibility of the density matrix, the conservation of energy plays a crucial r\^ole. 
Besides the theoretical interest, the analysis of an MCTDHF system with infinite (or non conserved) energy but finite mass is relevant. Indeed, to circumvent the possible degeneracy of the density matrix, physicists resort to  \textit{ad hoc} methods like perturbations of this matrix in order to ensure its invertibility. Typically, this is achieved as follows 
\begin{equation}\label{gamma-id}
\G_{\epsilon}=\G+\epsilon\,Id
\end{equation}
(see e.g. \cite{Scrinzi}), or by taking
\begin{equation}\label{gamma-eps}
\G_{\epsilon}=\G+\epsilon\,\exp(-\G/\epsilon)
\end{equation}
for small values of $\epsilon$ (see \cite{Meyer}). Note that in latter case vanishing eigenvalues 
are perturbed at order $\epsilon$ while the others are unchanged up to exponentially small  
errors in terms of $\epsilon$. Then the perturbed system reads for an $\epsilon>0 $
\begin{equation}\label{Seps}
\left\{
\begin{aligned}   i\,\frac{d C}{dt} &=\mathbb{K}[\Phi]\:C,     \\ 
 i\,\frac{\partial \Phi}{\partial t} &=\mathbf{H}\,\Phi
+\, \G_\epsilon(C)^{-1}\,(\mathbf{I}- \mathbf{P}_\Phi)\,\mathbb{W}[C,\Phi]\Phi, \\
 C(0)= C^0,&\quad \Phi(0)=\Phi^0.
 \end{aligned} 
  \right. 
\end{equation}
On the other hand, when a  laser  field is turned on, the Hamiltonian of the system is then time-dependent  which is a relevant configuration from the physical point of view (see \cite{Scrinzi} and Section~\ref{sec:extension} below). In such situation, the conservation of the energy fails and a recourse to alternative theories is necessary.  

However in both situations the $L^2$ norm (which corresponds to the electronic charge) is conserved and this justifies an $L^2$ analysis of the MCTDHF outside the energy space. Therefore the Strichartz estimates turn out to be a natural tool  in the same spirit as  Castella~\cite{Castella} and Zagatti~\cite{Zagatti}. In \cite{Saber-L2},  existence and uniqueness of global-in-time  mild solutions has been  obtained for $L^2$  initial data.  As in the previous section (and with the same notation)  the perturbed working equations  are written in ``Duhamel" form   
\begin{align*}
C(t)&=C(0) +\int_0^t  \mathbb{K}[\Phi(s)]\:C(s)ds\,,   \\
\Phi(t)&=S(t)\,\Phi^0-i\int_0^tS(t-s)\:U\:\Phi(s)\,ds\\
& \quad -i\int_0^t S(t-s)\: \G_\epsilon[C(s)]^{-1}\,(\mathbf{I}- \mathbf{P}_\Phi)\,\mathbb{W}[C,\Phi]\Phi\,ds\,,
\end{align*} 
where $S(t)=\exp[-\frac 12 i\,t\Delta]$ denotes the group of isometries generated by  $-\frac i2\Delta$ on $L^2(\mathbb{R}^3,\mathbb{C})$.  The potentials $U $ and $v=v(|x|)$ belong to $L^{\frac32} +L^\infty\,.$  

{}From the relation 
\[
\Vert S(t)\phi\Vert  _{L^\infty(\R^3)} \le \frac{1}{(4\pi\,t)^{3/2}} \, \Vert \phi\Vert_{L^1(\R^3)}
\]
and
\[
\Vert S(t)\phi\Vert _{L^2(\R^3)} = \Vert \phi\Vert_{L^2(\R^3)}
\]
one deduce by interpolation the so-called  Strichartz estimates
\[
\Vert S(t)\phi\Vert_{L^p(0,T;L^q(\R^3))} \le C(q) T^{\frac 3q -\frac 12}, 
\]
that hold for any  \textit{Strichartz pairs}  $(p,q)\in [2,+\infty]\times [2,6]$ with $\frac{2}{3p}=(\frac 1 2-\frac 1 q)$.  (Strichartz estimates for the endpoints $p=2$ and $q=6$ are more intricate and due to Keel and Tao \cite{KT}). 

Following    Zagatti \cite{Zagatti} and Castella   \cite{Castella},     the spaces
 \begin{equation*} 
 X_T= L^{\infty}(0,T;\C^r)\times \Big(L^{\infty}\big(0,T;L^2(\R^3) \big)\cap   L^p\big(0,T;L^q(\R^3)\big )\Big)^K\,,    
 \end{equation*}
are introduced for any Strichartz pairs. For  some $R>0$ and some $T>0$ small enough, the non-linear operator $(C,\Phi)\mapsto L(C,\Phi)$ which appears in the   Duhamel  integral 
 \begin{equation*}
 L(C,\Phi)(t)=\left(\begin{array}{c}\displaystyle \int_0^t   \K[\Phi (s)]\:C(s)\,ds  
 \\
\vspace{2mm}
\displaystyle \int_0^tS(t-s)\Big(U\, \Phi(s) 
+  \G_\epsilon(s)^{-1} (I-\mathbf{P}_\Phi)\:\mathbb W[C(s),\Phi(s) ]\:\Phi(s)\Big)\,ds\end{array}\right)
\end{equation*}
is a strict contraction in the ball
\[\left\{(C,\Phi) \in X_T\::\quad \Vert C \Vert_{\mathbb{C}^r} + \Vert \Phi\Vert_{L^{\infty}(0,T;L^2(\R^3))} +\Vert \Phi\Vert_{L^{p}(0,T;L^q(\R^3))}\leq R\right\}.\]  
%maps the ball of radius $R$ of $X_T$ into itself and is Lipschitz  continuous  with a Lipschitz constant bounded by
% \begin{equation*}
%    C(\G_\epsilon, |\!|U|\!|_{L^d},|\!|v|\!|_{L^d}) R^5 T^{\frac 3q -\frac 12} \label{uni}\,.
%\end{equation*}
Next using the conservation of the $L^2$ norms of the orbitals and the estimate  
\[\Vert \Phi\Vert_{L^{p}(0,T;L^q(\R^3))} \lesssim \:\Vert \Phi^0\Vert_{L^2(\R^3)}\]  
one follows the lines of Tsutsumi in \cite{Tsu}  to get   existence and uniqueness of a strong  solution in $X_{\infty} $ (see the details in \cite{Saber-L2}). This is summarized in the 
  \begin{prop} Let $\epsilon>0$. 
  For any initial data $(C_0,\Phi_0)\in \partial{\mathcal F}_{N,K}$ and for any Strichartz pairs $(p,q)$, the $\epsilon$-regularized working equations admit a unique strong solution
  $$(C_\epsilon(t),\Phi_\epsilon(t))\in  L^{\infty}(\R^{+};\C^r)\times \big(L^{\infty}(\R^{+};L^2(\R^3) )\cap  L_{\rm loc} ^p(\R^{+};L^q(\R^3) \big)^K\,$$
that lives in $\mathcal{F}_{N,K}$ for all $t\geq 0$. If  in addition $\Phi_0\in H^1(\R^3)^K$ then $\Phi_\epsilon(t)\in C^0(\R^{+};H^1(\R^3))^K$.
\end{prop} 
Eventually one expects that whenever the original solution is well-defined (with  a non degenerate density matrix $\G(t)$) on a time interval $0\le t < T^*$ it will be on the same interval the limit for $\epsilon\rightarrow 0$ of the solution of the perturbed working equations. This is the object of the following
%%%%%%%%%%
\begin{thm}  Let  $(C_0,\Phi_0)\in \mathcal{S}^{r-1}\times (H^1(\R^3))^K$.  Assume that the corresponding solution $(C(t),\Phi(t))$  to \eqref{working} is well-defined on $[0, T]$ and is such that  
\begin{equation}\label{borne-G}
\sup_{0\leq t\leq T}|\!| \G(t)^{-1}|\!| \le M< +\infty. 
\end{equation}
Then,  on the same time interval it is the limit in $\C^r\times H^1(\R^3)^K$ for $\epsilon \rightarrow 0$ of the solution $(C_\epsilon, \Phi_\epsilon)$ to the regularized problem \eqref{Seps} with same initial data.
\end{thm}
%%%%%%%%%%
\begin{proof} We first recall the obvious \textit{a posteriori} bounds
\begin{equation*}%\label{apriori-eps1}
\Vert C\Vert =\Vert C_\eps\Vert =1, \quad \Vert \Phi\Vert_{L^2} =\Vert \Phi_\eps\Vert_{L^2} =1
\end{equation*}
on $[0,T]$, and, as a consequence of \eqref{borne-G} and the energy conservation, 
\[
\max_{0\leq t\leq T}\Vert \Phi(t)\Vert_{H^1}\leq M'
\]
with $M'=M'(\mathcal{E}\big(\pi(C_0,\Phi_ 0)\big),M)$. We can also rely on the orthonormality of the orbitals in $\Phi$ and  $\Phi_{\eps}$. We introduce   the notation
\[
U=\left(\begin{array}{c} C  \\ \Phi \end{array}\right),\quad U_\epsilon=\left(\begin{array}{c} C_\epsilon  \\ \Phi_\epsilon \end{array}\right), \quad\mathcal A=\left(\begin{array}{c} 0  \\ {\mathbf H} \end{array}\right),\quad  \mathcal{B}_{(\eps)}(U)=\left(\begin{array}{c} \K[\Phi]\,C
 \\ 
\G_{(\eps)}^{-1}\,\B(U)\end{array}\right)
\]
where  
\[
\B(U)=(\mathbf{I}-\mathbf{P}_{\Phi})\,\mathbb{W}[C,\Phi]\Phi \]
and  where the index $(\eps)$ means that the claim  holds both  for  the regularized system and the initial one, uniformly in $\epsilon$. System~\eqref{Seps} can also be written in synthetic  form:
\begin{align}\label{duhamelU}
U(t)&= e^{-it\mathcal A}U_0-i \int_0^t e^{-i(t-s)\mathcal A}\mathcal{B}\big(U(s)\big)\,ds,
\\\label{duhamelUeps}
U_\epsilon(t)&= e^{-it\mathcal A}U_0-i \int_0^t e^{-i(t-s)\mathcal A} \mathcal{B}_\eps\big(U_\epsilon(s)\big)\,ds.
\end{align}
Since the initial $\Phi_\eps(0)=\Phi_0$ is in $H^1$ and since the regularized system propagates the regularity, $\Phi_\eps$ is in $H^1(\R^3)^K$ for all time.  
\vskip6pt
We fix $\epsilon>0$. We introduce a parameter $\eta>0$ to be made precise later and the set
\[
I_\epsilon=\big\{t\in [0,T]\,:\, \Vert U_\epsilon(t) -U(t)\Vert \leq \eta\big\}
\]
with $\Vert U\Vert=\Vert C\Vert +\Vert \Phi\Vert_{H^1}$. The mappings $t\mapsto U_{(\epsilon)}(t)$  being  continuous from $[0,T]$ to $X_T:= \C^r\times L^\infty\big(0,T; H^1(\R^3)^K\big)$,  the set $I_\epsilon$ is  closed and since $\Vert U_\epsilon(0)- U(0)\Vert=0$, there exists a maximal time $T_\epsilon>0$ in $I_\epsilon$ such that 
\[
\forall t\in [0,T_\epsilon],\quad \Vert U_\epsilon(t)-U(t)\Vert \leq \eta.\]
We now prove by contradiction that $T_\epsilon=T$.  Assume then $T_\eps<T$. \vskip6pt
Subtracting \eqref{duhamelU} to \eqref{duhamelUeps} and taking norms first yields to
\begin{align}%\label{diff-C}
\Vert C_\eps(t)-C(t)\Vert& \leq  \int_0^t \Vert C_\eps\Vert\,\Vert \K[\Phi_\eps]-\K[\Phi]\Vert+ \Vert C_\eps-C\Vert\,\Vert \K[\Phi]\Vert,
\nonumber \\ 
%&\leq &\int_0^t \kappa_1 \,\Vert \Phi_\eps-\Phi\Vert_{H^1}+ \kappa_0\,\Vert C_\eps-C\Vert\,\Vert \K[\Phi]\Vert\nonumber\\
%&\leq & \kappa_1\,\int_0^t \Vert \Phi_\eps-\Phi\Vert_{H^1}+\kappa_0\,M'\,\int_0^t \Vert C_\eps-C\Vert\nonumber\\
&\leq C(\eta)\,\int_0^t \Vert U_\eps-U\Vert\,ds
\label{eval-C}
\end{align}
for all $0\leq t\leq T_\eps$. Here and below $C(\eta)=C(M,\mathcal{E}\big(\pi(C_0,\Phi_ 0)\big), \eta)$ denotes a positive constant that may vary from line to line but that is independent of $\epsilon$ and continuous and non-decreasing with respect to  $\eta$. Indeed we use the fact that the non-linearity  $\Phi\mapsto \K[\Phi]$ is locally Lipschitz continuous in $H^1$ (Subsection~\ref{ssec:local}) together with the uniform bound 
\[
\max_{0\leq t\leq T_\epsilon} \Vert \Phi_{\epsilon}\Vert_{H^1}\leq M'+\eta.
\] 
On the other hand, we write 
\begin{align}
\Vert \Phi_\epsilon(t)-\Phi(t)\Vert&\lesssim \int_0^t \Vert \B(U_\eps)\Vert_{H^1}\,\Vert \G_\eps^{-1}-\G^{-1}\Vert+ \Vert \G^{-1}\Vert\, \Vert \B(U_\eps)-\B(U)\Vert_{H^1}\nonumber\\
&\leq  C(\eta)\, \int_0^t \Big(\Vert \G_\eps^{-1}-\G^{-1}\Vert+  \Vert U_\eps-U\Vert_{H^1}\Big)\,ds,
\label{diff-Phi2}
\end{align}
by  using  the local Lipschitz  bounds of $U\mapsto \B(U)$ given in Subsection~\ref{ssec:local}.  We now turn to the quantity $\Vert \G_\eps^{-1}-\G^{-1}\Vert$. Both regularization \eqref{gamma-id} and \eqref{gamma-eps} of the density matrix take the form :
\[
\G_\eps=\G(C_\epsilon)+\eps\,g(C_\eps)
\]
with $\Vert g(C_\eps)\Vert \leq \eps$. Then, 
\begin{align}
\Vert \G_\eps-\G\Vert &\leq \Vert \G(C_\eps)-\G(C)\Vert +\eps\nonumber\\
&\leq \kappa\,( \Vert C_\eps- C\Vert+\eps)\label{borne-G-eps}
\end{align}
by using  the obvious bound $\Vert\G(C)\Vert \lesssim \Vert C\Vert^2$ for  $C,\,C_\eps\in S^{r-1}$, where $\kappa$ only depends on $N$ and $K$. We now assume that 
\begin{equation}\label{apriori-eps}
\eps,\,\eta\leq \frac 1{4\,\kappa\,M},
\end{equation}
where $M$ is given in the statement of the theorem. Using
\[
\G_\eps =\Big(I-\big(\G-\G_\eps\big)\,\G^{-1}\Big)\,\G,
\]
we deduce
\begin{equation*}
 \G_\eps^{-1}= \G^{-1}\,\Big(I-\big(\G-\G_\eps)\,\G^{-1}\Big)^{-1}= \G^{-1}\,\sum_{n\geq 0} \Big(\big(\G-\G_\eps)\,\G^{-1}\Big)^{n}.
\end{equation*}
Therefore
\begin{align}
\Vert \G_\eps^{-1}-\G^{-1}\Vert &\leq \sum_{n\geq 1} \Vert \G-\G_\eps\Vert^n \;\Vert\G^{-1}\Vert^{n+1}\nonumber\\
&\leq  \sum_{n\geq 1}M^{n+1}\,\kappa^n\, (\Vert C_\eps-C\Vert+\eps)^n\nonumber
\end{align}
by using \eqref{borne-G-eps}. Hence 
\begin{align}
\Vert \G_\eps^{-1}-\G^{-1}\Vert &\leq M^2\,\kappa\,\big(\Vert C_\eps-C\Vert+\eps)\,\sum_{n\geq 0} M^{n}\,\kappa^n\, (\Vert C_\eps-C\Vert+\eps)^n\nonumber\\
&\leq 2\, M^2\,\kappa\,\big(\Vert C_\eps-C\Vert+\eps)\label{diff-G-eps}
\end{align}
since  $M\,\kappa\,\big(\Vert C_\eps(t)-C(t)\Vert+\eps)\leq \frac{1}{2} $  by \eqref{apriori-eps} and  for $t$ in $[0,T_\eps]$. Inserting  \eqref{diff-G-eps} in   \eqref{diff-Phi2} we get:
\begin{equation}\label{eval-Phi}
\Vert \Phi_\epsilon(t)-\Phi(t)\Vert\leq C(\eta)\,\int_0^t \big(\Vert U_\eps(s)-U(s)\Vert+\eps\big)\,ds.
\end{equation}
Eqn. \eqref{eval-Phi} together with \eqref{eval-C} finally leads to
\begin{equation}\label{eval-U}
\Vert U_\epsilon(t)-U(t)\Vert\leq C(\eta)\,\int_0^t\big(\Vert U_\eps(s)-U(s)\Vert+\eps\big)\,ds,
\end{equation}
for all $t\in [0,T_\eps]$. Eventually, thanks to Gronwall's inequality,
\begin{equation}\label{eval-diff-U}
\max_{0\leq t\leq T_\eps } \Vert U_\epsilon(t)-U(t)\Vert\leq \epsilon\, e^{C(\eta)\,T}.
\end{equation}
With $\eta$ as in \eqref{apriori-eps}, next  
\begin{equation}\label{eps-final}
\epsilon\leq \min\big(\frac 1{4\,\kappa\,M}, \frac \eta 2\, e^{-C(\eta)\,T}\big),
\end{equation}
we get
\[
\max_{0\leq t\leq T_\eps } \Vert U_\epsilon(t)-U(t)\Vert\leq \frac \eta 2
.\]
By continuity of $t\mapsto \Vert U_\epsilon(t)-U(t)\Vert$, we may then find $T_\epsilon'>T_\epsilon$ such that $\displaystyle \max_{0\leq t\leq T_\eps' } \Vert U_\epsilon(t)-U(t)\Vert\leq  \eta $. Hence the contradiction with the definition of $T_\epsilon$. Therefore, $I_\eps=[0,T]$ and, going back to \eqref{eval-diff-U} we obtain:
\begin{equation}
\max_{0\leq t\leq T} \Vert U_\epsilon(t)-U(t)\Vert\leq \epsilon\, e^{C(\eta)\,T},
\end{equation}
for say  $\eta=\frac 1{4\,\kappa\,M}$ and $\epsilon$  small enough, satisfying \eqref{eps-final}, whence the result.
\end{proof}
%%%%%%%%%%%%%%%%%%%%%%%%%%%%%%%%%%%%%%%%%%%%%%%%%%%%%%%%%%%%%%%%%%%%%%%%%%
In the forthcoming (and last) section we comment on straight extensions of the above analysis.

%%%%%%%%%%%%%%%%%%
\section{Extensions}\label{sec:extension}
%%%%%%%%%%%%%%%%%%
The present contribution is focused on the algebraic and functional analysis properties of the MCTDHF equations for fermions.  Multi-configuration approximations can  also be considered for symmetric wave-functions or also for wave-functions with no symmetry (see \textit{e.g. }\cite{Meyer, Lubich1}). The mathematical analysis of  the  equations which play the r\^ole of the ``working equations" of Section~\ref{sec:flow} is similar. On the other hand, the fermionic case is important by itself and leads to much better geometric  structure in terms of principal fiber bundle as described in Section~\ref{sec:stationary}. Hence our choice. Our results could be generalized to  general (symmetric)  $n$-body interactions as well including the $n$-body density matrices.    
%%%%%%%%%%%%%%%%
\subsection{Beyond Coulomb potentials.} Although above  results and proofs  are mainly detailed for  Coulomb potentials they carry through more general real-valued potentials. Indeed well-posedness results in $H^1$ and $H^2$ are  still valid for  $U$ and $v$  in the  class $ L^p(\R^3)+L^\infty (\R^3)$ with $p>3/2$, and $v\geq 0$.   These conditions ensure that $\mathcal{H}_N$ is self-adjoint in $L^2(\Omega^N)$, that the one-body operator $-\frac 12\Delta +U$ is  a semi-bounded self-adjoint in $L^2(\R^3)$ with domain $H^2(\R^3)$ and that the  Kato inequality holds  for the potential $U$. Under these assumptions,  the energy space is $\C^r\times H^1(\R^3)^K$  (respectively $\C^r\times H^1_0(\Omega)^K$ when $\Omega$ is a bounded domain) and  the propagator  $e^{-itH}$ is a one-parameter group of unitary operators in $H^2(\R^3)$ and in $H^1(\R^3)$.   

For the  global well-posedness  sufficient condition to hold true  (Theorem~\ref{mainth} and its corollary) further conditions on the potentials  are required  to ensure that the energy functional is weakly  lower semi-continuous  on the energy space.  Sufficient conditions are  (for example) $U\geq 0$ or $U_{-}$ (the negative part of $U$)  tending   to $0$ at infinity at least in a weak sense.
%%%%%%%%%%%%%%%%%%%%%%%%%% 
\subsection{Extension to time-dependent potentials.} One of the basic use of the MCTDHF is the simulation of ultra-short light pulses with matter \cite{Zang1}. Describing  this situation leads to the same type of equations but with the one-body Hamiltonian $\mathbf{H}$ being replaced by a one-body time dependent hamiltonian 
\[\mathbf{H}_{\omega,A}(t):=(i\nabla + A(t))^2 + \omega(t)\,U(x)\]
with $\omega(t)$ and $A(t)$ real, $A(0)=0$ and $U$ as in the above subsection.  A typical example is $A(t)=A_0\, \exp{\big(-(t/\tau)^2\big)}\sin(\alpha t)$ for some positive real parameters $A_0$, $\alpha$ and $\tau$ \cite{Zang1,Zang2}. This does not change neither  the algebraic and geometrical structure of the equations nor   the definition of the density matrix $\G$ nor  the notion of full-rank. The potential vector $A$ being independent of the $x$ variable the energy space is $H^1$. With convenient hypotheses (say  $\omega$ and  $A$ continuous, bounded with bounded derivatives), the results in Section~\ref{sec:analysis}  concerning local-in-time  $H^1$ well-posedness of the Cauchy problem remain valid. For generalization of the use of Strichartz estimates and the local $L^2$ well-posedness one should follow for example 
  \cite{Caz}. Since the energy is now time-dependent  extra hypothesis have to be introduced for the persistence of the full-rank assumption done in Section \ref{sec:global}.
  
Assume that $\omega(t)$ and  $A(t)$ take their values in a bounded set (the set of ``control" $\mathcal C$) and that their derivatives are  also bounded. The system $\mathcal{S}_0$ \eqref{S0} with $\mathbf{H}$ replaced by $\mathbf{H}_{\omega,A}(t)$ keeps on preserving  the constraints since Lemma~\ref{cons-contrainte} only relies on the self-adjointness of the Hamiltonian. Similarly  solutions to \eqref{S0} satisfy the Dirac--Frenkel variational principle. The energy is no longer conserved  by the flow. Indeed, following the lines of the proof of Corollary~\ref{energyconservation}, we have  
\[
\frac d{dt}\mathcal{E}\big(\Psi(t)\big)=\frac d{dt}\big\langle \mathcal{H}(t)\,\Psi(t)\big\vert\Psi(t)\big\rangle= \big(\omega'(t)+ 2\,A(t)\,A'(t)\big)\, \langle\Psi(t)\big\vert\Psi(t)\rangle, \]
with the prime denoting time derivatives. 
%or in the second case:
%\[\frac d{dt}\mathcal{E}\big(\Psi(t)\big)=\big\langle (A(t) \frac{d A(t)}{d t}+  \frac{d A(t)}{d t} A(t)\Psi(t)\big\vert\Psi(t)\big\rangle. \]
%and therefore in any case
However 
\[\mathcal{E}\big(\Psi(t)\big)= \omega(t)+2\,A(t)^2+\mathcal{E}\big(\Psi(0)\big)-\omega(0),\]
%\begin{equation}
%|\frac d{dt}\mathcal{E}\big(\Psi(t)\big)|\le h(t) \mathcal{E}\big(\Psi(t)\big)\hbox { with } \int_0^\infty h(t)dt <\infty
%\end{equation}
and  the energy  in controlled for any finite time, whence the existence of a maximal solutions in  $H^1$ as long as the matrix $\G\big(C(t)\big)$ remains invertible.  

To adapt the result concerning the global   full-rank hypothesis, we introduce the minimization problems for any real numbers $\bar\omega$ and $\bar A$ 
\[
\mathcal{I}_{\bar\omega,\bar A}(K)= \inf\Big\{\mathcal{E}_{\bar\omega,\bar A}(\Psi)\::\: \Psi \in  \mathcal{B}_{N,K}\Big\}
\]
with 
\[
\mathcal{E}_{\bar\omega,\bar A}(\Psi)= \left(\Big( \mathbf{H}_{\bar \omega,\bar A}\:\G +
\frac{1}{2}\mathbb{W}[C,\Phi]\Big)\Phi,\Phi\right)_{L^2(\Omega)^K}
\]
for $\Psi=\pi(C,\Phi)$.
%%$$
%%I_{\mathbf C} = \inf_{u\in \mathbf C, \Psi \in  \mathcal{B}_{N,(K-1)}} \frac{(\sum_i  -\frac12\Delta_{x_i} + u U(x_i)) + V )\Psi,\Psi)}{|\Psi|^2}
%%$$
%%or
%$$I_{\mathbf C} (K-1)= \inf_{A\in \mathbf C, \Psi \in  \mathcal{B}_{N,(K-1)}} \frac {(\sum_i (i\nabla_{x_i} + A(t))^2) + V )\Psi,\Psi)}{|\Psi|^2}
%$$
The global-in-time conservation of full-rank in Theorem~\ref{mainth} remains true under the hypothesis
\begin{align*}
\mathcal{E}_{\omega(0),A(0)}\big(\Psi(0)\big)&<\inf\big\{\mathcal{I}_{\bar \omega,\bar A}(K-1)\::\: \vert \bar \omega\vert \leq \Vert\omega\Vert_{L^\infty(\R^+)},\, \vert \bar A\vert \leq \Vert A\Vert_{L^\infty(\R^+)}\big\}\\
&\qquad-\Vert\omega\Vert_{L^\infty(\R^+)}-2\,\Vert A\Vert_{L^\infty(\R^+)}^2+\omega(0).
\end{align*}
There is a lot of room for improvement in the above argument. For example, if we assume  that, for all time,  the solution $\Psi=\pi(C,\Phi)\in \partial\mathcal{B}_{N,K}$ satisfies
\[
\langle \frac{\partial \mathcal H}{\partial t}\,\Psi(t)\big\vert\Psi(t)\rangle\leq h(t)\, \langle \mathcal{H}(t)\,\Psi(t)\big\vert\Psi(t)\rangle\]
%\fbox{Exemple? Chgt de variables ? }
for a given function $h$, then by the Gronwall lemma
\[
\langle \mathcal{H}(t)\,\Psi(t)\big\vert\Psi(t)\rangle-\langle \mathcal{H}(0)\,\Psi_0\big\vert\Psi_0\rangle\leq \exp\Big(\int_0^t h(s)\,ds\Big). \]   
The result of Theorem~\ref{mainth} remains true provided
\[
\mathcal{E}(\Psi_0)=\langle \mathcal{H}(0)\,\Psi_0\big\vert\Psi_0\rangle \leq \mathcal{I}(K-1)-\exp\Big(\int_0^{+\infty} h(s)\,ds\Big).\]

%%%%%%%%%%%%%%%%%%%%%%
\subsection{Discrete systems.} \label{sec:discrete} The emphasis has been but in particular for the functional analysis on the case when $\Omega = \R^3$  although in the first part we have described the problem in any open subset of $\R^3$. In fact all the formal and algebraic derivations can also be adapted to the case when $\Omega$ is a discrete set equipped with a discrete Lebesgue measure and in particular when $\Omega$ is a finite set~\cite{AML}.. Such situation is important for two reasons. On the one hand many models of quantum physics (the Ising model for instance) involve a discrete Hamiltonian defined on a discrete set. On the other hand  the discretization of the original problem  in view of any numerical algorithm leads to a discrete problem. 

Up to now only a rough   \textit{ a posteriori} error estimate has been proven. However if the MCTDHF algorithm is applied to a discrete model say of dimension $L$ then one always has   $K\leq  L$.  The error  formula~\eqref{error} shows that, for $K=L$,  the MCTDHF  algorithm is exact. It should be eventually observed that in general the two operations : - Discretization of the original $N$-particle problem and use of a MCTDHF approximation or - Use of a MCTDHF approximation and then discretization of the equations,  lead to different algorithms.
%%%%%%%%%%%%%%%%

%%%%%%%%%%% 
\section*{Appendix -- Proofs of technical lemmas in Subsection~\ref{gauge-transforms}}
\subsection*{Proofs of Corollary~\ref{tutile} and Lemma~\ref{utilG}}
For $\sigma$ and $\tau$ given and fixed $1\leq j\leq N$  it is convenient to denote by ${\mathbb U}_{\sigma, \tau(j)}$ the column vector in $\C^N$ with entries $\big({\mathbb U}_{\sigma(i), \tau(j)}\big)_{1\leq i\leq N}$ and by 
$$[{\mathbb U}_{\tau(1)}, {\mathbb U}_{\tau(2)},\ldots , {\mathbb U}_{\tau(N)}]_\sigma$$
the determinant composed with these vectors. With this notation (\ref{polar}) gives
\begin{equation}
i\frac{d{\mathbb U}_{\sigma, \tau(j)}}{dt}=\sum_{k=1}^K M_{k,\tau(j)}\,{\mathbb U}_{\sigma, k} \label{direct2}
\end{equation}
Differentiating  the relation
\begin{equation*}
{\mathbb U}_{\sigma, \tau}=[U_{\tau(1)}, U_{\tau(2)},\ldots , U_{\tau(N)}]_\sigma
\end{equation*}
and using the multi-linearity with respect to the column vectors and Eqn.~(\ref{direct2}) one obtains:
\begin{equation}
i\:\frac{d {\mathbb U}_{\sigma, \tau}}{dt}=\sum_{\substack{1\le k\le K\\1\le j \le N}} M_{k,\tau(j)} [U_{\tau(1)}, U_{\tau(2)},\ldots , U_{\tau(j-1)}, U_k, U_{\tau(j+1)},\ldots, U_{\tau(N)}]_\sigma.\label{direct3}
\end{equation}
On the other hand since $\mathbb U(t)$ is a flow of unitary matrices it is solution to  a differential equation of the following type:
\begin{equation} 
i\:\frac{d{\mathbb U}_{\sigma, \tau}}{dt}=\sum_{\tau'}[U_{\tau'(1)}, U_{\tau'(2)},\ldots , U_{\tau'(N)}]_{\sigma} {\tilde{\mathbb M}}_{\tau',\tau}\label{direct4}
\end{equation}
Identification of the coefficients of 
$$
[U_{\tau'(1)}, U_{\tau'(2)},\ldots , U_{\tau'(N)}]_{\sigma}
$$
gives, taking in account the number of permutation needed to change
$$\tau(1), \tau(2),\ldots \tau(j-1), k, \tau(j+1),\ldots \tau(N) \hbox{ into } \tau'(1),\tau'(2)\ldots , \tau'(N)$$
\[
{\tilde{ \mathbb M}}_{\tau',\tau}= \sum_{\substack{k\in \tau', j\in \tau\\ \tau'\setminus\{k\}=\tau\setminus\{j\}}} M_{k,j}(-1)^{\tau^{-1}(j)+ \tau'^{-1}(k)}.
\]
Let us now prove \eqref{truc3}. Let $\sigma,\tau\in \Sigma_{N,K}$. We first observe that 
\begin{equation}\label{Gdet}
\sum_{i=1}^N  \mathbf{G}_{x_i}\,\Phi_\sigma=\sum_{i=1}^N \phi_{\sigma(1)}\wedge\ldots\wedge\mathbf{G}\,\phi_{\sigma(i)}\wedge\ldots\wedge\phi_{\sigma(N)}.
\end{equation}
Now we use \eqref{ps-det} and the  Laplace  method  to develop a determinant with respect to the row that contains the terms involving $\mathbf{G}$ to get 
\begin{align*}
\sum_{i=1}^N  \big\langle\mathbf{G}_{x_i}\,\Phi_\sigma\big|\Phi_\tau\big\rangle
&=
\sum_{i=1}^N \big\langle \phi_{\sigma(1)}\wedge\ldots\wedge\mathbf{G}\,\phi_{\sigma(i)}\wedge\ldots\wedge\phi_{\sigma(N)}\big|\Phi_\tau\big\rangle\\
&=\sum_{i,j=1}^N(-1)^{i+j}
\langle \mathbf{G}\:\phi_{\sigma(i)},\phi_{\tau(j)}\rangle \,\delta_{\sigma\setminus\{\sigma(i)\}, \tau\setminus\{\tau(j)\}},
\end{align*}
in virtue of \eqref{ps-sigma-tau}. Hence \eqref{truc3} using \eqref{truc2} and  the definition of $M$. 
\hfill$\Box$
\vskip10pt
%%%%%%%%%
\subsection*{Proofs of Theorem~\ref{thm-gauge} and Theorem~\ref{gauge}.} 
Let $(C(t),\Phi(t))$ be a solution to $\mathcal{S}_0$ and let $\mathbf{G}$ be as in the statement of the theorem. With $M_{ij}=\langle \mathbf{G}\:\phi_i,\phi_j\rangle$ we define the family of unitary transforms $U(t)$ according to Lemma~\ref{utilG} and $d(U)(t)= \overline{\mathbb{U}}(t) $ is then given by Corollary~\ref{tutile}. We set $\mathbb{V}=\overline{\mathbb{U}}$,   $C'(t)=\V(t)\:C(t)$ and $\Phi'(t)=U(t)\Phi(t)$. Thanks to \eqref{basic1}, $\V$ solves 
\begin{equation}\label{basicV}
\left\{
\begin{aligned}
 i\frac{d{\mathbb V}}{dt}&=-{\mathbb V}\: \overline{{\mathbb M }},\\
 \mathbb{V}(0)&=d\big(U^0\big).
 \end{aligned}
 \right.
\end{equation} 
Then, for all $\sigma\in \Sigma_{N,K}$, 
\begin{align*}
 i\:\frac{d C'}{dt}&=i\:\frac{d\mathbb{V}}{dt}\,C+\mathbb{V}\: i\frac{d C}{dt}=-{\mathbb V}\: \overline{{\mathbb M }}\:\mathbb{V}^\star\:C'+\mathbb{V}\:\bigl\langle \mathcal{H}\:\Psi\:\big\vert\:\nabla_C\Psi\bigr\rangle\\
 &= -\mathbb{V}\:\overline{\mathbb{M}}\:\mathbb{V}^\star\:C'+\mathbb{V}\:\bigl\langle \mathcal{H}\:\Psi\:\big\vert\:\nabla_{C'}\Psi\,\mathbb{V}\bigr\rangle
\end{align*}
thanks to \eqref{transfo-grad} and \eqref{basicV}. On the one hand, since $\mathbb{V}$ is unitary, 
\[
\mathbb{V}\:\bigl\langle \mathcal{H}\:\Psi\:\big\vert\:\nabla_{C'}\Psi\,\mathbb{V}\bigr\rangle=\bigl\langle \mathcal{H}\:\Psi\:\big\vert\:\nabla_{C'}\Psi\bigr\rangle
.\]
On the other hand, when $M$  is obtained through $\mathbf{G}$, we get by a direct calculation from \eqref{truc3}
\[
\Big(\mathbb{V}\:\overline{\mathbb{M}}\:\mathbb{V}^\star\:C'\Big)_\sigma=\sum_{\tau}\Big\langle \sum_{i=1}^N\mathbf{G}_{x_i}\:\Phi_\tau'\big\vert \Phi'_\sigma\Big\rangle \:c'_\tau=\Big\langle \sum_{i=1}^N\mathbf{G}_{x_i}\:\Psi\Big\vert \Phi'_\sigma\Big\rangle.\]
Combining these two facts we get the first equation in $\mathcal{S}_\mathbf{G}$, namely 
\[
 i\:\frac{d C'}{dt}=\Big\langle\mathcal{H}\:\Psi\big\vert \nabla_{C'}\Psi\Big\rangle-\Big\langle \sum_{i=1}^N\mathbf{G}_{x_i}\:\Psi\big\vert \nabla_{C'}\Psi\Big\rangle.\]
We turn now to the equation satisfied by $\Phi'$. To simplify the notation we use the shorthand $\G$ for $\G(C)$ and $\G'$ for $\G(C')$ respectively. Then, using $\G'=U\:\G\:\U^{\star}$ and \eqref{polar}, we have 
\begin{align} 
i \:\G'\:\frac{\partial \Phi'}{\partial t} &= \G'\: i \:\frac{dU}{d t}\,\Phi+ \G'\:U\: i \:\frac{\partial \Phi}{\partial t}\nonumber\\
&= \G'\:UMU^{\star}\:\Phi'+ U\:\G\: i \:\frac{\partial \Phi}{\partial t}
\nonumber\\
&=\G'\:UMU^{\star}\:\Phi'+(\mathbf{I}-\mathbf{P}_{\Phi'}) \:U\:\nabla_\Phi \Psi^\star\bigl[\mathcal{H}\:\Psi\bigr]\nonumber \\
&=\G'\:UMU^{\star}\:\Phi'+(\mathbf{I}-\mathbf{P}_{\Phi'}) \:\nabla_{\Phi'}\Psi^\star\bigl[\mathcal{H}\:\Psi\bigr]\label{Phi'M}
\end{align}
thanks to \eqref{chainrule} and since clearly $\mathbf{P}_{\Phi'}=\mathbf{P}_{\Phi}$ for $\Span\{\Phi\}=\Span\{\Phi'\}$. It is easily checked that when $M$ is given through $\mathbf{G}$ we have 
\[
\big(UMU^{\star}\big)_{ij}=\langle\mathbf{G} \,\phi'_i,\phi'_j\rangle \]
and therefore 
\[
UMU^{\star}\:\Phi'=\mathbf{P}_{\Phi'}\,\mathbf{G}\,\Phi'.
\]
Hence \eqref{Phi'M} also writes 
\begin{equation*}
i\:\G' \:\frac{\partial  \Phi'}{\partial t}= \G' \:\mathbf{G} \Phi' + (\mathbf{I}-\mathbf{P}_{\Phi'})\:\nabla_{\Phi'}\Psi^\star\bigl[\mathcal{H}\:\Psi\bigr]-(\mathbf{I}-\mathbf{P}_{\Phi'})\:\G' \:\mathbf{G}\:\Phi'.\end{equation*}
We now check that, for all $1\leq i\leq N$, 
\[
(\mathbf{I}-\mathbf{P}_{\Phi'})\:\big(\G \:\mathbf{G}\:\Phi'\big)_i=(\mathbf{I}-\mathbf{P}_{\Phi'})\:\frac{\partial \Psi}{\partial \phi'_i}^\star\Bigl[\sum_{j=1}^N\mathbf{G}_{x_j}\:\Psi\Bigr]\,,
\]
thereby proving that 
\[
i\:\G' \:\frac{\partial  \Phi'}{\partial t}= \G' \:\mathbf{G}\: \Phi' + (\mathbf{I}-\mathbf{P}_{\Phi'})\:\nabla_{\Phi'}\Psi^\star\Bigl[\mathcal{H}\:\Psi-\sum_{i=1}^N\mathbf{G}_{x_i}\:\Psi\Bigr]
.\]
Indeed, for all $\xi\in \0$, using \eqref{eq:termsforVP} in Lemma~\ref{termsforVP} in \eqref{1} and using  \eqref{Gdet} in \eqref{2}, we have
\begin{align}
\big\langle\: (\mathbf{I}-\mathbf{P}_{\Phi})\:\big(\G \:\mathbf{G}\:\Phi\big)_i,\xi\big\rangle&=\sum_{k=1}^K\G_{ik}\:\big\langle \mathbf{G}\:\phi_k, (\mathbf{I}-\mathbf{P}_{\Phi})\:\xi\big\rangle\nonumber\\
&= \sum_{k=1}^K\Big\langle\: \frac{\partial \Psi}{\partial \phi_k}[\mathbf{G}\:\phi_j]\:|\:\frac{\partial \Psi}{\partial \phi_i}[(\mathbf{I}-\mathbf{P}_{\Phi})\:\xi]\:\Big\rangle\label{1}\\
&=\Big\langle\:\sum_{j=1}^N\mathbf{G}_{x_j}\:\Psi\:|\:\frac{\partial \Psi}{\partial \phi_i}[(\mathbf{I}-\mathbf{P}_{\Phi})\:\xi]\:\Big\rangle\label{2}\\
&=\Big\langle\: (\mathbf{I}-\mathbf{P}_{\Phi}) \:\frac{\partial \Psi}{\partial \phi_i}^\star\big[\sum_{j=1}^N\mathbf{G}_{x_j}\:\Psi\big]\:,\:\xi\:\Big\rangle\nonumber
\end{align}
by the definition \eqref{def-adjoint} of $\frac{\partial \Psi}{\partial \phi_i}^\star$; whence the result since $\xi$ is arbitrary in $\0$.
\hfill$\Box$

%%%%%%%%%%%%%%%%%%%%%%%%%%%%%%%%%%%%%%%%%%%%%%%
\section*{Acknowledgment}
This work was supported by the Austrian Science Foundation
(FWF) via the Wissenschaftkolleg ``Differential equations" (W17), by the Wiener Wissenschaftsfonds (WWTF
project MA 45) and the EU funded Marie Curie Early Stage Training Site DEASE (MEST-CT-2005-021122).
\vskip6pt
The authors  warmly acknowledge  Mathieu Lewin for a careful reading of a preliminary version of this work and for his valuable comments. They also would like to thank Alex Gottlieb for many discussions and suggestions. 

%%%%%%%%%%% REFERENCES

\end{document}